\documentclass[twocolumn]{aastex631}

\shorttitle{New Constraints on DMS and DMDS in K2-18~b using JWST MIRI}

\shortauthors{Madhusudhan et al.}

\graphicspath{{./}{figures/}}
\usepackage{comment}
\usepackage[T1]{fontenc}
\usepackage{amsmath}
\begin{document}

\title{New Constraints on DMS and DMDS in the Atmosphere of K2-18~b from JWST MIRI}

\author[0000-0002-4869-000X]{Nikku Madhusudhan}
 
\affiliation{Institute of Astronomy, University of Cambridge, Madingley Road, Cambridge CB3 0HA, UK}
\email{Correspondence: nmadhu@ast.cam.ac.uk}

\author[0000-0001- 6839-4569]{Savvas Constantinou}\altaffiliation{These authors contributed equally to this work; naming order is alphabetical.}\affiliation{Institute of Astronomy, University of Cambridge, Madingley Road, Cambridge CB3 0HA, UK}

\author[0000-0002-0931-735X]{M\aa ns Holmberg}\altaffiliation{These authors contributed equally to this work; naming order is alphabetical.}\affiliation{Space Telescope Science Institute, 3700 San Martin Drive, Baltimore, MD 21218, USA}

\author[0000-0002-2705-5402]{Subhajit Sarkar} \altaffiliation{These authors contributed equally to this work; naming order is alphabetical.}\affiliation{School of Physics and Astronomy, Cardiff University, The Parade, Cardiff CF24 3AA, UK}

\author[0000-0002-4487-5533]{Anjali A. A. Piette}
\affiliation{School of Physics and Astronomy, University of Birmingham, Birmingham, B15 2TT, UK}

\author[0000-0002-8837-0035]{Julianne I. Moses}
\affiliation{Space Science Institute, Boulder, CO 80301, USA}

\begin{abstract}
The sub-Neptune frontier has opened a new window into the rich diversity of planetary environments beyond the solar system. The possibility of hycean worlds, with planet-wide oceans and H$_2$-rich atmospheres, significantly expands and accelerates the search for habitable environments elsewhere. Recent JWST transmission spectroscopy of the candidate hycean world K2-18~b in the near-infrared led to the first detections of carbon-bearing molecules CH$_4$ and CO$_2$ in its atmosphere, with a composition consistent with predictions for hycean conditions. The observations also provided a tentative hint of dimethyl sulfide (DMS), a possible biosignature gas, but the inference was of low statistical significance. We report a mid-infrared transmission spectrum of K2-18~b obtained using the JWST MIRI LRS instrument in the $\sim$6-12~\textmu m range. The spectrum shows distinct features and is inconsistent with a featureless spectrum at 3.4-$\sigma$ significance compared to our canonical model. We find that the spectrum cannot be explained by most molecules predicted for K2-18~b with the exception of DMS and dimethyl disulfide (DMDS), also a potential biosignature gas. We report new independent evidence for DMS and/or DMDS in the atmosphere at 3-$\sigma$ significance, with high abundance ($\gtrsim$10 ppmv) of at least one of the two molecules. More observations are needed to increase the robustness of the findings and resolve the degeneracy between DMS and DMDS. The results also highlight the need for additional experimental and theoretical work to determine accurate cross sections of important biosignature gases and identify potential abiotic sources. We discuss the implications of the present findings for the possibility of biological activity on K2-18~b.
\end{abstract}


\section{Introduction} \label{sec:intro}

The discoveries of temperate exoplanets orbiting nearby stars and the advent of the James Webb Space Telescope \citep[JWST;][]{Gardner2006} are opening the possibility of detecting biosignatures in habitable exoplanets. While habitable exoplanets orbiting Sun-like stars are still beyond the reach of JWST, such planets orbiting smaller M dwarf stars are within the range of observability with JWST. A number of low-mass exoplanets orbiting M dwarfs have been observed in recent years, highlighting their diversity and both the challenges and opportunities in characterizing their atmospheres with JWST \citep[e.g.][]{Madhusudhan_2023_K218b, moran_high_2023, May2023, Lim2023, Alderson2024, holmberg_possible_2024, benneke_jwst_2024, damiano_lhs_2024, Scarsdale2024, Wallack2024, Alam2025}.

The Hycean paradigm developed in recent years has the potential to significantly expand and accelerate the search for life elsewhere \citep{Madhusudhan2021}. Hycean worlds are planets with habitable ocean-covered surfaces and H$_2$-rich atmospheres. Their lower densities, larger sizes and lighter atmospheres compared to Earth-like planets make hycean worlds more readily detectable and more conducive for atmospheric characterization. Similarly, their wider habitable zone compared to that of Earth-like planets also makes them more abundant, with over a dozen hycean candidates already identified \citep[e.g.][]{Madhusudhan2021,Fukui2022,Kawauchi2022,Evans2023,Piaulet2023}. 

Early observations with JWST have bolstered the promise of this new avenue, starting with the candidate hycean world K2-18~b. The planet has a mass of 8.63 $\pm$ 1.35 M$_\oplus$ and a radius of 2.61 $\pm$ 0.09 R$_\oplus$ \citep{Cloutier2019_K218b_mass, Benneke2019_K218b}, and orbits in the habitable-zone of an M2.5V star \citep{Montet2015, cloutier2017, sarkis2018}. The bulk parameters of the planet are consistent with a degenerate set of internal structures, including a hycean world, a mini-Neptune or a gas dwarf, i.e. a rocky planet with a thick H$_2$-rich atmosphere \citep{madhu2020}. Atmospheric observations are key to breaking the degeneracy.

A transmission spectrum of K2-18~b obtained with the Hubble Space Telescope WFC3/G141 spectrograph (1.1-1.7 $\mu$m) was initially used to infer the presence of water vapour (H$_2$O) in its atmosphere \citep{Benneke2019, Tsiaras2019, madhu2020}. However, significant degeneracies were found between potential absorption due to H$_2$O and that due to methane (CH$_4$) \citep{Blain2021, Bezard2022} or contributions from stellar heterogeneities \citep{Barclay2021}. Transmission spectroscopy of K2-18~b with JWST led to the first detections of carbon-bearing molecules, CH$_4$ and carbon dioxide (CO$_2$), at 5-$\sigma$ and 3-$\sigma$ significance, respectively, in its H$_2$-rich atmosphere \citep{Madhusudhan_2023_K218b}. The high sensitivity and wide wavelength range of the JWST spectrum helped resolve the previous CH$_4$-H$_2$O degeneracy, resulting in a strong detection of CH$_4$ and nondetection of H$_2$O. The nondetection of H$_2$O was consistent with the low photospheric temperature retrieved, implying H$_2$O condensation at the altitudes probed by the transmission spectrum \citep{Madhusudhan_chem_2023}.

The retrieved atmospheric composition of K2-18~b also provided important constraints on its internal structure. The detections of CH$_4$ and CO$_2$ at significant abundances, along with the nondetections of ammonia (NH$_3$) and carbon monoxide (CO) and the overall high CO$_2$/CO ratio, are consistent with prior predictions for a hycean atmosphere \citep{Hu_photo21, Tsai2021, Madhusudhan_chem_2023}. Other scenarios requiring a deep H$_2$-rich atmosphere, such as a mini-Neptune or a gas dwarf discussed above, are inconsistent with the retrieved abundances \citep{Madhusudhan_2023_K218b}. For example, a mini-Neptune scenario \citep[e.g.,][]{wogan_jwst_2024} is incompatible with most of the retrieved abundances, especially the low NH$_3$ and high CO$_2$/CO ratio \citep{glein_geochemical_2024, Cooke2024}. Similarly, models considering NH$_3$ depletion due to magma oceans \citep{shorttle_distinguishing_2024} in the gas dwarf scenario were found to be inconsistent with mass and density constraints, among other factors, while also being inconsistent with the retrieved composition \citep{Rigby_towards}. Therefore, presently, the atmospheric abundances of K2-18~b are best explained by a hycean world scenario and are incompatible with mini-Neptune or gas dwarf scenarios requiring a deep H$_2$-rich atmosphere. 

Open questions remain on the possibility of habitable conditions on K2-18~b. While the atmospheric composition is consistent with predictions for hycean conditions \citep{Madhusudhan_2023_K218b} and a large water inventory in the interior \citep{Yang_Hu_24, Luu2024}, the nature of the possible ocean beneath the H$_2$-rich atmosphere is unknown. A habitable liquid water ocean requires an adequate albedo (A$_{\rm B}$) due to clouds/hazes \citep{piette_temperature_2020,  Madhusudhan2021}, with the latest theoretical estimate of the required albedo being A$_{\rm B}$$>$0.5-0.6 \citep{leconte_3d_2024}, similar to that assumed for candidate hycean worlds \citep{Madhusudhan2021}. A cloud/haze-free atmosphere would render the surface too hot to be habitable and/or have water in a supercritical state \citep{madhu2020, piette_temperature_2020, Scheucher2020, innes_runaway_2023, Pierrehumbert2023, leconte_3d_2024}. While the required albedo may be consistent with the evidence for clouds/hazes reported at the day-night terminator of K2-18~b \citep{Madhusudhan_2023_K218b} and is within the range of A$_{\rm B}$ of 0.3-0.8 known for atmospheres of most solar system planets, the dayside albedo of K2-18~b has not been measured directly. 

On the other hand, recent studies have also indicated the potential for biotic conditions on K2-18~b. The CH$_4$ on K2-18~b may be contributed, partly or predominantly, from biogenic sources, similar to CH$_4$ from  methanogenic bacteria in the Earth's atmosphere \citep{Madhusudhan_chem_2023, Madhusudhan_2023_K218b, wogan_jwst_2024, Cooke2024}. In particular, the detection of abundant CH$_4$ alongside CO$_2$ in a shallow H$_2$-rich atmosphere is more easily explained by an inhabited hycean scenario than an uninhabited case \citep{wogan_jwst_2024, Cooke2024}. The CH$_4$-CO$_2$ pair has also been proposed as a promising biosignature for Earth-like habitable exoplanets, as may have been the case for the early Earth \citep{Krissansen-Totton2018}. However, the prospect of abiotically produced CH$_4$ through atmospheric chemistry cannot be ruled out in the uninhabited hycean scenario for K2-18~b \citep{Cooke2024}. Another potential  indication of biological activity was suggested by a weak ($\lesssim$2-$\sigma$) inference of dimethyl sulfide (DMS) in K2-18~b with previous JWST observations \citep{Madhusudhan_2023_K218b}.

The tentative inference of DMS in K2-18~b opens an important debate on the possible presence of life on K2-18~b. On the one hand, the low detection significance highlights the challenges in detecting such molecules. In the case of the previous JWST observations of K2-18~b, the detection significance of DMS depended on the relative offsets between the spectra observed from different detectors on the JWST NIRISS and NIRSpec instruments, ranging from 2.4-$\sigma$ with no offsets to below 1-$\sigma$ for two offsets \citep{Madhusudhan_2023_K218b}. Another challenge is the strong degeneracy between the spectral features of DMS near 3.3 \textmu m and 4.3 \textmu m with strong features of CH$_4$ and CO$_2$ at overlapping wavelengths, as well as potential contributions from other hydrocarbons with strong features in the 3-5 \textmu m range \citep{Madhusudhan_2023_K218b, Tsai24_Bio}. On the other hand, a confident detection of a molecule like DMS would serve as a more robust biosignature than molecules like CH$_4$, which are more easily detectable but may be present in abundance through abiotic chemistry. 

The robustness of DMS as a biosignature in H$_2$-rich environments has been proposed extensively in the literature both for rocky planets \citep{DomagalGoldman2011, Seager2013a, Catling2018, Schwieterman2018} and hycean worlds \citep{Madhusudhan2021}. Despite the low detection significance, the reported abundance constraints of DMS are physically plausible for realistic levels of biogenic sources \citep{Madhusudhan_2023_K218b, Tsai24_Bio}. In particular, DMS mixing ratios as high as 10$^{-2}$ are possible in K2-18~b for high biogenic fluxes of sulfur-based biosignature gases above $\sim$20$\times$ Earth levels \citep{Tsai24_Bio}. 

In this work, we conduct an independent search for molecular species, including DMS, in K2-18~b in a different wavelength range using the JWST MIRI spectrograph. As discussed above, the previous tentative inference of DMS in K2-18~b was made using a near-infrared transmission spectrum in the 1-5 $\mu$m range obtained with the JWST NIRISS and NIRSpec instruments. However, the evidence for DMS was affected by potential flux offsets between the different detectors \citep{Madhusudhan_2023_K218b}. Therefore, an independent search for DMS and other such molecules using a different instrument in a complementary spectral range is invaluable to assess both the significance of prior findings and to provide an independent line of evidence. Mid-infrared spectroscopy with JWST provides a promising avenue in this direction. In addition to providing a complementary spectral window ($\sim$5-12 \textmu m) to previous observations, this wavelength range also encompasses strong spectral features of DMS and several other biosignature gases \citep{DomagalGoldman2011, Seager2013b, Schwieterman2018, Tsai24_Bio}.

We present a mid-infrared transmission spectrum of K2-18~b with JWST, the first for a habitable-zone exoplanet. This allows for an independent search for DMS and other biosignature gases in the atmosphere of K2-18~b as discussed above. In what follows, we present our observations, data reduction and light curve analyses in section~\ref{sec:obs}. We discuss our retrieval approach and present the atmospheric inferences obtained from the transmission spectrum in section~\ref{sec:retrieval}. We summarize our results and discuss the implications in section~\ref{sec:discussion}.

\section{Observations and Data Reduction} 
\label{sec:obs}

We report a mid-infrared transmission spectrum of K2-18~b using the JWST MIRI low-resolution spectrograph (LRS) \citep{Kendrew2015, Bouwman2023}. The observations were conducted as part of JWST GO Program 2722 (PI: N. Madhusudhan).
The observations were made in the slitless prism configuration with the F560W filter and the FASTR1 readout pattern. The target acquisition was conducted on the science target, the host star K2-18, which is a M2.5V dwarf star with J mag of 9.763 \citep{cloutier2017}.  The science exposure was performed between 23:13:29 UTC on April 25 2024 and 05:04:37 UTC on April 26 2024, for a duration of 5.85 hours with an in-transit duration of 2.68 hours and the remaining time providing the out-of-transit baseline.  The science observation consists of a total of 5095 integrations with 25 groups per integration.   
The maximum fraction of saturation across the detector was calculated to be 51\% using the JWST Exposure Time Calculator.
The spectrum spans a wavelength range of $\sim$5-12  \textmu m with an average native resolution of R $\sim$ 100. No high gain antenna (HGA) movement was reported during the observations.  
We carry out the data reduction and light-curve analysis in two parallel efforts, using independent pipelines and sensitivity analyses, as described below.

\begin{figure}
	\includegraphics[width=0.48\textwidth]{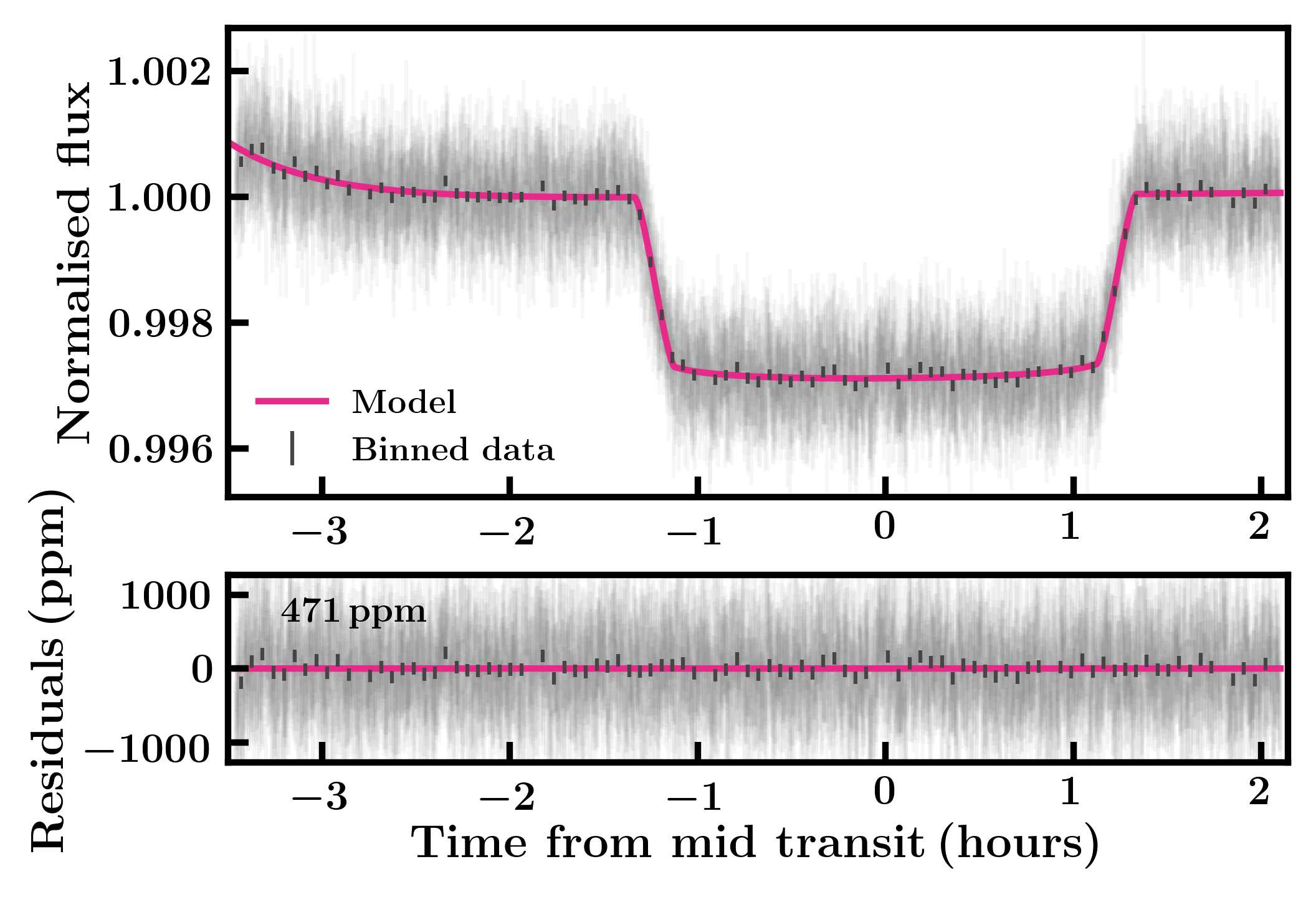}
    \caption{MIRI LRS white light curve of K2-18~b. The light curve is based on the time series spectroscopic data binned between 4.8-10~\textmu m and discarding the first 250 integrations. The top panel shows the white light curve with and without binning, together with the median model fit. The lower panel shows the residuals after subtracting the model. The standard deviation of the residuals without binning is 471 ppm -- corresponding to 1.25$\times$ the expected noise level from photon noise and read noise.}
    \label{fig:wlc_miri} 
\end{figure}

\begin{table*}
\centering
\begin{tabular}{lcccc}
\hline \hline
Parameter & \texttt{JExoRES} & \texttt{JexoPipe} \\ \hline
Mid-transit time, T$_0$ (BJD - 2400000.5) & $60426.128786_{-0.000082}^{+0.000082}$ &  $60426.128853_{-0.000080}^{+0.000081}$   \\ 
Inclination, $i$ ($^{\circ}$) & $89.5598_{-0.0066}^{+0.0068}$ & $89.5600_{-0.0067}^{+ 0.0067}$   \\ 
Normalised semi-major axis, $a / R_*$ & $80.34_{-0.45}^{+0.47}$ & $   80.39_{-0.47}^{+0.46}$   \\ 
Planet-to-star radius ratio, $R_\mathrm{p} / R_*$ & $0.05332_{-0.00018}^{+0.00017}$ & $0.05352_{-0.00019}^{+0.00018}$  \\ 
First limb-darkening coefficient, $u_1$ & $0.037_{-0.028}^{+0.056}$ & $0.033 _{-0.024}^{+0.046}$   \\ 
Second limb-darkening coefficient, $u_2$ & $0.114_{-0.105}^{+0.047}$ & $0.071_{-0.058}^{+0.046}$   \\ \hline
\end{tabular}
\caption{Parameters estimated from the white light curve analysis of our JWST MIRI LRS observation of K2-18 b. The results are shown from the two independent data analyses. In both cases, the period was fixed to 32.940045 days \citep{Benneke2019_K218b} and wavelengths between 4.8-10 \textmu m were used.
}
\label{tab:wlc_params}
\end{table*}

\subsection{JExoRES Pipeline} \label{sec:JExoRES}
We employ a modified version of the \texttt{JExoRES} pipeline \citep{holmberg2023, Madhusudhan_2023_K218b} adapted to MIRI LRS. First, we perform Stages~1-2 using the JWST Science Calibration Pipeline \citep{Bushouse2020} to perform calibrations and the detector ramp fitting. In Stage 1, we perform the data quality initialization, electromagnetic interference (EMI) correction, saturation flagging, first and last frame flagging, linearity correction, reset switch charge decay (RSCD) correction, dark current subtraction, and ramp fitting. For this, we use both the standard linearity and RSCD corrections and a custom approach detailed in Appendix~\ref{app:linearity}, as pursued by several recent works \citep{Kempton2023, grant_jwst-tst_2023, dyrek_so2_2024}. Overall, we find that the approach to non-linearity correction has a minimal effect on the transmission spectrum of K2-18~b, as discussed in  Appendix~\ref{app:linearity} and illustrated in Figure~\ref{fig:linearity}. Next, we perform the flat-field correction in Stage~2. 

In Stage 3, we first use the gain reference file to convert the flux from DN/s to e$^-$/s. We then search for cosmic ray hits by performing sigma-clipping on the time-series of each pixel by first removing a running median of seven integrations and using a threshold of 7-$\sigma$. In addition, we also mask neighboring pixels surrounding detected outliers as well as pixels which have been flagged with any issue during Stages~1-2. We then correct for the background by subtracting the mean of the flux outside the trace, for each detector column and integration. To estimate the background, we use columns with pixel numbers 11-30 and 44-63, while not including bad pixels or outliers. Finally, we extract the spectra with optimal extraction \citep{horne_optimal_1986} using the median of all integrations as a model for the point spread function for each spectral channel. For this, we use an aperture of 9 pixels. A box extraction approach produces very similar results, as shown in Figure~\ref{fig:box}. We reject additional outliers during optimal extraction and then in the light curves themselves using sigma-clipping with a running median filter. 

Next, we fit the light curves using the \texttt{batman} transit model \citep{kreidberg_batman_2015} and nested sampling with \texttt{MultiNest} \citep{Feroz2009}, assuming a circular orbit with a period from \cite{Benneke2019_K218b}. We first construct a white light curve by integrating the flux between 4.8-10~\textmu m. Using this wavelength range gave slightly lower (5\%) scatter compared to the full 4.8-12~\textmu m range. For fitting the light curve we use normally distributed priors on the normalised semi-major axis $a/R_*$ and the inclination $i$ from the weighted average of the NIRISS and NIRSpec fits from \cite{Madhusudhan_2023_K218b}. These orbital parameters are also consistent with pre-JWST measurements \citep{Tsiaras2019, Benneke2019_K218b}. For the limb darkening, we use the quadratic law with the parameterisation and priors by \cite{Kipping2013}. We use wide uniform priors on all other parameters. For the baseline trend, we adopt an exponential and a linear component: $ F_{obs}(t) = F_{\mathrm{out}}\,(1 + \alpha \tau + \gamma e^{-\tau / \epsilon})\,  F_{\mathrm{transit}}(t)$, where $\tau$ is the time since the start of the observation, $F_{\mathrm{transit}}$ is the \texttt{batman} transit model, and $F_{\mathrm{out}}$, $\alpha$, $\gamma$ and $\epsilon$ are trend parameters. Furthermore, we mask the first 250 integrations (1034 seconds) to remove the strongest effect of the detector settling. Figure~\ref{fig:wlc_miri} shows the MIRI white light curve of K2-18~b together with our model fit. Moreover, we also fit for an error inflation parameter to account for additional white noise. The parameters measured from the white light curve fitting are shown in Table~\ref{tab:wlc_params}.  

\begin{figure*}
	\includegraphics[width=1.0\textwidth]{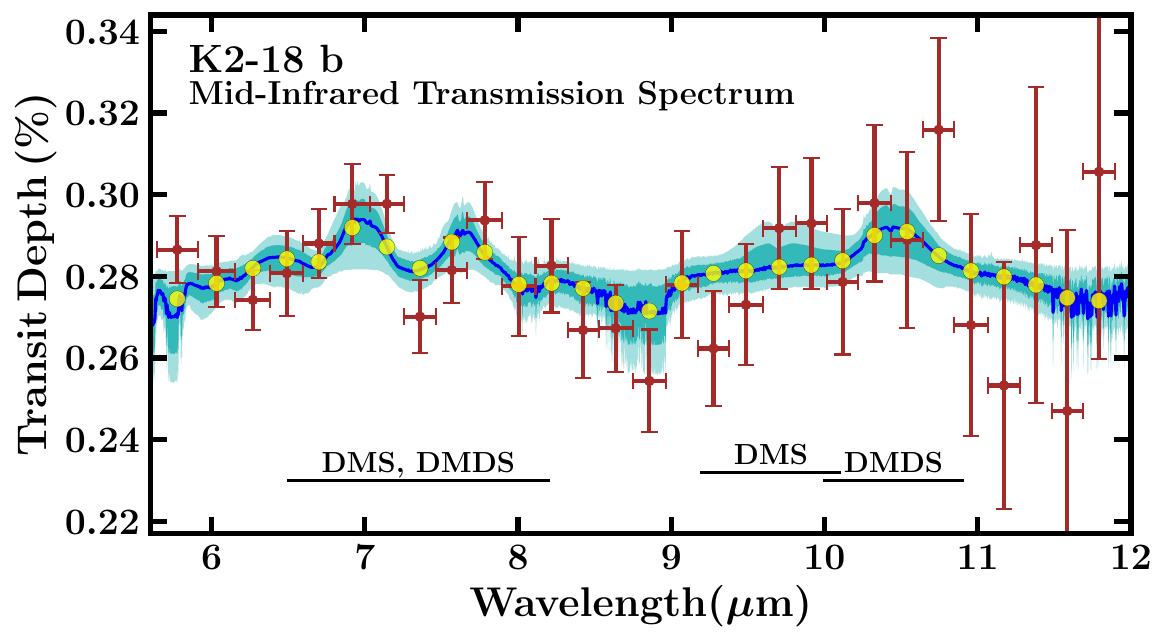}
    \caption{The mid-infrared transmission spectrum of K2-18~b obtained with the JWST MIRI LRS instrument. The data points with error bars (in brown) show the observed spectrum as described in section~\ref{sec:JExoRES}. The horizontal errorbars correspond to the spectral bin width. The dark blue curve denotes the median retrieved spectral fit, while the two lighter shaded regions denote the 1- and 2-$\sigma$ intervals. The prominent features of DMDS and DMS are identified. Both molecules have overlapping spectral features between 6.8-8 \textmu m, with broader features between $\sim$9-10 \textmu m for DMS and $\sim$10-11 \textmu m for DMDS. The individual spectral contributions of these molecules are shown in Figure~\ref{fig:contributions}.}
    \label{fig:spectral_fit} 
\end{figure*}

Finally, we perform the spectroscopic light curve fitting, fixing $a/R_*$, $i$, the mid-transit time, and the two limb-darkening coefficients to the values obtained from the white light curve. We bin the light curves in wavelength prior to fitting, nominally with a width of 0.2~\textmu m or 5 pixels, whichever contains the most pixels, as described in Appendix~\ref{app:binning}. A minimum bin width of 0.2~\textmu m is in line with the bin widths used in previous works, ranging between 0.15-0.5~\textmu m \citep[e.g.][]{Bouwman2023, grant_jwst-tst_2023, Bell2024, Powell2024}. Due to instability of the spectrum below 5.6~\textmu m as a function of binning, we disregard this part from further analysis to be conservative, as discussed in Appendix \ref{app:binning}. We again mask the first 250 integrations. We explore different choices of trends in Appendix~\ref{app:trends} and find that an exponential and a linear component produce a stable transmission spectrum when varying the number of integrations to mask at the start. Furthermore, we find evidence for time-correlated noise on the time scale of minutes, with increasing amplitude toward shorter wavelengths. Accounting for this additional noise using Gaussian processes (GPs), as described in Appendix~\ref{app:noise}, we find that the transit depth uncertainty increases by around 30\% at 6~\textmu m, and less at longer wavelengths. We also test different assumptions for the limb darkening and find that it does not significantly alter the transmission spectrum, as shown in Appendix~\ref{app:limb_darkening}. The resulting MIRI transmission spectrum of K2-18~b is shown in Figure~\ref{fig:spectral_fit}.

\begin{figure*}
	\includegraphics[width=1.0\textwidth]{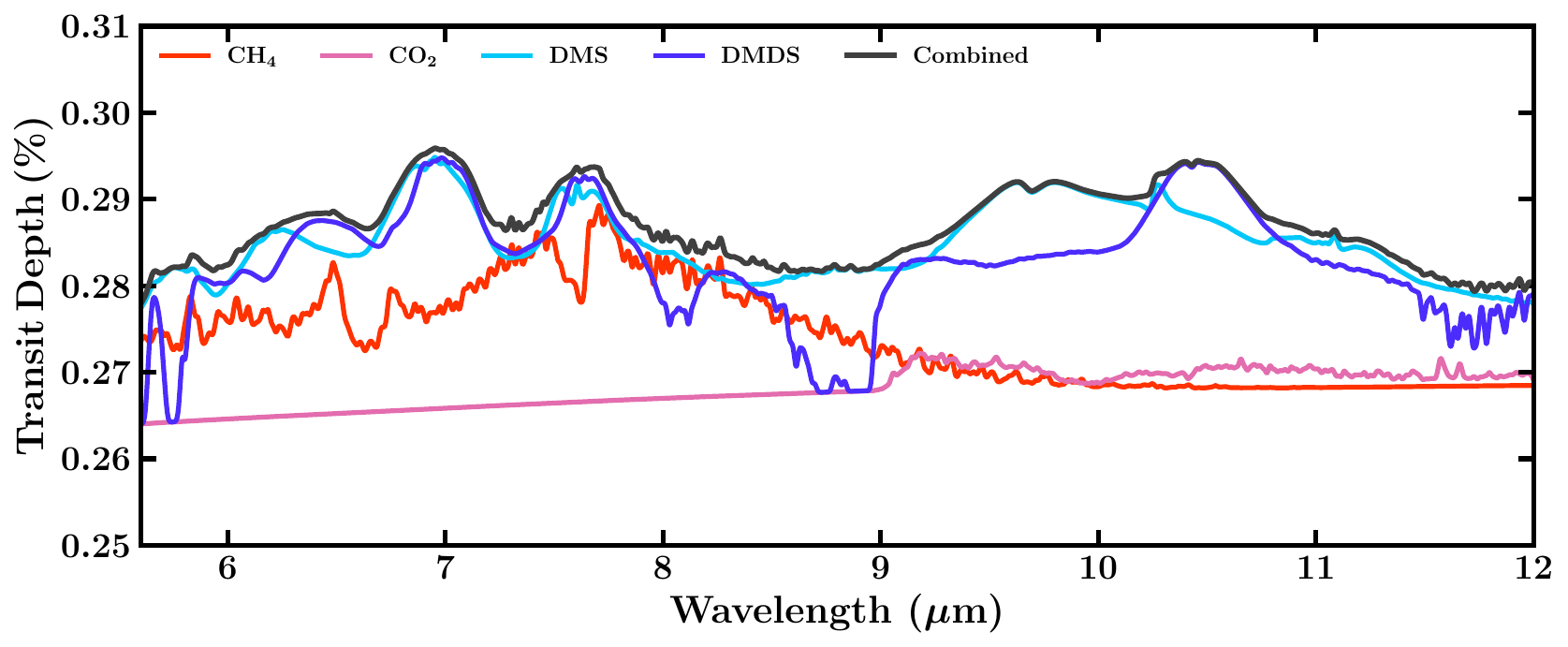}
    \caption{Spectral contributions of notable chemical species in the MIRI band. Each curve denotes the spectral contribution of a particular molecule to the model spectrum, as denoted in the legend. The mixing ratios of DMS and DMDS are set to a representative value of $5 \times 10^{-4}$, while CH$_4$ and CO$_2$ are set to $10^{-2}$, consistent with constraints obtained from previous near-infrared observations \citep{Madhusudhan_2023_K218b}. The black line denotes the resulting transmission spectrum with all molecular contributions combined. The individual spectral contributions of these molecules are shown in Figure \ref{fig:contributions}.}
    \label{fig:contributions} 
\end{figure*}

\subsection{JexoPipe Pipeline} \label{sec:JexoPipe}
We use a second pipeline, \texttt{JexoPipe}, a JWST data reduction framework previously applied to NIRSpec Prism and G395H data \citep{Sarkar2024} and adapted here for MIRI LRS. Starting with the .uncal files, in stage 1 we utilise the official JWST Science Calibration Pipeline steps in the following sequence: group scale, data quality array initialisation, EMI correction, saturation flagging, first frame, last frame, reset, linearity correction, RSCD correction, dark subtraction, jump detection, ramp fitting and gain scale. Stage 1 ends with integration level .rateints files. After combining all stage 1 segments into a single .rateints file, we begin stage 2 by applying the JWST calibration pipeline steps: assign WCS and flat field. We then apply  a custom bad pixel flagging step \citep{Sarkar2024}.  This step applies NaN values to all pixels that have abnormal DQ flags as well as to 3-$\sigma$ outliers found on a row-by-row basis in each integration image.
Next, we perform background subtraction per integration, using pixel columns 10-14 and 60-66. We apply an outlier mask, and then subtract the mean of each row from all pixel values in that row.

We then apply a custom bad pixel correction step using a combination of temporal and spatial interpolation as described in \cite{Sarkar2024}. Next, we detect remaining outliers on each integration by comparison with a rolling median (of 20 contiguous integrations), replacing pixel values $\pm$ 5-$\sigma$ from the rolling median with the median value. For stage 3, 1-D spectral extraction, for each integration, we apply an aperture of 9 pixel columns centred on the spectrum maximum and then perform optimal extraction \citep{horne_optimal_1986}. We obtain the wavelength for each pixel row from the mean of the wavelengths in that row after application of the aperture.

We use the 1-D spectra time series to construct a white light curve  (between 4.8-10 \textmu m). We exclude the first 250 integrations, where the systematic trend is most extreme. We identify outliers on the white light curve, $\pm$ 2.5-$\sigma$ from a rolling median, and replace the 1-D spectra corresponding to these outliers with linearly interpolated spectra from adjacent integrations. We scale the error bars on the light curve points such that the average error bar equals the observed standard deviation of the scatter in the out-of-transit residuals. We use \texttt{emcee} \citep{foreman-mackey_emcee_2013} to perform a Markov Chain Monte Carlo parameter estimation of the white light curve, fitting for a transit model with quadratic limb-darkening generated by \texttt{pylightcurve} \citep{tsiaras_2016_pylightcurve} multiplied by a systematic trend consisting of an exponential term and a linear term (as in section \ref{sec:JExoRES}).  In the white light curve, we fit for $R_p/R_*$, mid-transit time, $a/R_*$, $i$,  quadratic limb-darkening coefficients and four parameters for the trend. Uniform priors are used except for $a/R_*$ and $i$, where we apply Gaussian priors based on values in \cite{Madhusudhan_2023_K218b} and use the Kipping parameterisation \citep{Kipping2013} for limb-darkening priors. We fix the period to 32.940045 days \citep{Benneke2019_K218b}, the argument of periastron to 90$^o$ and the eccentricity to 0. The white light 
curve parameter estimates are given in Table \ref{tab:wlc_params}.  

We bin the spectral light curves following the same prescription as in section \ref{sec:JExoRES}.  We fit the spectral light curves in a similar manner to the white light curve, with the exception of the mid-transit time, $a/R_*$ and $i$ which we fix to the white light values. We also fix the limb darkening coefficients to those from the white light fit. To check the sensitivity of the transmission spectrum to the treatment of limb darkening, we also adopted fixed wavelength-dependent limb darkening coefficients from ExoTic-LD \citep[Kurucz model;][]{Grant2024_exoticLD}. In doing so, we find that the spectra from the two treatments are consistent to well within the 1-$\sigma$ uncertainties.

For each spectral light curve fit, the free parameters are $R_p/R_s$ and the four systematic trend parameters.  To account for time-correlated noise, we obtain the $ \beta $ factor for each spectral light curve using the `time-averaging' method to produce Allan deviation plots \citep{Winn2007,Cubillos2017}. The spectral light curve fits are then re-run after inflating the uncertainties on the original light curves by the corresponding $ \beta$ value. The final transmission spectrum is shown in Figure \ref{fig:pipelines}.

\subsection{Robustness of the Transmission Spectrum}

\begin{figure*}
	\includegraphics[width=1.0\textwidth]{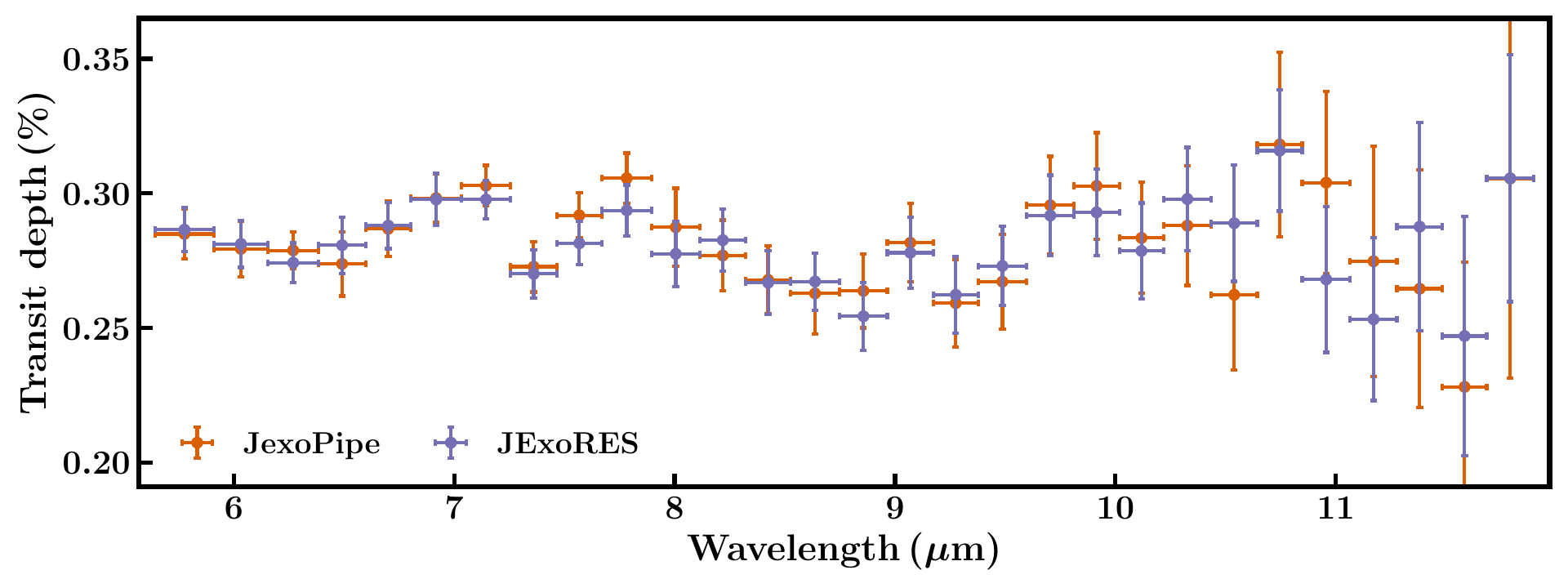}
    \caption{Demonstration of the stability of the MIRI transmission spectrum of K2-18~b using two independent data reduction pipelines. The spectra from the \texttt{JExoRES} and \texttt{JexoPipe} pipelines are shown in purple and orange, respectively, and are further described in sections~\ref{sec:JExoRES} and~\ref{sec:JexoPipe}. }
    \label{fig:pipelines} 
\end{figure*}

The transmission spectra obtained from the two pipelines show good agreement, as shown in Figure ~\ref{fig:pipelines}, with all the data points agreeing within the 1-$\sigma$ uncertainties. While cross-checks between pipelines are useful to ensure consistency, it is also important to check the robustness and sensitivity of the transmission spectrum obtained using a given pipeline \citep{Holmberg2024}. In the present work, we conduct robustness tests with both pipelines across various considerations in the data reduction and analyses. We discuss some of these results in Appendix \ref{Robustness of Transmission Spectra}. We demonstrate the robustness of the resultant transmission spectra to a wide range of parameters and assumptions in data reduction and light curve analyses. These include different treatments for the non-linearity correction, spectral extraction, spectral binning and limb-darkening, and different trends in light curve fitting, as well accounting for correlated noise.

\begin{figure*}
	\includegraphics[width=1.0\textwidth]{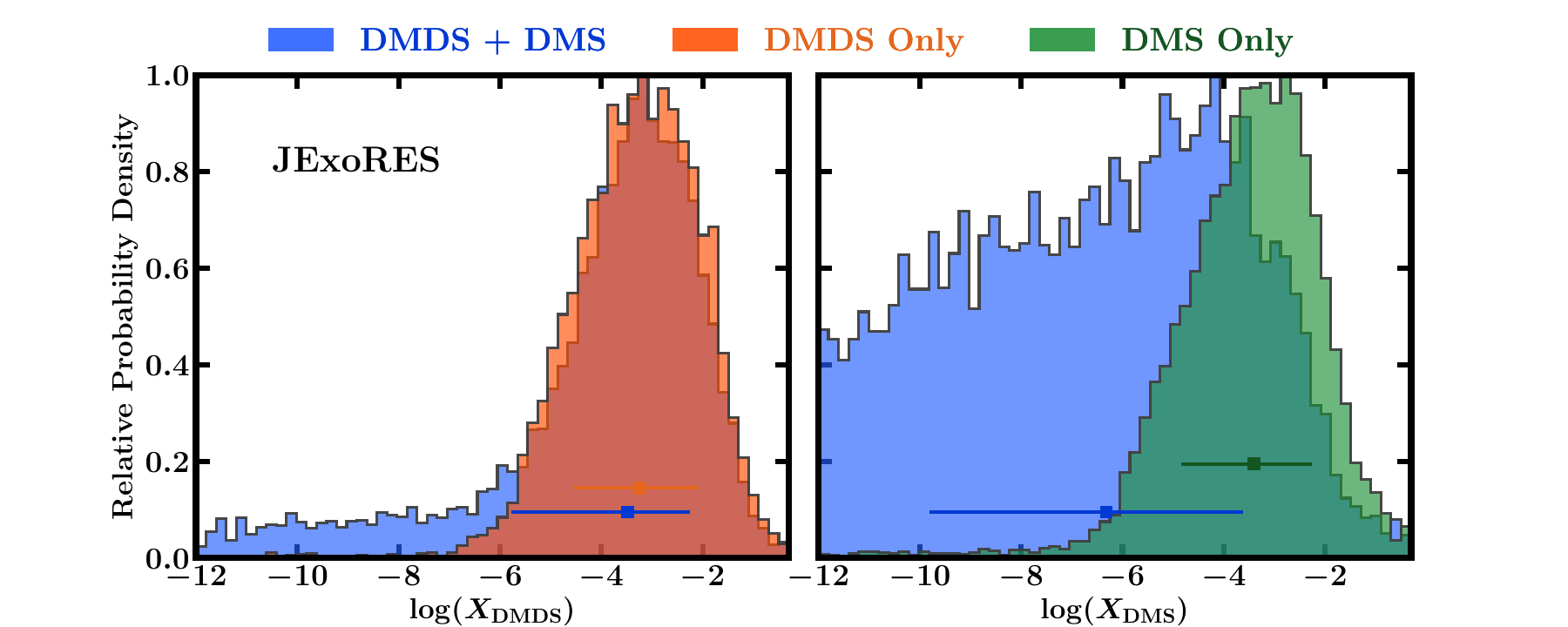}
    \includegraphics[width=1.0\textwidth]{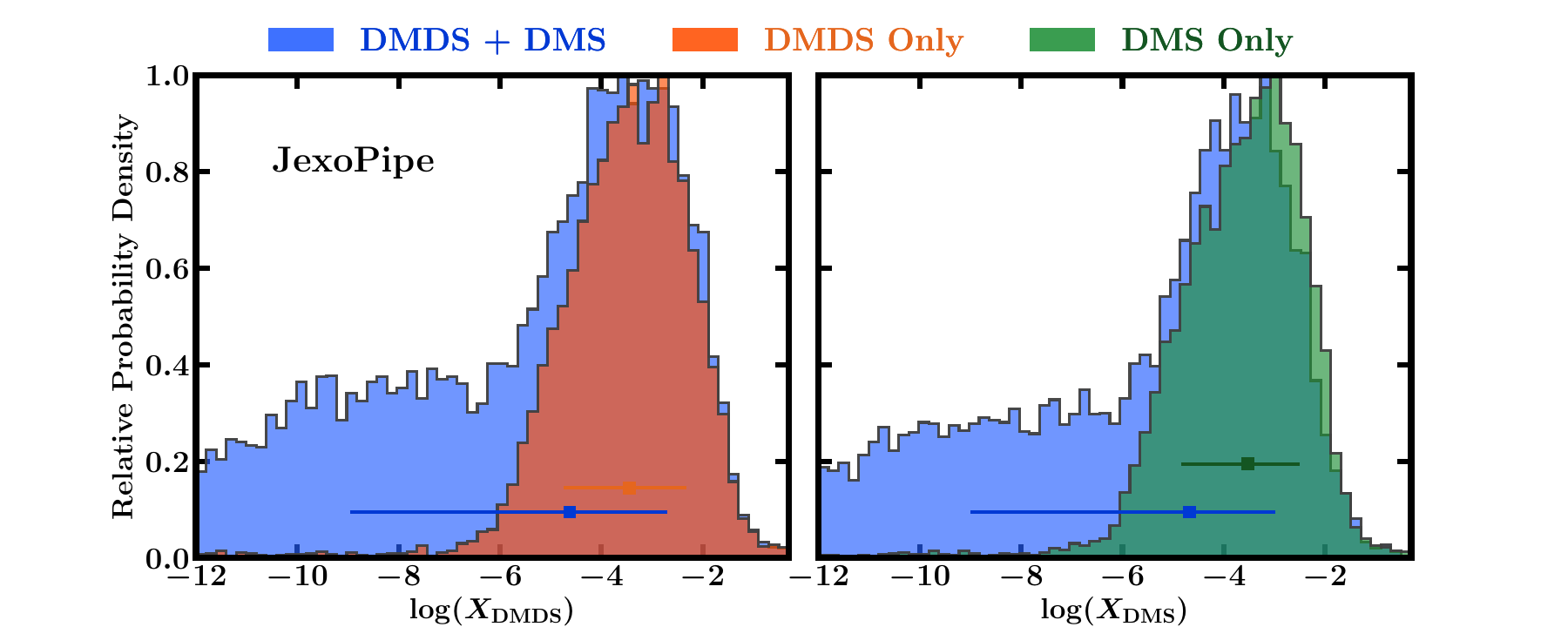}
    \caption{Retrieved posterior probability distributions for DMDS and DMS from our canonical retrievals described in Section \ref{sec:retrieval},  using data from the \texttt{JExoRES} (top) and \texttt{JexoPipe} (bottom) pipelines. The posteriors shown in blue correspond to the canonical retrieval containing both DMDS and DMS, along with CH$_4$ and CO$_2$. The other two cases show the retrievals with only one of the two molecules, DMDS or DMS, included in the canonical retrieval with all other parameters remaining unchanged. The orange distribution corresponds to the retrieval with only DMDS present and the green distribution corresponds to the retrieval with only DMS present.}
    \label{fig:posteriors} 
\end{figure*}

The resulting spectra are consistent within the uncertainties across all these cases, as shown in the Appendix \ref{Robustness of Transmission Spectra}. The corresponding atmospheric constraints are also consistent across these cases, as shown in Table~\ref{tab:abundances}. We also conduct atmospheric retrievals with spectra from both \texttt{JExoRES}  and \texttt{JexoPipe} to ensure consistency in the overall conclusions. The abundances and detection significances obtained from each reduction are summarised in Table \ref{tab:abundances} and posterior distributions are shown in Figure \ref{fig:posteriors}. This is further discussed below.

\begin{table*}
\centering

\caption{Retrieved abundance estimates and detection significances from the MIRI transmission spectrum of K2-18~b.}

\begin{tabular}{lcccccr}
\hline \hline
Canonical Model & Data & DMS & DMDS & $\ln$($\mathcal{Z}$) & $\ln$($B$) & Detection Significance \\
\hline 
DMS Only   & JExoRES & $-3.42_{-1.44}^{+1.16}$ & - & 215.45 & 2.86 & DMS (2.9 $\sigma$) \\
DMDS Only  & JExoRES & - & $-3.25_{-1.30}^{+1.17}$ & 216.40 & 3.81 & DMDS (3.2 $\sigma$) \\
DMS + DMDS & JExoRES$^1$ & $<$ -2.44 & $-3.48_{-2.27}^{+1.24}$ & 216.40 & 3.80 & DMS+DMDS (3.2 $\sigma$) \\

DMS Only   & JexoPipe & $-3.53_{-1.33}^{+1.03}$ & - & 211.18 & 3.18 & DMS (3.0 $\sigma$) \\
DMDS Only  & JexoPipe & - & $-3.45_{-1.30}^{+1.13}$ & 211.18 & 3.17 & DMDS (3.0 $\sigma$) \\
DMS + DMDS & JexoPipe & -4.68$_{-4.34}^{+1.72}$ & $-4.61_{-4.36}^{+1.93}$ & 211.59 & 3.59 & DMS+DMDS (3.2 $\sigma$) \\
\hline
\multicolumn{7}{c}{Effect of Trend$^{2}$}\\
\hline
DMS + DMDS & JExoRES (Exp+Linear1)$^1$ & $<$ -2.44 & $-3.48_{-2.27}^{+1.24}$ & 216.40 & 3.81 & DMS+DMDS (3.2 $\sigma$) \\
DMS + DMDS & JExoRES (Exp+Linear2) & $<$ -1.81 & $-3.67_{-3.51}^{+1.59}$ & 216.40 & 2.89 & DMS+DMDS (2.9 $\sigma$) \\
DMS + DMDS & JExoRES (Exp+Quadratic) & $<$ -2.42 & $-3.70_{-2.82}^{+1.32}$ & 212.92 & 2.85 & DMS+DMDS (2.9 $\sigma$) \\
DMS + DMDS & JExoRES (Quadratic) & $-4.56_{-3.90}^{+1.52}$ & $-4.89_{-4.24}^{+1.87}$ & 214.50 & 4.22 & DMS+DMDS (3.4 $\sigma$) \\
\hline
\multicolumn{7}{c}{Masking of Potential Shadow Region$^{3}$}\\
\hline
DMS + DMDS & JExoRES ($\lambda$ < 10.5 \textmu m) & $<$ -2.10 & $-3.88_{-4.24}^{+1.56}$ & 168.44 & 3.51 & DMS+DMDS (3.1 $\sigma$) \\
DMS + DMDS & JExoRES ($\lambda$ < 10 \textmu m) & $<$ -2.15 & $-3.46_{-2.06}^{+1.23}$ & 153.20 & 3.09 & DMS+DMDS (3.0 $\sigma$) \\
\hline
\multicolumn{7}{c}{Accounting for Time-correlated Noise}\\
\hline
DMS + DMDS & JExoRES (no GP) & $<$ -2.16 & $-3.72_{-3.14}^{+1.38}$ & 215.94 & 4.08 & DMS+DMDS (3.3 $\sigma$) \\
DMS + DMDS & JExoRES (GP)$^\star$ & $<$ -2.44 & $-3.48_{-2.27}^{+1.24}$ & 216.40 & 3.81 & DMS+DMDS (3.2 $\sigma$) \\
\hline
\multicolumn{7}{c}{Effect of Binning$^4$}\\
\hline
DMS + DMDS & JExoRES (0.2 \textmu m) & $<$ -2.16 & $-3.72_{-3.14}^{+1.38}$ & 215.94 & 4.08 & DMS+DMDS (3.3 $\sigma$) \\
DMS + DMDS & JExoRES (0.4 \textmu m) & $<$ -2.44 & $-2.88_{-1.97}^{+0.94}$ & 112.72 & 3.98 & DMS+DMDS (3.3 $\sigma$) \\
DMS + DMDS & JExoRES (0.8 \textmu m) & $<$ -2.87 & $-3.06_{-1.34}^{+0.93}$ & 63.71 & 3.79 & DMS+DMDS (3.2 $\sigma$) \\
\hline
\end{tabular}

\vspace{1mm}
\begin{flushleft}
{\bf Note:} The first column shows the canonical model used with DMS and/or DMDS included. The second column shows the spectrum used from one of the two pipelines \texttt{JExoRES} and \texttt{JexoPipe}, along with any specific treatments in the data reduction or light curve analysis. The DMS and DMDS abundances in the third and fourth columns are shown as $\log_{10}$ of the volume mixing ratios. The retrieved abundances are reported as median values with the 1-$\sigma$ uncertainties for cases with a well defined peak in the posterior or as 95\% upper-limits otherwise. The model evidence is shown for each case as $\ln$($\mathcal{Z}$), where $\mathcal{Z}$ is the Bayesian evidence. $\ln$($B$) refers to the natural logarithm of the Bayes factor, comparing the canonical model with a model with DMS and/or DMDS removed. The corresponding molecules are shown in the last column with their detection significances in parentheses. We note that there is a typical uncertainty of $\sim$0.1 $\sigma$ in the calculation of the detection significance using MultiNest. GP indicates analysis using Gaussian processes to account for time-correlated noise, as outlined in Appendix~\ref{app:noise}. \\

 $^{1}$These correspond to our canonical case presented in Figure~\ref{fig:spectral_fit}. \\

 $^{2}$The trend refers to systematic trend used in the light curve analysis, as explored in Figure~\ref{fig:trends}. For the two Exp+Linear cases, we remove the first 250 and 500 integrations for Linear1 and Linear2, respectively. \\

 $^{3}$We do not find a strong discontinuity in the detector behaviour past 10.6~\textmu m, called the shadow region, sometimes found in MIRI LRS data \citep{Bell2024}. Nevertheless, we conduct retrievals with different wavelength limits for robustness.\\

 $^{4}$These cases correspond to the spectra presented in Figure~\ref{fig:binning}, which explore the effect of binning. These cases do not use the GP model. For the 0.2~\textmu m case, we used a minimum bin width of 5 pixels, representing our nominal binning as discussed in Section \ref{sec:JExoRES}.
\end{flushleft}

\label{tab:abundances}
\end{table*}

\section{Atmospheric Retrieval} \label{sec:retrieval}
We conduct atmospheric retrievals to derive the properties of the day-night terminator region of the atmosphere of K2-18~b using the MIRI transmission spectrum. We follow the retrieval approach of the previous analyses of JWST observations of K2-18~b \citep{Madhusudhan_2023_K218b}. We perform atmospheric retrieval using the AURA retrieval framework \citep{pinhas2019}, as implemented in recent works \citep{Constantinou2023,Madhusudhan_2023_K218b, Constantinou2024}. The terminator of K2-18~b is modelled as a plane-parallel atmosphere in hydrostatic equilibrium with a non-uniform thermal structure, described using the parametric temperature profile of \citet{Madhusudhan2009} with six free parameters. The transmission spectrum is computed considering radiative transfer along the slant path length across the terminator. 

The opacity is contributed by molecular line absorption as well as collision-induced absorption due to H$_2$-H$_2$ and H$_2$-He \citep{Borysow1988,orton2007,abel2011,richard2012}. We consider molecular line opacities from prominent molecules possible in temperate H$_2$-rich  atmospheres, as discussed below, with the volume mixing ratio of each molecule being a free parameter in the model. The atmospheric composition is treated as uniform across the observable photosphere. 

We consider a wide range of molecules to explain the features in the observed spectrum. Across the different models considered in this work we explore 20 molecules, as discussed below and in Appendix \ref{app:retrievals}. 
The model also includes extinction from clouds/hazes, following the parametric prescription outlined in \citet{pinhas2019}, involving four free parameters corresponding to a cloud-top pressure, amplitude and slope of a Rayleigh-like haze, and a coverage fraction. Other free parameters in the model include a reference pressure ($P_{\rm ref}$) corresponding to the white light planet radius, using the stellar radius from \cite{Benneke2019_K218b}, and a constant vertical offset added to the spectrum from a given instrument, in this case MIRI LRS. We note that the offset is degenerate with $P_{\rm ref}$. Nevertheless, we include both parameters to maintain uniformity in analyses with prior works, and in case an offset is needed that is greater than the allowed pressure range for $P_{\rm ref}$. In total, the number of free parameters in a typical model is $N$ = $N_{\rm mol}$ + 12, where $N_{\rm mol}$ is the number of molecules.  

The Bayesian inference and parameter estimation is conducted using the nested sampling algorithm \citep{Skilling2004} through the MultiNest implementation \citep{Feroz2009, Buchner2014}. Across our retrievals, we use 2000 live points which ensures fine sampling of the parameter space and accurate estimation of the Bayesian evidence. For robustness, we also perform several of the canonical retrievals discussed below using the \texttt{UltraNest} implementation \citep{Buchner2021}, as used in the VIRA retrieval framework, the most recent variant of AURA \citep{Constantinou2024}.

\subsection{Retrieval Setup and Model Selection}
\label{sec:retrieval_setup}

The present transmission spectrum is the first mid-infrared ($\sim$6-12 \textmu m) transmission spectrum of a candidate hycean world ever observed. Given the wide range of molecules that could contribute spectral features in this range, we follow a hierarchical Bayesian approach to determine the preferred model to explain the observations. We carry out the model selection in two stages, with the model parameters in the retrieval at the second stage being informed by the first stage, as discussed below. At the end, the preferred model given the data is chosen by comparing the Bayesian evidence between the two stages.

We first conduct a maximal retrieval using the \texttt{JExoRES} spectrum, with an atmospheric model that includes a substantially expanded set of molecules over previous atmospheric retrievals of K2-18~b. We consider 20 molecules, including 11 molecules considered in previous work \citep{Madhusudhan_2023_K218b} and 9 new ones. The previous molecules include H$_2$O, CH$_4$, NH$_3$, CO, CO$_2$, HCN, (CH$_3$)$_2$S (DMS), CH$_3$Cl, CS$_2$, OCS and N$_2$O. The molecules H$_2$O, CH$_4$, NH$_3$, CO, CO$_2$, and HCN have been predicted to be the prominent C-, O- and N-bearing species in temperate H$_2$-rich atmospheres, including hycean worlds \citep[e.g.,][]{Madhusudhan2021, Yu2021, Hu_photo21}. The remaining molecules (CH$_3$)$_2$S (DMS), CH$_3$Cl, CS$_2$, OCS, N$_2$O have been predicted as observable potential biosignatures in H$_2$-rich atmospheres of habitable super-Earths \citep{segura2005, DomagalGoldman2011, Seager2013a} and hycean worlds \citep{Madhusudhan2021}. 

The new molecules include other prominent sulfur-based molecules, hydrocarbons, and potential biomarkers \citep[e.g.,][]{Schwieterman2024, Tsai24_Bio, sousa-silva2020}: H$_2$S, SO$_2$, C$_2$H$_2$, C$_2$H$_4$, C$_2$H$_6$, (CH$_3$)$_2$S$_2$ (dimethyl disulfide; DMDS), CH$_3$SH, PH$_3$ and CH$_3$OH. Of these molecules, DMDS, CH$_3$SH, C$_2$H$_6$ and PH$_3$ are particularly relevant as potential biomarkers both in hycean as well as terrestrial-like atmospheres as suggested in recent studies \citep[e.g.][]{sousa-silva2020, Schwieterman2024, Tsai24_Bio}. Of all the molecules considered, DMS, DMDS, CH$_3$SH, CH$_3$Cl and N$_2$O are expected to be the most reliable biosignatures in hycean conditions, whereas others may have alternate abiotic sources \citep{segura2005, DomagalGoldman2011, Seager2013a, Schwieterman2024, Tsai24_Bio}.

The retrieval obtains a good fit to the spectrum with the model being inconsistent with a featureless spectrum, i.e. a flat line, at 3-$\sigma$ significance. However, an inspection of the posteriors reveals only one well-constrained peak for the molecule DMDS, while all the other molecules are unconstrained, as shown in Appendix \ref{app:retrievals}. We also do not find significant constraints for cloud/haze parameters that could affect the mid-infrared spectrum.
 
We further conduct a retrieval without DMDS and find a well-constrained peak in the posterior for DMS. We find that DMS replaces DMDS due to the strong degeneracy between the two molecules in the MIRI wavelength range where prominent peaks are evident in the data, as shown in Figure~\ref{fig:spectral_fit}. The spectral contributions due to the two molecules are shown in Figure.~\ref{fig:contributions}. Thus, the retrievals highlight the degeneracy between DMS and DMDS, with a higher preference for DMDS from this dataset. On removing both molecules from the retrieval we find no significant constraints on any of the remaining molecules. The model without DMS or DMDS does not provide a good fit to the data, with the maximal model preferred over this model at 2-$\sigma$. Moreover, the model without DMS and DMDS is only marginally favoured over a flat spectrum below 2-$\sigma$ significance. The data therefore show notable evidence for only two molecules, DMDS and DMS.

Informed by the above retrieval, we arrive at a canonical model that includes DMS and DMDS. We additionally include CH$_4$ and CO$_2$, which have been detected in K2-18~b previously in the near-infrared \citep{Madhusudhan_2023_K218b}. In total, our canonical model includes 16 free parameters: 4 molecular mixing ratios, 6 for the $P$-$T$ profile, 4 for the cloud parameters, 1 for $P_{\rm ref}$ and 1 for the offset on the MIRI data. The retrieval with this model provides a comparable fit to the data as the maximal model above, but with a higher Bayesian evidence due to the lower number of unconstrained free parameters. The canonical model is preferred over the maximal model by 2.2-$\sigma$ and over a featureless spectrum, i.e., a flat line, by $\geq$ 3.4-$\sigma$. For computing the Bayesian evidence for the one-parameter model of a featureless spectrum, we assume uniform priors on the model transit depth with a range of $\pm$$N$ ppm relative to the measured white light transit depth. Considering the span of transit depths in the observed spectrum of $\sim$600 ppm, we explore values of $N$ between 600-2000 ppm. Over this range, the canonical model is preferred over a featureless spectrum at 3.4-3.8 $\sigma$, with higher preference obtained for wider priors, i.e., larger $N$. We adopt the minimum preference of 3.4-$\sigma$ as a conservative estimate. 

As with the maximal model, we find that only DMDS is constrained by this dataset while other parameters remain unconstrained. We also recover the degeneracy between DMDS and DMS, such that when DMDS is excluded from the model DMS is recovered. In what follows, we present the detection significances of the molecules and atmospheric constraints with the canonical model using the spectra obtained from both pipelines. 

\subsection{Atmospheric Constraints} \label{sec:atm_constraints}

The spectrum provides important insights into the atmospheric composition of K2-18~b. The retrieved spectral fit along with the 1- and 2-$\sigma$ contours using the \texttt{JExoRES} spectrum is shown in Figure \ref{fig:spectral_fit}. As noted above, the present MIRI spectrum spans a wavelength range of $\sim$5.8-12 \textmu m which encompasses strong spectral features of several prominent molecules expected in temperate H$_2$-rich atmospheres, as shown in Figure \ref{fig:contributions} and \ref{fig:additional_contributions}. However, as discussed above, we are unable to explain the observed features with most of the 20 prominent species considered, with the maximal retrieval finding no significant evidence for 18 of the 20 species. The only species with notable evidence are DMDS and DMS. Both molecules have a similar double-peak feature between 6.8-8 \textmu m, with additional broad peaks around 9.8 \textmu m and 10.5 \textmu m for DMS and DMDS, respectively, which are consistent with the observed spectrum. Additionally, the amplitudes of the observed spectral features, of $\sim$300-400 ppm, in the present MIRI band are significantly larger compared to the $\sim$200 ppm amplitudes seen in previous near-infrared ($\sim$1-5 \textmu m) observations with the NIRISS and NIRSpec instruments \citep{Madhusudhan_2023_K218b}. While previous observations detected strong spectral features of CH$_4$ and CO$_2$ in the near-infrared, neither of them are detectable in the present spectrum, as can be seen in Figure ~\ref{fig:additional_contributions}. While CO$_2$ does not have strong spectral features in the MIRI LRS range, those of CH$_4$ are dwarfed by stronger contributions due to DMS and/or DMDS. As such, the contributions of both CO$_2$ and CH$_4$ at previously reported mixing ratios ($\sim$1\%) are insufficient to explain the present data in the MIRI band, as can be seen in Figure \ref{fig:contributions}. Instead, contributions from DMDS and DMS at volume mixing ratios of $\sim$10$^{-5}$-10$^{-3}$ readily explain the data due to their strong absorption cross sections in the MIRI band. 

The spectrum provides notable constraints on the presence and abundance of DMDS and DMS in the atmosphere, as shown in Figure \ref{fig:posteriors}. Due to the degeneracy between the spectral features of DMDS and DMS in the MIRI band, as shown in Figure \ref{fig:contributions}, it is difficult to robustly distinguish between them at the current data quality. Considering the canonical model, where both DMDS and DMS are present, DMDS is preferred at a somewhat higher significance ($\sim2 \sigma$) compared to DMS for the \texttt{JExoRES} spectrum. The presence of DMDS in the model is favored at 2-$\sigma$ significance compared to the model without DMDS. DMS is not well constrained, and the model with DMS is preferred over the model without DMS at only 1-$\sigma$ significance. The lack of either molecule in the model is compensated by the other due to the degeneracy in their spectral features as discussed above. However, the combination of DMDS and DMS together is favored at 3.2-$\sigma$ over a model with neither molecule included (Table \ref{tab:abundances}). 

On the other hand, retrievals with models including only one of the two molecules at a time show relatively high detection significances for both molecules individually. For example, considering the canonical model with only DMDS included, i.e. DMS removed, we find that the detection significance for DMDS is 3.2-$\sigma$. Similarly, the canonical model with only DMS included, i.e. DMDS removed, provides a detection significance for DMS of 2.9$\sigma$. Therefore, the present data provide evidence for the presence of DMDS and/or DMS at 2.9-3.2 $\sigma$ significance. 

The retrieved posterior distributions for the volume mixing ratios of DMDS and DMS shown in Figure \ref{fig:posteriors} reflect the degeneracy between the two species that is apparent from their spectral contributions as discussed above. When both molecules are included in the model, the spectral contribution of DMDS dominates over that of DMS. The posterior of DMDS shows a clear peak, albeit with a low-abundance tail due to the degeneracy with DMS. The abundance of DMDS is retrieved to be $\log(X_\mathrm{DMDS})$ = -3.48$^{+1.24}_{-2.27}$ while DMS is unconstrained. Here, $X_\mathrm{DMDS}$ is the volume mixing ratio of DMDS. In retrievals with only one of the two molecules present, the posterior of that molecule shows a single well-constrained peak with no low-abundance tail. In such retrievals, the abundances are retrieved to be $\log(X_\mathrm{DMDS})$ = -3.25$^{+1.17}_{-1.30}$ and $\log(X_\mathrm{DMS})$ = -3.42$^{+1.16}_{-1.44}$, as shown in Table~\ref{tab:abundances}. Lastly, we repeat the canonical retrieval using the UltraNest nested sampling implementation, and find consistent abundance estimates and detection significances to our standard retrievals using the MultiNest implementation.

We also conduct another canonical retrieval with our other spectrum obtained with the \texttt{JexoPipe} pipeline. We find that the detection significances and abundance estimates of DMS and DMDS are consistent with those obtained using the \texttt{JExoRES} spectrum, as shown in Table~\ref{tab:abundances}. We find the combined detection significance of DMDS and DMS to be 3.2-$\sigma$, similar to that obtained with the canonical retrieval with the \texttt{JExoRES} data. The abundances are slightly lower but still consistent. For this canonical model, where DMS and DMDS are both included, the posteriors for DMS and DMDS are comparable, as shown in Figure \ref{fig:posteriors}. Using the individual DMS-only or DMDS-only cases, we find evidence for each of the molecules at 3-$\sigma$ significance and similar abundance estimates to the corresponding \texttt{JExoRES} cases: $\log(X_\mathrm{DMDS})$ = $-3.45^{+1.13}_{-1.30}$ and $\log(X_\mathrm{DMS})$ = $-3.53^{+1.03}_{-1.33}$. Therefore, across all our retrievals we detect DMDS and/or DMS at 3-$\sigma$ significance.

The observations do not provide significant constraints on any of the other molecules or atmospheric properties but are consistent with previous inferences from the near-infrared JWST spectrum \citep{madhusudhan_carbon-bearing_2023}. Specifically, we do not obtain significant constraints on CH$_4$ or CO$_2$ which have previously been detected. However, the upper-limits we retrieve for the abundances of both molecules are consistent with their previous abundance estimates. The contributions from both molecules at previously measured abundances of $\sim$1\% each are insufficient to explain the observed spectral features in the present spectral range with JWST MIRI (6-12 \textmu m). While CO$_2$ has no strong spectral features in this range, those of CH$_4$ are significantly weaker compared to those of DMDS and/or DMS, as shown in Figure \ref{fig:contributions}. The abundances we retrieve for DMS and/or DMDS are also consistent with the constraints for DMS reported in previous work.

Similarly, we do not find significant evidence for spectral contributions due to clouds/hazes in the MIRI band and none of the cloud parameters are constrained. We find the photospheric pressure corresponding to the white light radius to be $\log$($P_{\rm ref}$/bar) = $-4.32_{-0.93}^{+1.15}$, with no significant offset to the data retrieved. The offset is retrieved to be 12$^{+51}_{-58}$ ppm, which is consistent with zero within the 1-sigma uncertainties. The photospheric temperature at 1 mbar at the terminator is also weakly constrained to be $422_{-133}^{+141}$ K, which is somewhat higher but consistent with previous estimate in the near-infrared to within the 1-$\sigma$ uncertainties.

For additional robustness, we conduct retrievals on the transmission spectra obtained using a wide range of assumptions for the data reduction and analyses. Using \texttt{JExoRES}, we explore differences in the treatment of systematic trends and time-correlated noise, as well as different wavelength limits and the effect of spectral binning, as discussed in the Appendix. Across all of these cases, we find consistent abundance constraints and detection significances of DMDS and/or DMS, ranging from 2.9-3.4 $\sigma$, as outlined in Table~\ref{tab:abundances}. 

\begin{figure*}
    \centering
    \includegraphics[width=\textwidth]{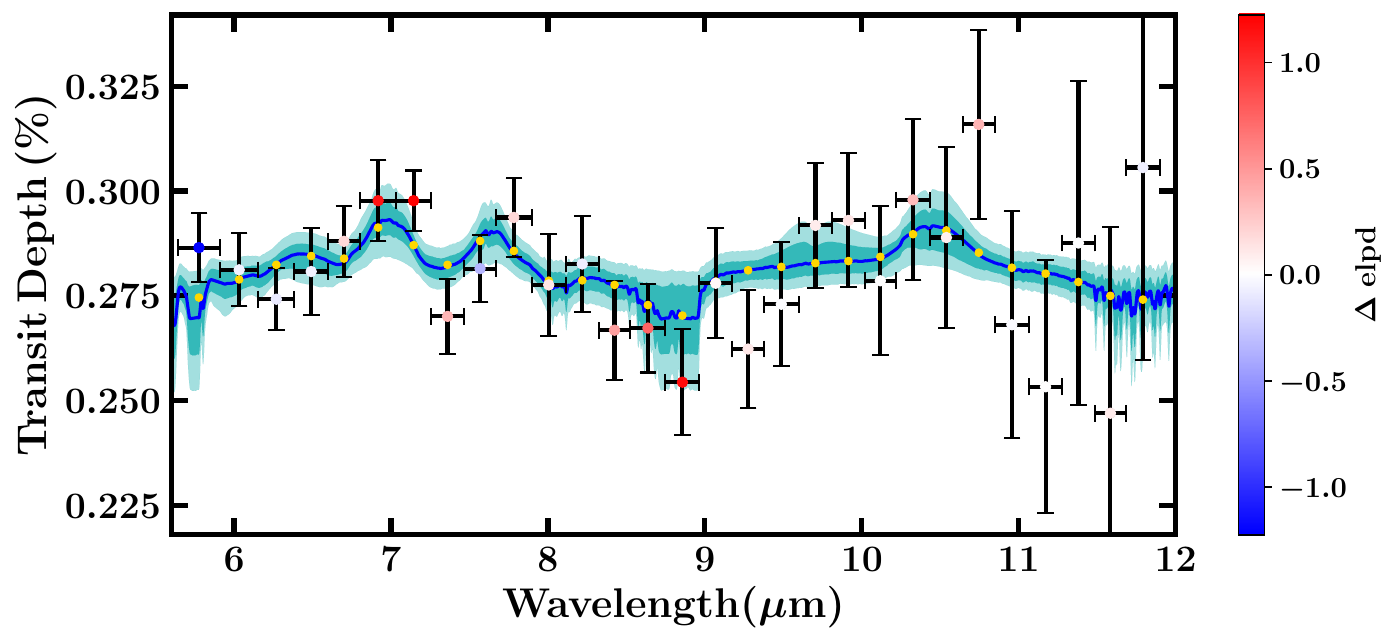}
    \caption{Leave-one-out analysis for the MIRI transmission spectrum of K2-18~b. The black data points with error bars show the observed transmission spectrum obtained using the \texttt{JExoRES} pipeline. The center shading of each datapoint denotes the corresponding elpd difference ($\Delta$elpd) between the canonical atmospheric model with DMS and DMDS, and one with both molecules absent. Redder points denote a high and positive $\Delta$elpd, indicating that the canonical model is more capable of predicting them if they are absent from the fit than the atmospheric model without DMS and DMDS. Similarly bluer points denote larger negative $\Delta$elpd values, indicating a preference against DMS and DMDS, while white points denote $\Delta$elpd values close to zero. Gold points denote the median retrieved spectrum binned for each datapoint, while the dark and light cyan shaded region denotes the retrieved 1-$\sigma$ and 2-$\sigma$ contours, respectively.}
    \label{fig:delta_elpd}
\end{figure*}

\subsection{Leave-one-out Analysis}
\label{sec:leave-one-out}
As a further robustness check for our findings, we carry out leave-one-out cross-validation using the \texttt{JExoRES} data. We follow \citet{Vehtari2017} and \citet{Welbanks2023}, calculating the expected log pointwise predictive density  (elpd) for each data point. For each data point $d_{i}$, we carry out an atmospheric retrieval with all other points considered and that data point removed. We do not make any approximations, e.g. Pareto-smoothed importance sampling, and instead directly compute the elpd of each datapoint by carrying out a retrieval considering all other data points. Using the retrieved posterior samples, we compute the probability density of the $i^{\mathrm{th}}$ point $d_{i}$, given a retrieval using model $\mathcal{M}$ on all other data $d_{-i}$ resulting in a total of $S$ samples of the parameter space for each retrieval. The elpd metric provides an indication of how well each data point can be predicted by the model and therefore quantifies the performance of the model. The metric is given by:

\begin{equation} 
\begin{aligned}
    \mathrm{elpd}_{i, \mathcal{M}} &= \mathrm{log}\left( p(y_i | y_{-1}, \mathcal{M}) \right) \\[0.3em]
    &= \mathrm{log} \left( \frac{1}{S} \sum^{S}_{j =1} p(y_i | \theta_j, \mathcal{M})  \right),
\end{aligned}
\end{equation} 

where the parameters corresponding to the $j^{\mathrm{th}}$ sample in each retrieval are denoted as $\theta_j$. We perform the above calculation for two atmospheric models: the canonical model described in Section \ref{sec:retrieval} and one with DMS and DMDS absent. As such we run a total of two retrievals per data point. The difference between the elpd values of the two models, $\Delta \mathrm{elpd}_i$, indicates their relative performance. Conversely, in this case $\Delta \mathrm{elpd}_i$ serves as an indication  for the relative contributions of the data points towards the present detection of DMDS and DMS. The pointwise $\Delta \mathrm{elpd}_i$ is shown in Figure \ref{fig:delta_elpd}. We find that most points have $\Delta \mathrm{elpd}_i$ values above zero, indicating that the detection of DMDS and DMS combined is driven by multiple data points to various extents. $\Delta \mathrm{elpd}_i$ is highest for two data points near 7~\textmu m and another two points just under 9~\textmu m. This indicates that these points are better predicted by a model that includes DMDS and DMDS than a model without the two molecules. Conversely, the data point below 6~\textmu m has the most negative $\Delta \mathrm{elpd}_i$, indicating that it is better predicted by a model without DMS and DMDS. Overall, our leave-one-out analysis indicates that the present detection is unlikely to be the result of localized systematics or individual outlier points.

\section{Summary and Discussion} \label{sec:discussion}

We report a mid-infrared transmission spectrum of the candidate hycean world K2-18~b observed with the JWST MIRI LRS instrument, the first for a habitable-zone sub-Neptune. The spectrum shows multiple spectral features between $\sim$6-11 \textmu m which are best explained by a combination of DMDS and DMS in the atmosphere, both molecules uniquely produced by life on Earth and predicted as promising biosignatures in habitable exoplanets \citep{Domagal-goldman2011, Catling2018, Schwieterman2018, Seager2013b, Tsai24_Bio}. We detect the combination of DMDS and/or DMS at a significance of 2.9-3.2 $\sigma$ across the canonical retrievals reported in this work. In the absence of DMDS, the spectral features can also be explained to a large extent by DMS, and vice versa, due to degeneracy between the spectral features of DMS and DMDS in the mid-infrared. In retrievals considering only one of the two molecules (DMS or DMDS) in the canonical model, DMS is retrieved at 2.9-3.0 $\sigma$ significance and DMDS is retrieved at 3.0-3.2 $\sigma$ significance.

The observations provide limited constraints on other atmospheric properties. In the absence of DMS and DMDS, no significant detections or abundance constraints are obtained on any of the remaining molecules considered in the retrievals. The strong DMDS and/or DMS features dominate over potential molecular contributions from any other species in this wavelength range. We find high abundances of DMDS and/or DMS with volume mixing ratios of $\gtrsim10^{-5}$ (10 ppmv) within the 1-$\sigma$ uncertainties, and a MIRI photospheric temperature of $422^{+141}_{-133}$~K at 1~mbar for the canonical retrieval. The DMS abundance and photospheric temperature are consistent with but somewhat higher than those derived from previous NIRISS and NIRSpec observations in the 1-5 \textmu m region \citep{Madhusudhan_2023_K218b}. On the other hand, both the retrieved mixing ratios and temperature are strongly dependent on the available cross-sections of DMDS and DMS which are obtained for an Earth-like atmosphere at nearly STP conditions, using N$_2$ as a broadener \citep{HITRAN2016}. Therefore, the derived abundances and temperatures may be impacted by the assumed collisional broadening factors and/or the adopted molecular cross-sections, in general. 

Overall, our findings continue to raise the prospects of possible biological activity on K2-18~b and motivate new experimental and theoretical work for detailed characterisation of its atmospheric properties. In what follows, we discuss the implications and future directions.

\subsection{Biosignature on a Hycean World}
\label{sec:bsig_hycean}
Our findings provide new independent evidence for the possibility of a biosphere on K2-18~b. As discussed in Section~\ref{sec:intro}, the detection of carbon-bearing molecules, CH$_4$ and CO$_2$, and nondetections of NH$_3$ and CO \citep{Madhusudhan_2023_K218b}, are consistent with prior predictions for a hycean world and inconsistent with mini-Neptune or gas-dwarf scenarios \citep{Cooke2024, Rigby_towards}. The previous observations also showed tentative hints of DMS, a possible biosignature molecule in the atmosphere \citep{Madhusudhan_2023_K218b}, but the evidence was weak. This new independent evidence for the presence of DMS and/or DMDS at 2.9-3.2 $\sigma$ adds to the tentative inference of DMS reported previously and bolsters the chances of a biosphere on K2-18~b.

Recent photochemical modeling of K2-18~b suggests that large quantities of DMS and DMDS, with mixing ratios up to 10$^{-2}$ can accumulate in the atmosphere for assumed biogenic oceanic fluxes of these gases of $\gtrsim$20$\times$ Earth levels \citep{Tsai24_Bio}. 
Such quantities are consistent with our current MIRI results for DMDS and DMS and with previous abundance constraints of DMS \citep{Madhusudhan_2023_K218b}. These otherwise photochemically fragile molecules survive in the \cite{Tsai24_Bio}~models under high surface-flux emission conditions: (1) because of the favorable ultraviolet spectral energy distribution of M-dwarf stars \citep[see also][]{segura2005, DomagalGoldman2011, Seager2013a}, (2) because DMS and DMDS can eventually self-shield, as well as shield molecules such as CO$_2$ from photolysis when their column density is high enough, significantly cutting down on their photochemical destruction pathways \citep{Tsai24_Bio}, and (3) because the models assume that biogenically emitted DMS and DMDS do not have a return sink to the ocean, allowing the abundance to build up over time. Our present MIRI abundance estimates of DMS and/or DMDS at mixing ratios of $\sim$10$^{-5}$-10$^{-3}$ could, therefore, imply very strong biological activity on the planet in comparison to Earth.  We note however, that DMDS could potentially condense on K2-18~b \citep[see][]{Sagan1971, Khare1978, VonNiederhausern2006}, depending on tropopause temperatures, restricting its gas-phase abundance in the stratospheric region that is probed in the transit observations. Therefore, abundances significantly higher than $\sim 10^{-4}$ - 10$^{-3}$ may be difficult to achieve if the stratosphere is too cold.

On the other hand, it is possible that the abundance estimates derived in our work are strongly influenced by uncertainties in spectral parameters and cross-sections of these molecules used in the models. The derived abundance and temperature are strongly dependent on the absorption cross-section of a molecule detected. For both DMDS and DMS, only limited cross sections are available in the literature \citep{HITRAN2016}, which are derived experimentally assuming an Earth-like N$_2$-rich background atmosphere at nearly STP conditions. It is possible that the collisional broadening factors and cross-sections may be different for an H$_2$-rich background gas at lower pressures probed in the present transmission spectroscopy. Therefore, our results highlight the acute need for laboratory and theoretical work to derive high-fidelity absorption cross sections for these and other biosignature molecules to enable their robust detection and abundance estimates in habitable exoplanets.

\subsection{False Positives}

As with any potential biosignature molecule, it is important to ask whether there can be abiotic sources of DMS and DMDS in a temperate H$_2$-rich atmosphere that could explain the observations. Both DMS and DMDS have been predicted to be robust biosignatures for Earth-like planets as well as planets with H$_2$-rich atmospheres, including super-Earths and hycean worlds \citep{DomagalGoldman2011, Seager2013a, Catling2018, Schwieterman2018, Madhusudhan2021}. Their identification as robust biosignatures is due to the fact that on Earth, both molecules are uniquely produced by life (particularly marine biota) in small quantities of $\lesssim$1 ppb by volume and are not supplied by abiotic photochemistry. Nevertheless, here we explore some potential alternatives.

Experimental studies have demonstrated the feasibility of forming several organosulfur compounds abiotically using ultraviolet irradiation or electric discharges of gaseous mixtures containing H$_2$S and CH$_4$ \citep[e.g.,][]{Raulin1975, He2020, Vuitton2021, Reed2024}. In particular, some of these studies have demonstrated the production of both DMS \citep{Raulin1975, Reed2024} and DMDS \citep{Sagan1971, Khare1978} in gas mixtures containing both CH$_4$ and H$_2$S, arguing in favor of their possible abiotic production in reduced planetary atmospheres. However, both DMS and DMDS are highly reactive and have very short lifetimes in the above experiments (i.e., a few minutes) and in the Earth's atmosphere (i.e., between a few hours to $\sim$1 day), due to various photochemical loss mechanisms \citep[e.g.][]{Seager2013b}. Thus, the resulting DMS and DMDS mixing ratios in the current terrestrial atmosphere are quite small (typically $\lesssim$1 ppb), despite continual resupply by phytoplankton and other marine organisms. 

Therefore, sustaining DMS and/or DMDS at over 10-1000 ppm concentrations in steady state in the atmosphere of K2-18~b would be implausible without a significant biogenic flux. Moreover, the abiotic photochemical production of DMS in the above experiments requires an even greater abundance of H$_2$S as the ultimate source of sulfur --- a molecule that we do not detect --- and requires relatively low levels of CO$_2$ to curb DMS destruction \citep{Reed2024}, contrary to the high reported abundance of CO$_2$ on K2-18~b \citep{Madhusudhan_2023_K218b}.

Another recent study has reported evidence for the presence of DMS on the comet 67P/Churyumov{\-}–Gerasimenko \citep{Hanni2024}, motivating the possibility of a potential abiotic source through cometary delivery to the exoplanet's atmosphere. Comets are also known to contain other ices and sublimated gases that might be considered biosignatures if present in an exoplanet atmosphere, including O$_2$, CH$_4$, and CH$_3$Cl \citep{Bieler2015, Fayolle2017, Rubin2019}. However, cometary delivery is implausible to explain such molecules in planetary atmospheres due to the insufficient contribution that trace species within a comparatively small comet would add to the much more massive planetary atmosphere \citep[e.g.][]{Leung2022,Court2012,Felton2022}. Moreover, molecules such as DMS and DMDS would be shock heated during a hypervelocity cometary impact and its related plume splashback phase, which would reset the bulk composition to simpler molecules that are more stable in thermochemical equilibrium at high temperatures and (for the splashback phase) low pressures \citep{zahnle1996}.  Therefore, DMS and DMDS would not be delivered at any significant measurable levels through such a process. We also note that DMDS has not been reported in the comet 67P. In summary, cometary impacts cannot be responsible for the relatively large DMS and/or DMDS abundance inferred from the present MIRI retrievals.

\subsection{Future Directions}
\label{sec:future} 
A conclusive identification of a biosignature necessitates a robust assessment of various factors, including the robustness of the detections, the environmental context, and potential false positives \citep[e.g.][]{Meadows2022, Schwieterman2018}. It is widely recognized that the detection of a biosignature is unlikely to be instantaneous or unambiguous in the first instance, rather relying on continued accumulation of evidence and addressing the above factors \citep{Meadows2022}. Our study is the first formal step in this direction, building on the first possible hints of DMS reported in our previous work \citep{Madhusudhan_2023_K218b} and further evidence of DMDS and/or DMS with a higher significance seen in the present observations. Finally, our present detection based on multiple spectral features with a different instrument in a different spectral range from previous work provides an important independent line of evidence in this direction. 

Further work is needed to robustly verify the current findings. More observations are required to robustly demonstrate the repeatability of our present findings, rule out potentially unaccounted for instrumental systematics, as well as increasing the detection significances. While DMDS and DMS best explain the current observations, their combined detection significance is $\sim$3-$\sigma$ which is at the lower end of robustness typically required for scientific evidence. The significance can be readily increased to a 4-5 $\sigma$ level by a modest amount of additional JWST time, e.g. between 1-3 additional transits with MIRI, i.e. only $\sim$8-24 hours. Secondly, while we have explored 20 prominent molecules to fit the spectrum, our search may still not be fully exhaustive. Therefore, future studies could investigate other potential molecules that could explain the data. At the same time, as discussed above, new experimental and theoretical studies are needed to determine accurate absorption cross sections for DMDS and DMS, and other potential biosignature molecules, for conditions relevant to candidate hycean worlds like K2-18~b. Future laboratory experiments and/or theoretical modeling is also needed to fully explore possible photochemical mechanisms for producing DMS and DMDS in dry, methane-rich, reduced environments to address potential abiotic sources of these molecules.

Overall, our findings present an important step forward in the search for signatures of life on exoplanets. However, robustly establishing both the veracity of the present findings and their possible association with life on K2-18~b needs a dedicated community effort in multiple directions --- observational, theoretical and experimental. Observations with JWST are already demonstrating that possible hycean worlds indeed significantly expand and accelerate the search for life elsewhere. The central question now is whether we are prepared to identify the signatures of life on such worlds. The opportunity is at our doorstep. 

{\it Acknowledgements:} This work is based on observations made with the NASA/ESA/CSA James Webb Space Telescope as part of Cycle 1 GO Program 2722 (PI: N. Madhusudhan).  We thank NASA, ESA, CSA, STScI and everyone whose efforts have contributed to the JWST, and the exoplanet science community for the thriving current state of the field. This work is supported by the UK Research and Innovation (UKRI) Frontier Grant (EP/X025179/1), PI: N. Madhusudhan. N.M. thanks Tony Roman and Sara Kendrew at STScI for their help with planning our JWST observations. J.M. acknowledges support from JWST-GO-02722, which was provided by NASA through a grant from the Space Telescope Science Institute, which is operated by the Association of Universities for Research in Astronomy, Inc., under NASA contract NAS 5-03127. We thank the anonymous reviewer for their valuable comments on the manuscript. 

This work was performed using resources provided by the Cambridge Service for Data Driven Discovery operated by the University of Cambridge Research Computing Service (\url{www.csd3.cam.ac.uk}), provided by Dell EMC and Intel using Tier-2 funding from the Engineering and Physical Sciences Research Council (capital grant EP/P020259/1), and DiRAC funding from STFC (\url{www.dirac.ac.uk}).

{\it Author Contributions:} N.M. conceived, planned and led the project. N.M. led the JWST proposal with contributions from S.C., S.S., A.P. and J.M. N.M., S.S. and M.H planned the JWST observations. N.M., M.H. and S.S. conducted the data reduction and spectroscopic analyses. N.M. and S.C. conducted the atmospheric retrievals. N.M. and J.M. conducted the theoretical interpretation. N.M. led the writing of the manuscript with contributions and comments from all authors. 

{\it Data Availability:} The spectra obtained with the \texttt{JExoRES} and \texttt{JexoPipe} pipelines, along with the retrieved posterior distributions for the canonical model retrievals are available at \url{https://osf.io/gmhw3/}. The JWST data presented in this article were obtained from the Mikulski Archive for Space Telescopes at the Space Telescope Science Institute. The specific observations analyzed can be accessed via \dataset[doi: 10.17909/rx29-yw62]{https://doi.org/10.17909/rx29-yw62}.

{\it Facilities:} JWST (MIRI)

\appendix

\section{Robustness of Transmission Spectra}
\label{Robustness of Transmission Spectra}

\subsection{Comparison Between Pipelines} \label{app:pipeline_comparison}

In this work, we thoroughly test several data-reduction methods and assumptions, such as the effect of the non-linearity correction, choice of binning, detrending, and limb-darkening, to produce a robust transmission spectrum. We also apply two independently developed data reduction pipelines to check for consistency, using \texttt{JExoRES} and \texttt{JexoPipe}, as described in Sections~\ref{sec:JExoRES} and~\ref{sec:JexoPipe}, respectively. Figure~\ref{fig:pipelines} demonstrates the stability of the MIRI transmission spectrum of K2-18~b when comparing these two pipelines.  Table~\ref{tab:wlc_params} provides the parameters from the white light curve fitting with both pipelines, yielding results consistent within 1-$\sigma$.

\subsection{Non-linearity Correction} \label{app:linearity}

\begin{figure}
	\includegraphics[width=1.0\textwidth]{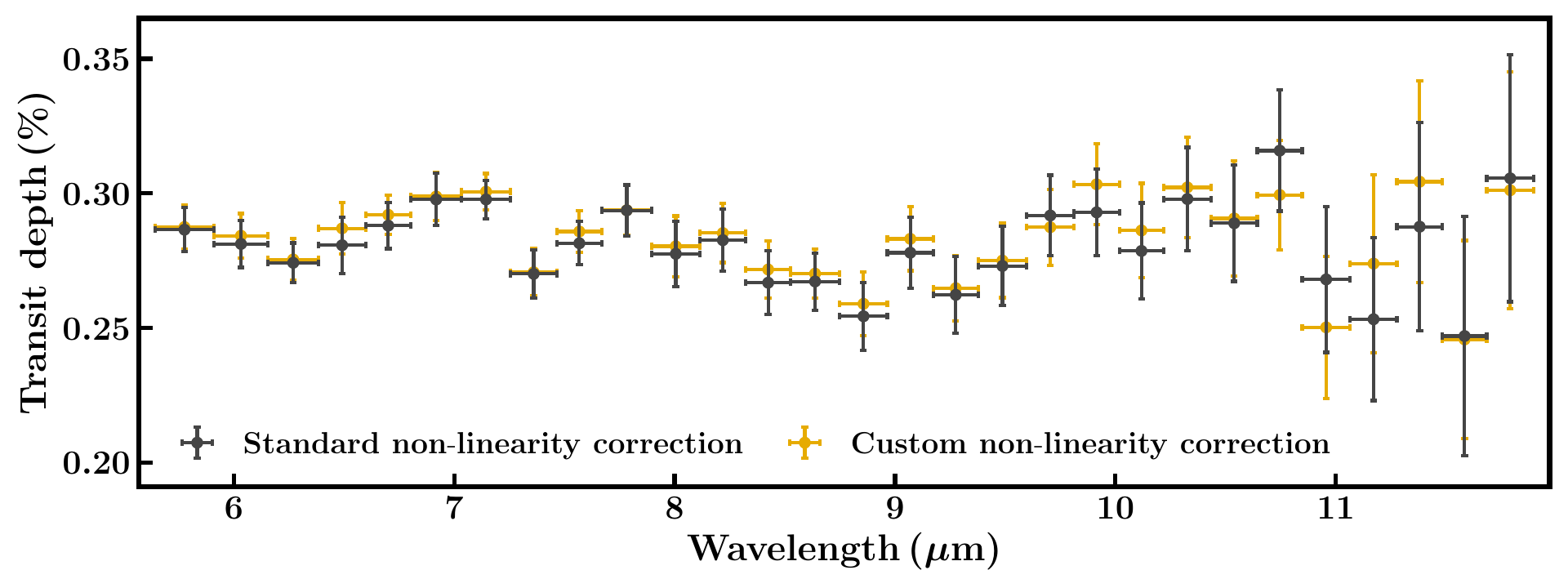}
    \caption{ Effects of different non-linearity corrections on the MIRI transmission spectrum of K2-18~b. The spectra with the standard and our custom non-linearity corrections are shown in dark grey and yellow, respectively. }
    \label{fig:linearity} 
\end{figure}

We explored how differences in non-linearity correction can affect the transmission spectrum, as pursued in recent works \citep{Kempton2023, grant_jwst-tst_2023, dyrek_so2_2024}. As the baseline case, we use the standard non-linearity corrections, RSCD correction and dark subtraction. Note that the RSCD correction step does not alter the raw data. Instead it applies a mask to the first few groups -- removing these from the ramp fitting. On the other hand, the dark reference file includes both the dark signal and the RSCD effect, which are corrected by applying the dark subtraction step. For our custom approach, we empirically derive our own RSCD and non-linearity correction. Using the average of the last 500 (out-of-transit) integrations of our observation, accounting for outliers, we first fit the ramp of each pixel using
\begin{equation}
    f(t) = p_1 + p_2 t + p_3 t^2 - p_4 \exp{(- t / p_5)}\,,
\end{equation}
where $t$ is the time since the start of an integration. Here, $p_1$ corresponds to an overall offset/bias, $p_2$ is the linear part of the signal which we use to represent the true flux \citep{Morrison2023}, $p_3$ is the quadratic non-linear component representing the change in response as charge accumulates, $p_4$ is the strength of the RSCD signal, and $p_5$ is the time scale of the RSCD signal. For this fit, we exclude the first and last group, as done in the main analysis described in Sections~\ref{sec:JExoRES} and \ref{sec:JexoPipe}. Next, we subtract 
the RSCD correction, including the constant term, i.e.,
\begin{equation}
    f_{\mathrm{reset}}(t) = p_1 - p_4 \exp{(- t / p_5)}\,,
\end{equation}
from the data before applying our custom non-linearity correction. We derive the parameters for the non-linearity correction using a fourth-degree polynomial $h$, defined as
\begin{equation}
    h(p_2 t + p_3 t^2) = p_2 t\,, \hspace{5mm} h(x) = \sum_{i=0}^4 a_i x^i\,,
\end{equation}
where $a_i$, being pixel-dependent, is used to populate the custom non-linearity reference file. Because we have already subtracted the RSCD signal, we do not apply the dark subtraction step in this scenario.

Following this custom non-linearity correction approach, we perform the rest of the analysis as outlined in Sections~\ref{sec:JExoRES}.Figure~\ref{fig:linearity} shows the transmission spectrum of K2-18~b using the standard and custom non-linearity correction. Overall, we find that the spectrum is largely insensitive to the choice of non-linearity correction. We note that other data sets may be more sensitive to the non-linearity correction if the flux level is higher (for the present observation, it is below 51\% saturation) or for targets with a larger transit depth than K2-18~b.

\subsection{Spectral Extraction Method} \label{app:extraction}

\begin{figure}
	\includegraphics[width=1.0\textwidth]{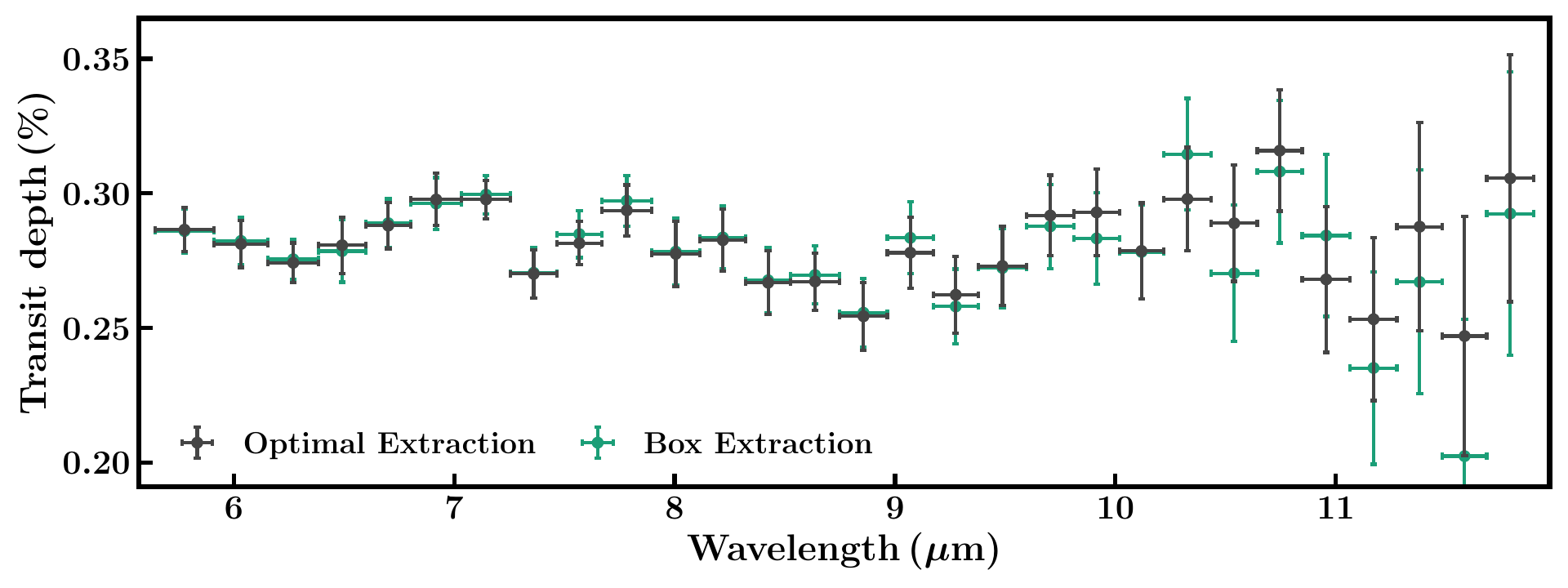}
    \caption{ Comparisons between different spectral extraction methods. The dark grey points show the MIRI transmission spectrum of K2-18~b using the optimal extraction method while the green points show the spectrum using box extraction.} 
    \label{fig:box} 
\end{figure}

We also check if the choice of spectral extraction method significantly affects the MIRI transmission spectrum of K2-18~b. Nominally, we adopt the optimal extraction method \citep{horne_optimal_1986}, as outlined in Sections~\ref{sec:JExoRES} and \ref{sec:JexoPipe}, which is commonly used in the literature \citep[e.g,][]{Bouwman2023, Kempton2023, dyrek_so2_2024, Bell2024}. Additionally, we also perform box extraction, where each pixel in the aperture is given equal weighting. For this, we used an aperture of 9 pixels, the same as for the optimal extraction. We find that the transmission spectrum is consistent between the two methods, as shown in Figure~\ref{fig:box}. The box extraction method produces somewhat larger uncertainties towards the red part of the spectrum compared to optimal extraction (around 15\% larger), as expected given the lower throughput in this region.

\subsection{Spectral Binning} \label{app:binning}
\begin{figure}
	\includegraphics[width=1.0\textwidth]{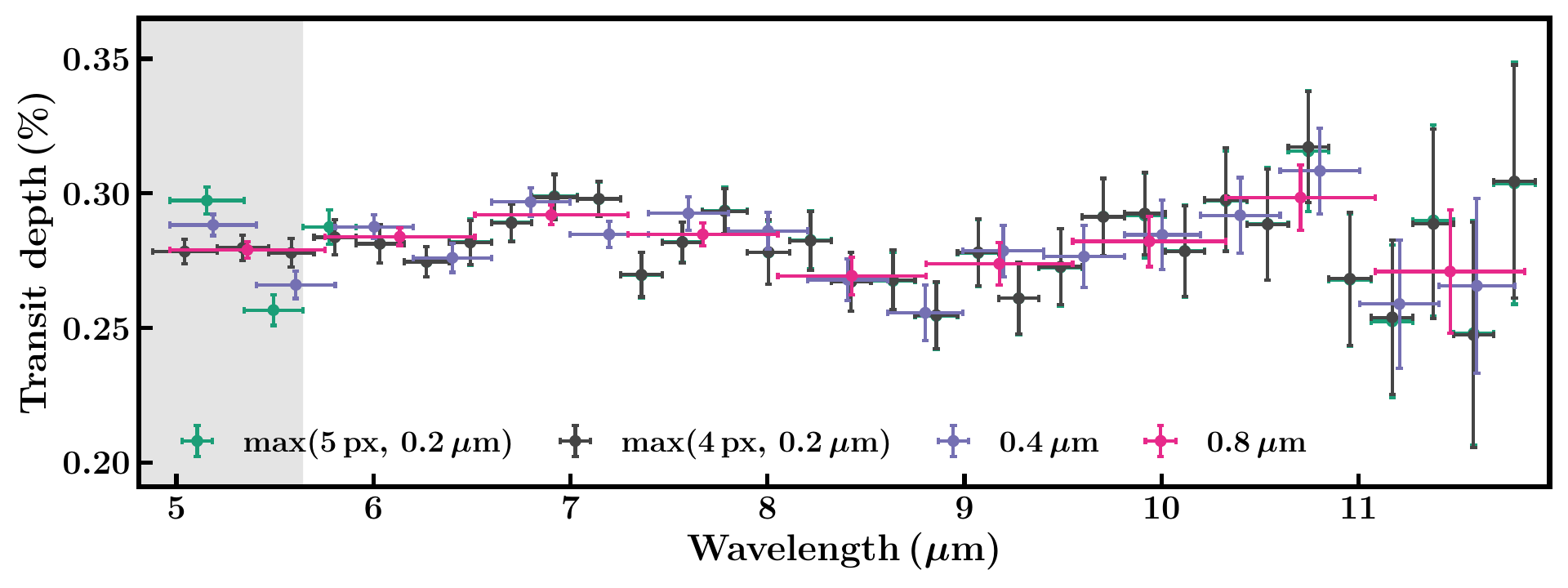}
    \caption{ MIRI transmission spectrum of K2-18~b using different bin widths. For our analysis, we do not consider the spectrum below 5.6~\textmu m to be conservative, given that the choice of binning affects the spectrum in this region. Note that these spectra did not use GPs, as discussed in Appendix~\ref{app:noise}, thus the error bars may be somewhat underestimated.}
    \label{fig:binning} 
\end{figure}

Previous work has shown that transmission spectra obtained with MIRI LRS can be excessively noisy when using small spectroscopic bins \citep{Bell2024}. Therefore, recent works have adopted wide bins with typical widths of 0.15-0.5~\textmu m \citep[e.g.,][]{Bell2024, Powell2024, Welbanks2024, Schlawin2024}. In this work, we test the sensitivity of our transmission spectrum of K2-18~b to different bin sizes, as shown in Figure~\ref{fig:binning}. We adopt four different bin widths: maximum of 4 pixels or 0.2~\textmu m (whichever contains the most pixels), maximum of 5 pixels or 0.2~\textmu m, 0.4~\textmu m, and 0.8~\textmu m. We limit the bin widths to at least 4 or 5 pixels to average over potential systematic effects. Note that we only use whole pixel rows to avoid introducing correlations between different spectral channels. We find that our transmission spectrum is generally consistent between the different bin widths, apart from a small region below around 5.6~\textmu m. Specifically, in this region, the spectrum is not consistent between the 4- and 5-pixel bin widths, i.e. the green and grey data points in Figure~\ref{fig:binning}. For this reason, we do not include the data below 5.6~\textmu m in our atmospheric retrievals. Nominally, we select a maximum of 5 pixels or 0.2~\textmu m as our canonical binning for both \texttt{JExoRES} and \texttt{JexoPipe}. We also conduct atmospheric retrievals with varying bin widths using our canonical model. We find consistent results between the retrievals with detection significances for DMS+DMDS between 3.2-3.3 $\sigma$ across the different bin widths, as shown in Table~\ref{tab:abundances}.
 
\subsection{Trends in Light Curve Fitting} \label{app:trends}
\begin{figure}
	\includegraphics[width=1.0\textwidth]{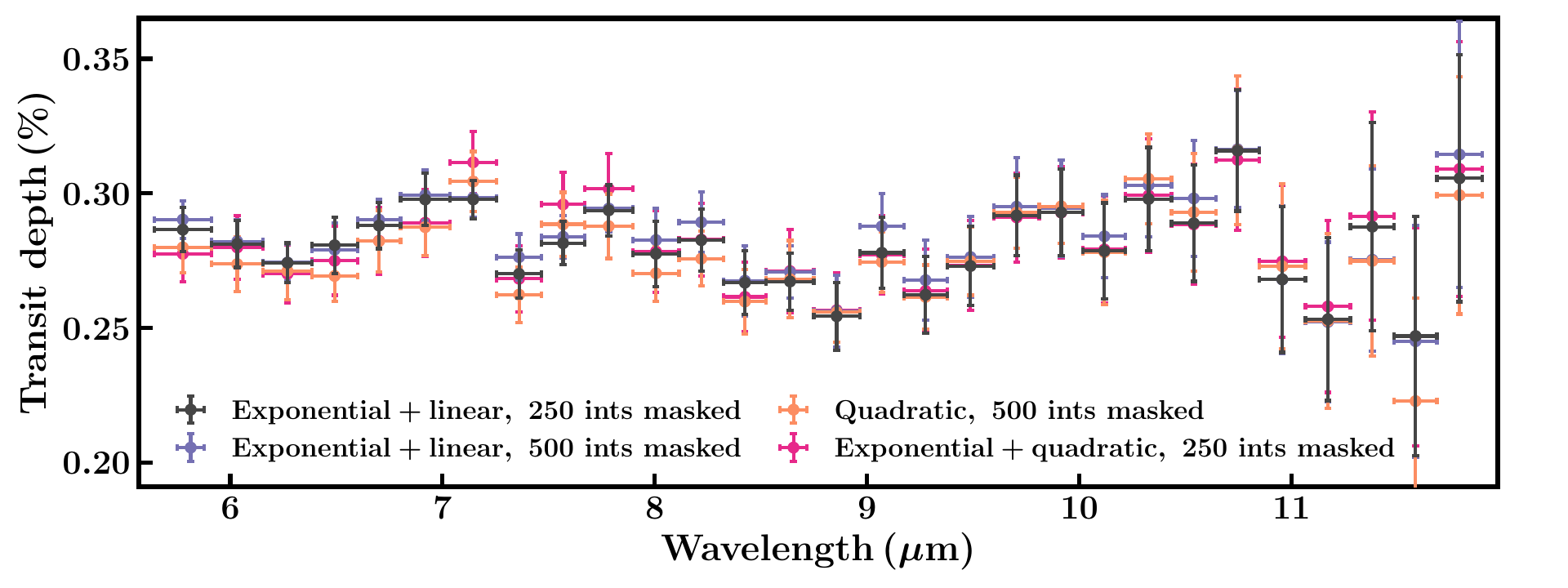}
    \caption{The MIRI transmission spectrum of K2-18~b obtained using different 
    treatments of the detector settling effect. We consider an exponential + a linear trend (dark grey and purple), an exponential + a quadratic trend (pink), and a quadratic trend alone (orange), as well as different numbers of integrations to mask at the start of the observation. 
    }
    \label{fig:trends} 
\end{figure}

MIRI time-series observations are affected by detector settling in which there is an exponential-like ramp in the light curves at the beginning of the observation \citep{Bouwman2023, Greene2023}. This effect can be seen in the white light curve of our observation, as shown in Figure~\ref{fig:wlc_miri}. A common approach to treat this effect is to mask some number of integrations at the start of the observation and/or to account for the detector settling when fitting the light curves \citep[e.g.,][]{Bouwman2023, Greene2023, Kempton2023, grant_jwst-tst_2023, dyrek_so2_2024, Bell2024}, for example, by including an exponential trend. For this reason, we investigate the impacts of different treatments of this effect on our MIRI transmission spectrum of K2-18~b. To model the light curves, we use the following systematic model:
\begin{equation}
    F_{obs}(t) = F_{\mathrm{out}}\,(1 + \alpha \tau + \beta \tau^2 + \gamma e^{-\tau / \epsilon})\,  F_{\mathrm{transit}}(t)\,,
\end{equation}
where $\tau$ is the time since the start of the observation, $F_{\mathrm{transit}}$ is the \texttt{batman} transit model, and $F_{\mathrm{out}}$, $\alpha$, $\beta$, $\gamma$, and $\epsilon$ are trend parameters. We consider three different trend prescriptions: an exponential and a linear trend ($\beta = 0$, as described in Section~\ref{sec:JExoRES}), an exponential and a quadratic trend, and a quadratic trend alone ($\gamma = 0$). Note that for the white light curve, we obtain a good fit to the data using the first option. Furthermore, we nominally remove 250 integrations (17 minutes) at the start of the observation to remove the worst part of the detector settling effect. However, for the quadratic trend, we mask 500 integrations (34 minutes) at the start to remove even more of the settling effect. Figure~\ref{fig:trends} shows the MIRI transmission spectrum of K2-18~b obtained using different trends, showing general agreement between these scenarios. For our canonical case with an exponential and a linear trend, we also consider a wider range of the number of masked integrations, between 125-750 integrations, and find that the spectrum is consistent across these cases (as shown in Figure~\ref{fig:trends} for 250 and 500 masked integrations).

\subsection{Limb Darkening} \label{app:limb_darkening}
\begin{figure}
	\includegraphics[width=1.0\textwidth]{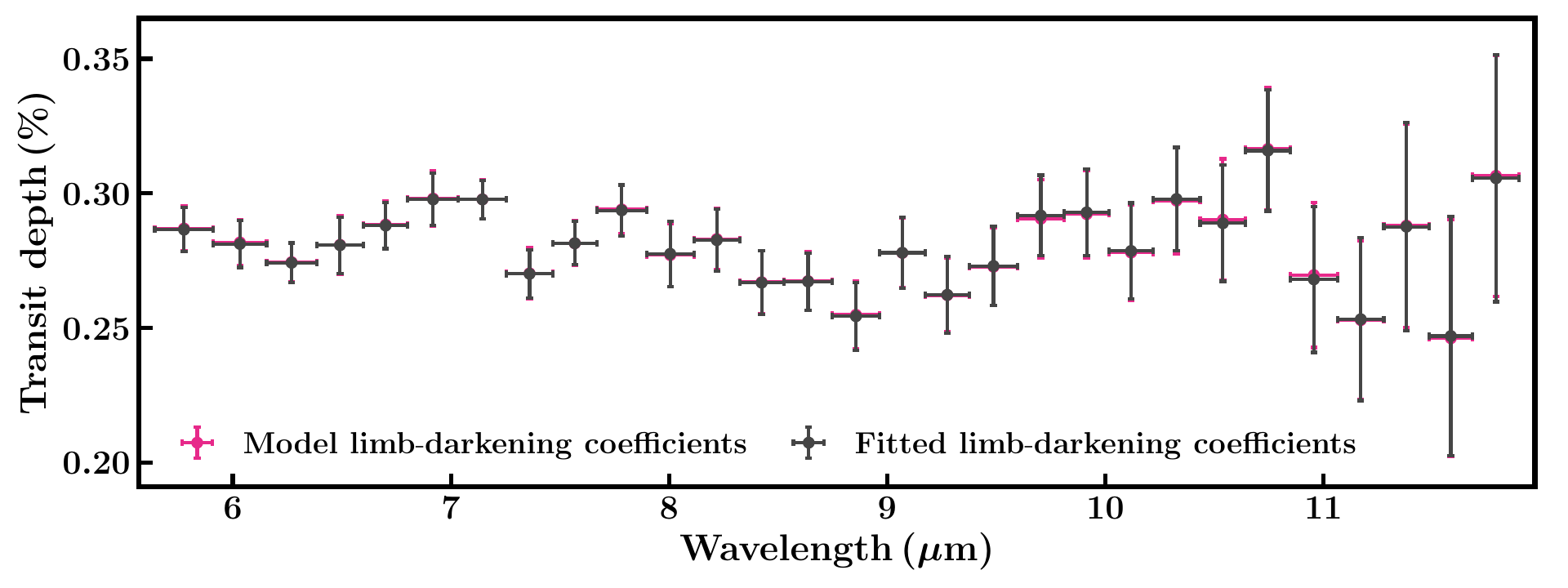}
    \caption{ Comparison of different limb darkening assumptions on the MIRI transmission spectrum of K2-18~b. The wavelength-dependent model limb darkening case is shown in pink, while the wavelength-independent empirical limb darkening case is shown in dark grey. The model limb darkening coefficients were obtained using an ATLAS9 model \citep{Castelli2003}, computed via ExoCTK \citep{bourque2021}.} 
    \label{fig:limb_darkening} 
\end{figure}

In our canonical data reduction, for both \texttt{JExoRES} and \texttt{JexoPipe}, we fix the quadratic limb-darkening coefficients to the values estimated from the corresponding white light curves, as given in Table~\ref{tab:wlc_params}. This is motivated by the small and approximately wavelength-independent limb darkening in the mid-infrared, between 5-12~\textmu m. To test this assumption, we generate a transmission spectrum of K2-18~b using wavelength-dependent model limb-darkening coefficients obtained via ExoCTK \citep{bourque2021}. For this, we use an ATLAS9 model \citep{Castelli2003} -- predicting $u_1 = 0.028 \pm 0.008$ and $u_2 = 0.107 \pm 0.011$ when integrating over the MIRI LRS bandpass (5-10 \textmu m) -- in agreement with our white light curve estimates. Figure~\ref{fig:limb_darkening} shows that the transmission spectrum of K2-18~b does not significantly differ between our two approaches, highlighting the robustness of the spectrum with respect to the effect of limb darkening.

\subsection{Time-correlated Noise} \label{app:noise}

\begin{figure}
	\includegraphics[width=0.495\textwidth]{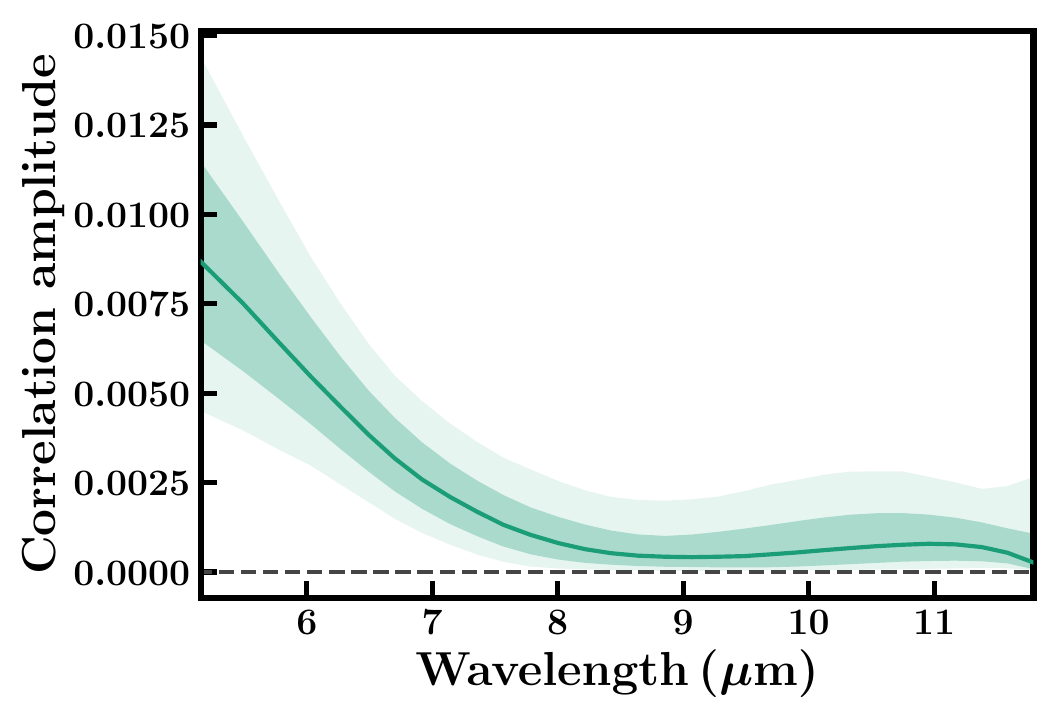}
    \includegraphics[width=0.495\textwidth]{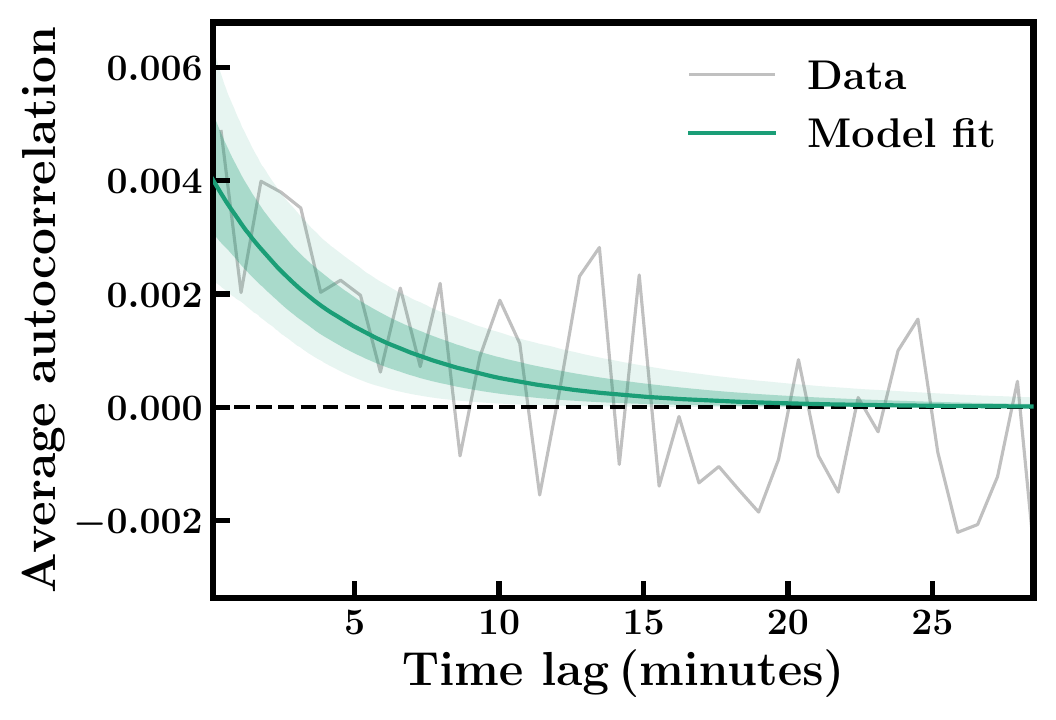}
    \caption{ Illustration of the correlation amplitude $\rho$ (left) and average autocorrelation function (right). The median fit is shown in solid green, while the 1-$\sigma$ and 2-$\sigma$ regions are shown in two lighter shades of green. }
    \label{fig:noise} 
\end{figure}

We find evidence for time-correlated noise when analysing the light curves. By inspecting the autocorrelation functions as a function of wavelength, we find that the time-correlated noise has a time scale of a few minutes and that the amplitude of this noise increases towards shorter wavelengths. To account for this, we fit a model of the noise to the residuals of the light curves after subtracting the fitted transit models and while masking the first hour of the observation. Based on the exponential-like shape of the autocorrelation functions, we employ the following wavelength-dependent GP kernel:

\begin{equation} \label{eq:kernel}
    \kappa_\lambda(t_i, t_j) = \sigma_\lambda^2 \, \delta_{ij} + \rho(\lambda)\,  \sigma_\lambda^2 \, e^{-|t_i-t_j| / \eta}\,,
\end{equation}

where $\sigma_\lambda$ is the standard deviation of a spectral channel with central wavelength $\lambda$, $\rho(\lambda)$ is a smooth function describing the amplitude of the time-correlated noise, and where $ \eta$ is the time scale of this noise (assumed to be wavelength-independent). We parameterise $\rho(\lambda)$ as a cubic polynomial, defined by four points that are equally spaced in wavelength. This gives us five parameters in total to describe the kernel. To estimate these parameters, we add up the log-likelihoods from each spectral channel, as computed via \texttt{celerite} \citep{Foreman-Mackey2017}, and use \texttt{MultiNest} \citep{Feroz2009} to obtain the posterior distributions. Figure~\ref{fig:noise} shows the estimated $\rho(\lambda)$ together with the average autocorrelation function.

\begin{figure}
	\includegraphics[width=1.0\textwidth]{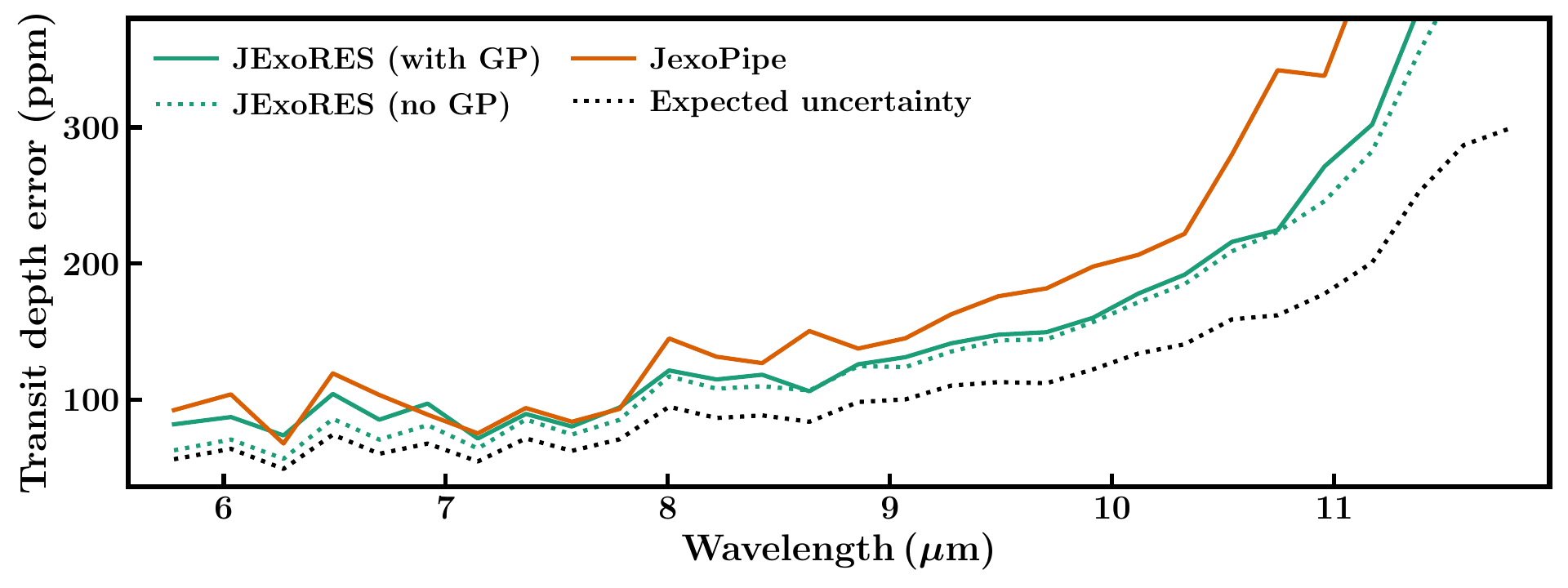}
    \caption{ Transit depth uncertainties of our MIRI transmission spectrum of K2-18~b with 0.2~\textmu m bins. The solid and dotted green lines show the uncertainties for \texttt{JExoRES}, with and without using the GP model, respectively, as discussed in Appendix~\ref{app:noise}.  The solid orange line shows the uncertainties obtained with \texttt{JexoPipe}, as also discussed in Appendix~\ref{app:noise}. The dotted line shows the expected uncertainties, obtained by propagating the photon and read noise.}
    \label{fig:spectrum_uncertainty} 
\end{figure}

Next, we use the median $\rho(\lambda)$ and $\eta$ to re-fit the light curves with the above kernel, again using \texttt{celerite} to evaluate the likelihood. For this, we do not use an error inflation parameter as that is already taken into account by $\sigma_\lambda$. We find that the center points of the spectrum (that is, the median values of the transit-depth posterior distributions) remains very similar to the case without using the kernel in (\ref{eq:kernel}), while the transit depth uncertainties increase by around 0-30~\%, reaching a maximum at the shortest wavelengths. Figure~\ref{fig:spectrum_uncertainty} shows the transit depth uncertainties with and without our GP model. Given these findings, we use the GP case as our canonical\texttt{JExoRES} spectrum, which has more conservative error bars. 

As an additional robustness check, we also account for time-correlated noise using a different method in the \texttt{JexoPipe} pipeline, applying the `time-averaging' method \citep{Winn2007,Cubillos2017}.  Here we obtain Allan deviation plots for the residuals from each spectral light curve, and find the average $ \beta$ factor for bins between 5 and 20 minutes duration, where $ \beta$ is the ratio between the observed fractional noise to the expected value for uncorrelated noise.  Including only $ \beta$ values $ \geq 1$, we find a median $ \beta$ of 1.14 across all the spectral bins used.  To account for correlated noise, we then inflate the errors on the light curve data points by these factors and repeat the light curve fits. This gives the final \texttt{JexoPipe} spectrum used for retrieval analysis with conservative error bars. Figure \ref{fig:spectrum_uncertainty} shows the final transit depth uncertainties.

\begin{figure*}
    \centering
    \includegraphics[width=\textwidth]{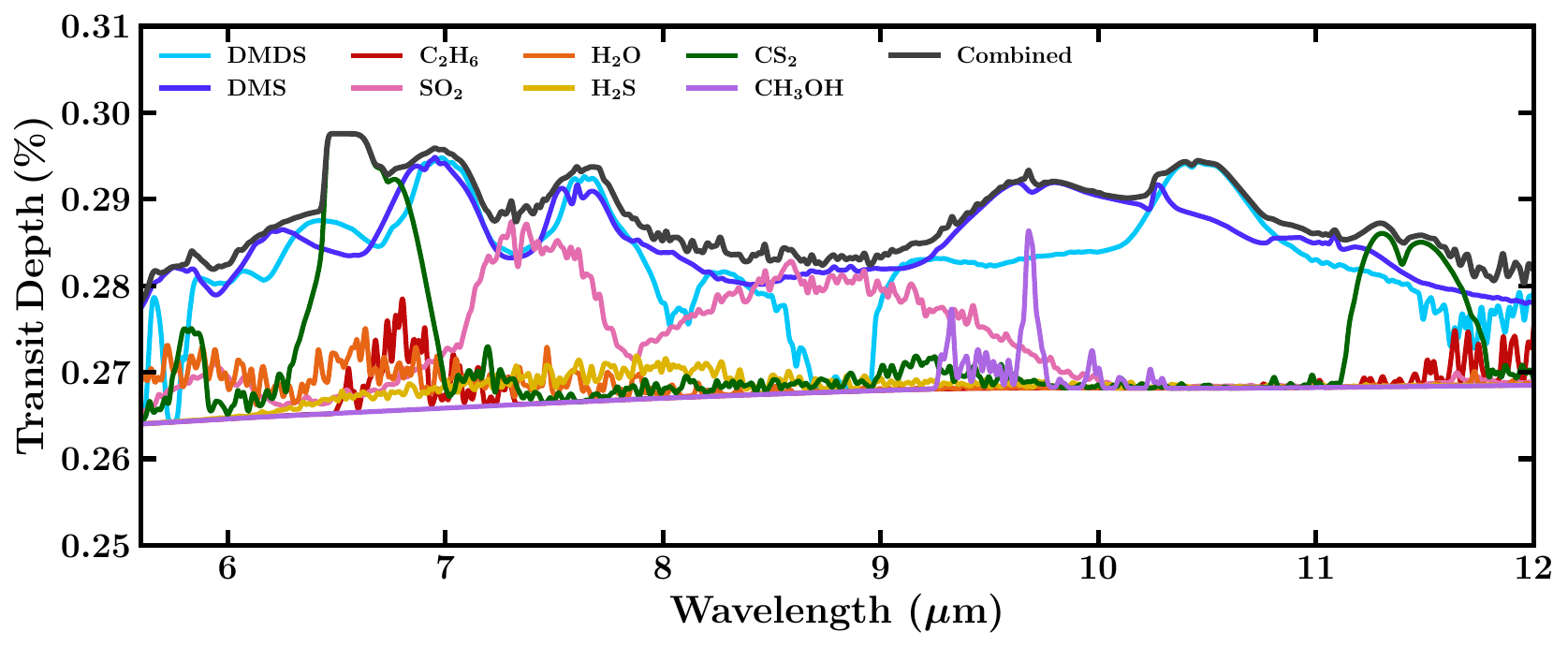}
    \caption{Spectral contributions from DMDS and DMS, also shown in Figure \ref{fig:contributions}, as well as several other molecular species with significant cross-sections in the mid-infrared considered in the maximal retrieval described in Section \ref{sec:retrieval_setup}. All mixing ratios are set to $5 \times 10^{-4}$. }
    \label{fig:additional_contributions}
\end{figure*}

\begin{figure*}
    \centering
    \includegraphics[width=\textwidth]{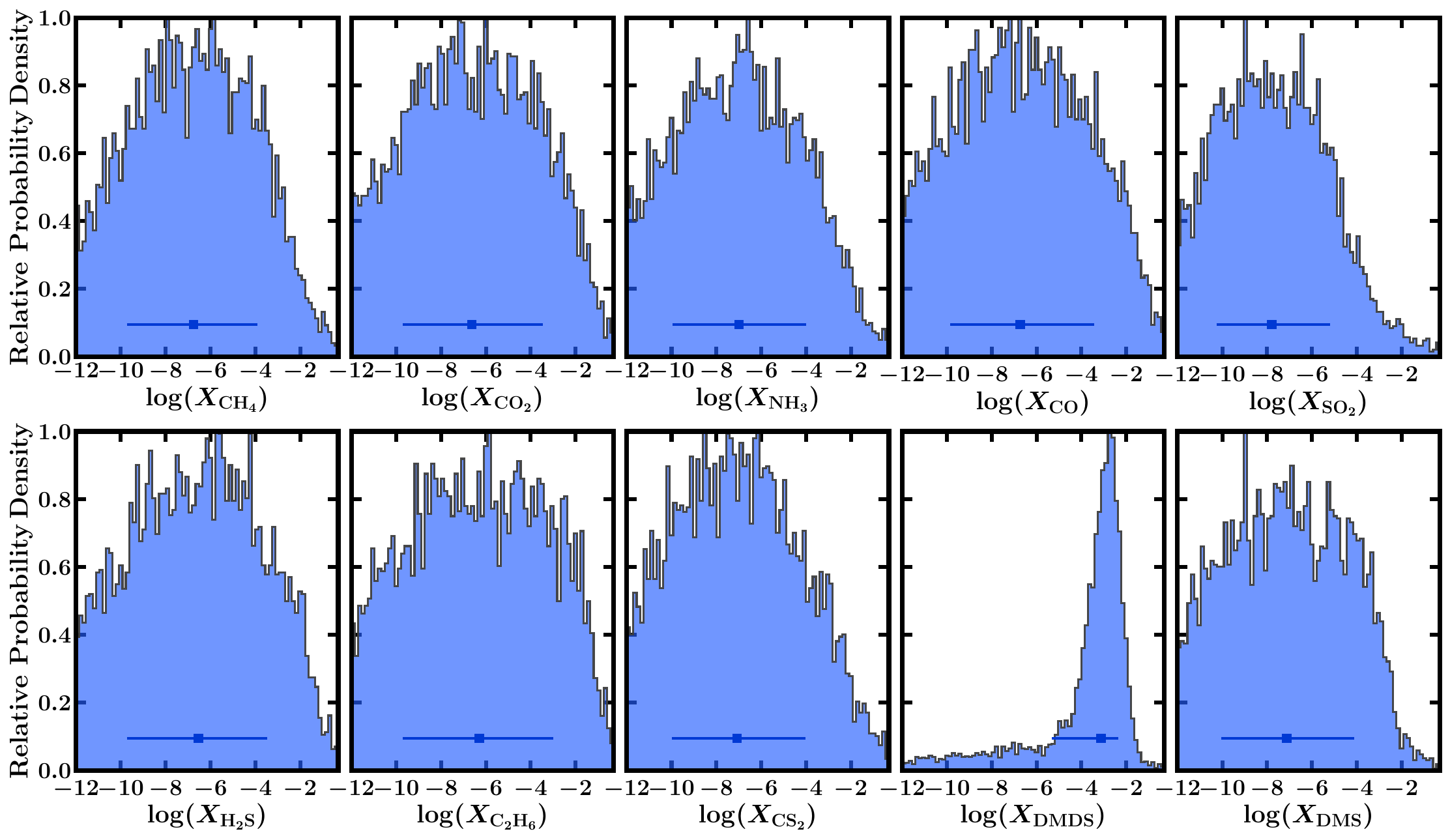}
    \caption{ Posterior probability distributions for a number of molecular species included in the maximal retrieval described in Section \ref{sec:retrieval_setup}. We find that only 1 out of the 20 species considered, DMDS, has a well-constrained posterior as shown. When DMDS is removed from the model a peak in the posterior is found for DMS.}
    \label{fig:additional_posteriors}
\end{figure*}

\section{Retrieval Setup}
\label{app:retrievals}

\begin{table*}
\small
\def\arraystretch{1.4}
\caption{Retrieved atmospheric parameters and corresponding prior probability distributions for the canonical retrieval with the \texttt{JExoRES} spectrum.}
\begin{tabular}{lclccc} \hline \hline
    Parameter & Bayesian Prior  & Description & DMS + DMDS & DMS Only & DMDS Only\\[0.5mm]
    \hline
    $\mathrm{log}(X_\mathrm{DMDS})$ & $\mathcal{U}$(-12, -0.3) & Mixing ratio of DMDS &                     $-3.48_{-2.27}^{+1.24}$ &       - &                             $-3.25_{-1.30}^{+1.17}$\\[0.5mm]
    $\mathrm{log}(X_\mathrm{DMS})$ & $\mathcal{U}$(-12, -0.3) & Mixing ratio of DMS &                       $-6.35_{-3.47}^{+2.75}$ &       $-3.42_{-1.44}^{+1.16}$ &       -\\[0.5mm]
    $\mathrm{log}(X_\mathrm{CH_4})$  & $\mathcal{U}$(-12, -0.3) & Mixing ratio of CH$_4$    &             $-6.66_{-3.22}^{+3.22}$ &       $-6.89_{-3.25}^{+3.23}$ &       $-6.82_{-3.17}^{+3.23}$\\[0.5mm]
    $\mathrm{log}(X_\mathrm{CO_2})$ & $\mathcal{U}$(-12, -0.3) & Mixing ratio of CO$_2$   &               $-6.42_{-3.47}^{+2.75}$ &       $-6.72_{-3.34}^{+3.53}$ &       $-6.59_{-3.35}^{+3.40}$\\[0.5mm]

    $T_0 / \mathrm{K} $ &             $\mathcal{U}$(100, 500)&  Temperature at 0.1 $\mu$bar&                $354_{-126}^{+129}$ &           $370_{-145}^{+131}$ &           $344_{-131}^{+128}$\\[0.5mm]
    
    $T_{1 \mathrm{mbar}} / \mathrm{K}$ & - & Temperature at 1 mbar &                                        $422_{-133}^{+141}$ &           $438_{-157}^{+145}$ &             $402_{-129}^{+143}$\\[0.5mm]
    
    $\alpha_1 / \mathrm{K}^{-\frac{1}{2}}$  &   $\mathcal{U}$(0.02, 2.00) & $P$-$T$ profile curvature &     $1.16_{-0.50}^{+0.50}$ &        $1.19_{-0.53}^{+0.50}$ &        $1.20_{-0.50}^{+0.50}$\\[0.5mm]
    $\alpha_2/ \mathrm{K}^{-\frac{1}{2}}$&      $\mathcal{U}$(0.02, 2.00)  & $P$-$T$ profile curvature &    $1.12_{-0.54}^{+0.58}$ &        $1.12_{-0.60}^{+0.55}$ &        $1.11_{-0.55}^{+0.59}$\\[0.5mm]

    $\mathrm{log}(P_1/\mathrm{bar})$   &        $\mathcal{U}$(-6, 1) & $P$-$T$ profile region limit &       $-2.53_{-1.25}^{+1.11}$ &       $-2.53_{-1.24}^{+1.10}$ &       $-2.43_{-1.28}^{+1.08}$\\[0.5mm]
    $\mathrm{log}(P_2/\mathrm{bar})$  &         $\mathcal{U}$(-6, 1) & $P$-$T$ profile region limit &       $-4.45_{-0.98}^{+1.28}$ &       $-4.50_{-0.98}^{+1.27}$ &       $-4.40_{-1.03}^{+1.33}$\\[0.5mm]
    $\mathrm{log}(P_3/\mathrm{bar})$   &        $\mathcal{U}$(-2, 1)& $P$-$T$ profile region limit &        $-0.84_{-0.65}^{+0.53}$ &       $-0.88_{-0.67}^{+0.58}$ &       $-0.84_{-0.63}^{+0.54}$\\[0.5mm]
    
    $\mathrm{log}(P_\mathrm{ref}/\mathrm{bar})$&$\mathcal{U}$(-6, 0)& Reference pressure at R$_\mathrm{P}$ &$-4.32_{-0.93}^{+1.15}$ &       $-4.27_{-0.96}^{+1.30}$ &       $-4.22_{-1.03}^{+1.27}$\\[0.5mm]
    
    $\mathrm{log}(a)$  & $\mathcal{U}$(-4, 10)& Rayleigh enhancement factor &                               $2.23_{-4.49}^{+3.92}$ &        $2.51_{-4.01}^{+4.46}$ &        $2.94_{-4.36}^{+4.34}$\\[0.5mm]
    $\gamma$   & $\mathcal{U}$(-20, 2)& Scattering slope &                                                  $-9.51_{-6.42}^{+6.42}$ &       $-10.01_{-6.35}^{+6.49}$ &      $-9.93_{-6.43}^{+6.63}$\\[0.5mm]
    $\mathrm{log}(P_\mathrm{c}/\mathrm{bar})$  & $\mathcal{U}$(-6, 1)& Cloud top pressure &                 $-2.20_{-1.96}^{+1.92}$ &       $-1.90_{-2.02}^{+1.86}$ &       $-2.21_{-1.94}^{+1.87}$\\[0.5mm]
    $\phi$  & $\mathcal{U}$(0, 1)& Cloud/haze coverage fraction &                                           $0.49_{-0.30}^{+0.31}$ &        $0.45_{-0.30}^{+0.32}$ &        $0.46_{-0.29}^{+0.31}$\\[0.5mm]
    
    $\delta_\mathrm{MIRI} / \mathrm{ppm}$  & $\mathcal{U}$(-100, 100)& MIRI dataset offset &                $-12_{-58}^{+51}$  &          $22_{-63}^{+48} $&              $-12_{-61}^{+52}$\\[0.5mm]
    \hline
\end{tabular}
\newline
\footnotesize{\textbf{Note.} The quoted parameter values denote the retrieved median and 1-$\sigma$ intervals. The $T_{1 \mathrm{mbar}}$ values are inferred from the retrieved PT-profile parameter constraints.
}
\label{tab:retrieval_priors}
\end{table*}

\begin{figure*}
    \centering
    \includegraphics[width=\textwidth]{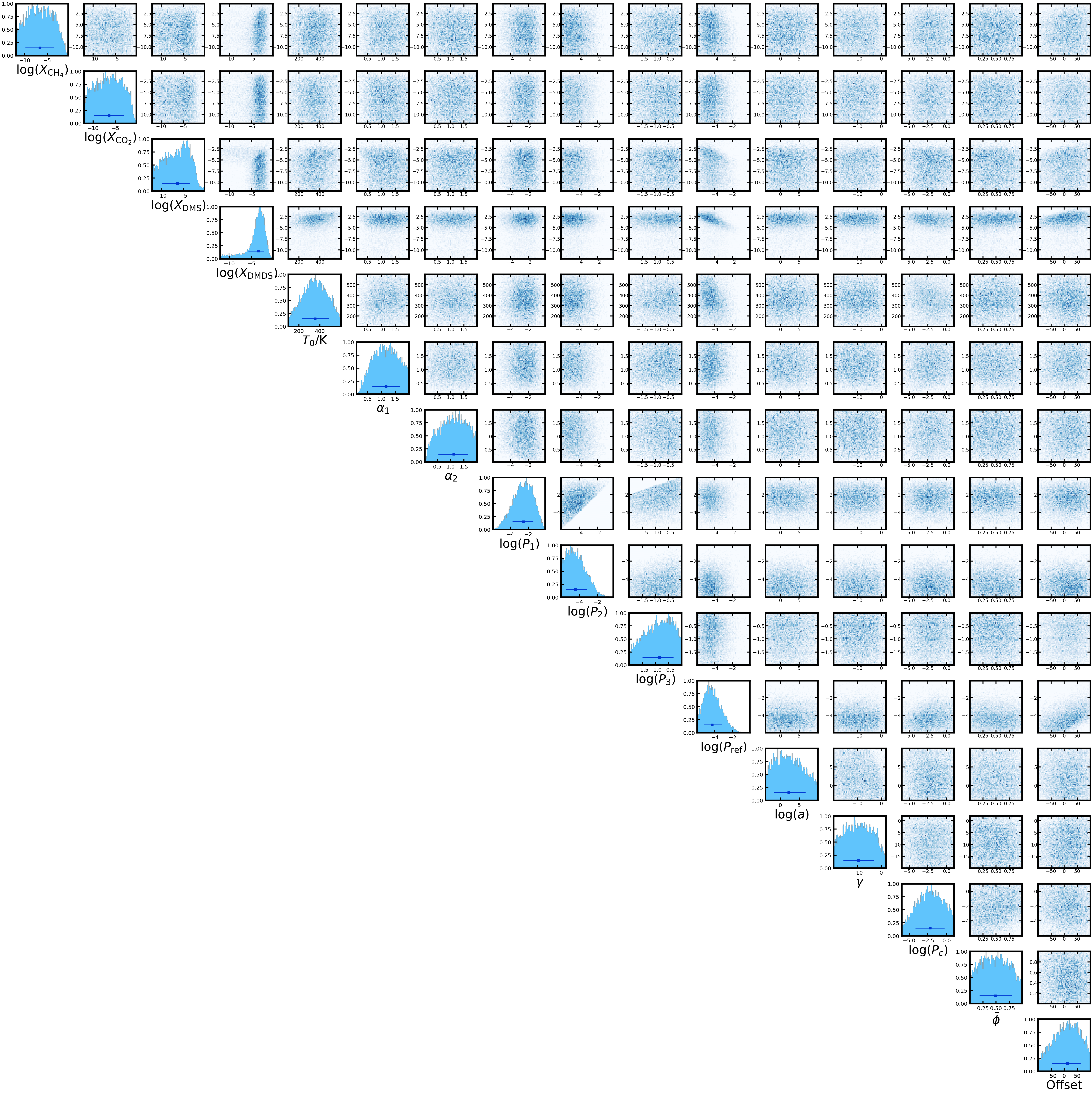}
    \caption{Posterior probability distribution for the canonical retrieval using the \texttt{JExoRES} spectrum. Diagonal panels show the posterior probability distribution for each parameter and off-diagonal panels show the pairwise correlations. Horizontal errorbars denote the median and 1-$\sigma$ intervals for each parameter.}
    \label{fig:full_posterior}
\end{figure*}

The present retrievals consider atmospheric opacity contributions from several molecular species. As discussed in section~\ref{sec:retrieval_setup}, for the maximal retrieval we consider opacity contributions from 20 notable molecular species. We compute the absorption cross sections from line list data, following \citet{Gandhi2017} and  \citet{Gandhi2020}, for CH$_4$ \citep{Yurchenko2014}, CO$_2$ \citep{Tashkun2015}, H$_2$O \citep{Barber2006, Rothman2010}, NH$_3$ \citep{Yurchenko2011}, CO \citep{Li2015}, SO$_2$ \citep{Underwood2016}, HCN \citep{Barber2014}, H$_2$S \citep{Azzam2016, Chubb2018}, CH$_3$OH \citep{Harrison2012}, C$_2$H$_2$ \citep{c2h2_mir, HITRAN2020}, C$_2$H$_4$ \citep{c2h4_mir, HITRAN2020}, C$_2$H$_6$ \citep{Reed2015}, CH$_3$Cl \citep{ch3cl_1, ch3cl_2}, OCS \citep{ocs_1, ocs_2, ocs_3, ocs_4, ocs_5, ocs_6, ocs_7}, N$_2$O \citep{n2o_2} and PH$_3$ \citep{HITRAN2020}. For DMS, DMDS, CH$_3$SH and CS$_2$, we consider the \textsc{HITRAN} \citep{dms_cs2_2,HITRAN2016,dms_cs2_1} absorption cross-sections at 1~bar and 298~K, similarly to \citet{Madhusudhan2021}.
Table \ref{tab:retrieval_priors} shows the priors and retrieved constraints for all free parameters of the canonical model, and the models without DMDS or DMS. Figure \ref{fig:full_posterior} shows the corresponding posterior distribution for all parameters of the canonical atmospheric model.

Figure \ref{fig:additional_contributions} shows spectral contributions from a number of the molecules listed above with significant cross-sections in the mid-infrared, which are included in the maximal atmospheric model described in Section \ref{sec:retrieval_setup}. Figure \ref{fig:additional_posteriors} shows the posterior distributions retrieved for 10 of the 20 molecular species considered in the maximal atmospheric model. Only 1 of the 20 molecules, DMDS, is well constrained by the data for this retrieval. Figure \ref{fig:full_posterior} shows the posteriors for all the parameters for the canonical retrieval discussed in Sections \ref{sec:retrieval_setup}
and \ref{sec:atm_constraints}, also showing a peak in the posterior for DMDS. When DMDS is removed from the model a peak in the posterior is found for DMS, as shown in Figure \ref{fig:posteriors}.

\bibliography{references.bib, refs1}

\newcommand{\noop}[1]{}
\begin{thebibliography}{}
\expandafter\ifx\csname natexlab\endcsname\relax\def\natexlab#1{#1}\fi
\providecommand{\url}[1]{\href{#1}{#1}}
\providecommand{\dodoi}[1]{doi:~\href{http://doi.org/#1}{\nolinkurl{#1}}}
\providecommand{\doeprint}[1]{\href{http://ascl.net/#1}{\nolinkurl{http://ascl.net/#1}}}
\providecommand{\doarXiv}[1]{\href{https://arxiv.org/abs/#1}{\nolinkurl{https://arxiv.org/abs/#1}}}

\bibitem[{{Abel} {et~al.}(2011){Abel}, {Frommhold}, {Li}, \& {Hunt}}]{abel2011}
{Abel}, M., {Frommhold}, L., {Li}, X., \& {Hunt}, K. L.~C. 2011, Journal of Physical Chemistry A, 115, 6805, \dodoi{10.1021/jp109441f}

\bibitem[{{Alam} {et~al.}(2025){Alam}, {Gao}, {Adams Redai}, {Wallack}, {Wogan}, {Aguichine}, {Dattilo}, {Alderson}, {Batalha}, {Batalha}, {Kirk}, {L{\'o}pez-Morales}, {Meech}, {Moran}, {Teske}, {Wakeford}, \& {Wolfgang}}]{Alam2025}
{Alam}, M.~K., {Gao}, P., {Adams Redai}, J., {et~al.} 2025, \aj, 169, 15, \dodoi{10.3847/1538-3881/ad8eb5}

\bibitem[{{Alderson} {et~al.}(2024){Alderson}, {Batalha}, {Wakeford}, {Wallack}, {Aguichine}, {Teske}, {Adams Redai}, {Alam}, {Batalha}, {Gao}, {Kirk}, {L{\'o}pez-Morales}, {Moran}, {Scarsdale}, {Wogan}, \& {Wolfgang}}]{Alderson2024}
{Alderson}, L., {Batalha}, N.~E., {Wakeford}, H.~R., {et~al.} 2024, \aj, 167, 216, \dodoi{10.3847/1538-3881/ad32c9}

\bibitem[{Auwera \& Fayt(2006)}]{ocs_4}
Auwera, J.~V., \& Fayt, A. 2006, Journal of Molecular Structure, 780-781, 134, \dodoi{10.1016/j.molstruc.2005.04.052}

\bibitem[{Auwera {et~al.}(2007)Auwera, Moazzen-Ahmadi, \& Flaud}]{c2h4_mir}
Auwera, J.~V., Moazzen-Ahmadi, N., \& Flaud, J.-M. 2007, Astrophysical Journal, 662, 750, \dodoi{10.1086/515567}

\bibitem[{{Azzam} {et~al.}(2016){Azzam}, {Tennyson}, {Yurchenko}, \& {Naumenko}}]{Azzam2016}
{Azzam}, A. A.~A., {Tennyson}, J., {Yurchenko}, S.~N., \& {Naumenko}, O.~V. 2016, \mnras, 460, 4063, \dodoi{10.1093/mnras/stw1133}

\bibitem[{{Barber} {et~al.}(2014){Barber}, {Strange}, {Hill}, {Polyansky}, {Mellau}, {Yurchenko}, \& {Tennyson}}]{Barber2014}
{Barber}, R.~J., {Strange}, J.~K., {Hill}, C., {et~al.} 2014, \mnras, 437, 1828, \dodoi{10.1093/mnras/stt2011}

\bibitem[{{Barber} {et~al.}(2006){Barber}, {Tennyson}, {Harris}, \& {Tolchenov}}]{Barber2006}
{Barber}, R.~J., {Tennyson}, J., {Harris}, G.~J., \& {Tolchenov}, R.~N. 2006, \mnras, 368, 1087, \dodoi{10.1111/j.1365-2966.2006.10184.x}

\bibitem[{{Barclay} {et~al.}(2021){Barclay}, {Kostov}, {Col{\'o}n}, {Quintana}, {Schlieder}, {Louie}, {Gilbert}, \& {Mullally}}]{Barclay2021}
{Barclay}, T., {Kostov}, V.~B., {Col{\'o}n}, K.~D., {et~al.} 2021, \aj, 162, 300, \dodoi{10.3847/1538-3881/ac2824}

\bibitem[{{Bell} {et~al.}(2024){Bell}, {Crouzet}, {Cubillos}, {Kreidberg}, {Piette}, {Roman}, {Barstow}, {Blecic}, {Carone}, {Coulombe}, {Ducrot}, {Hammond}, {Mendon{\c{c}}a}, {Moses}, {Parmentier}, {Stevenson}, {Teinturier}, {Zhang}, {Batalha}, {Bean}, {Benneke}, {Charnay}, {Chubb}, {Demory}, {Gao}, {Lee}, {L{\'o}pez-Morales}, {Morello}, {Rauscher}, {Sing}, {Tan}, {Venot}, {Wakeford}, {Aggarwal}, {Ahrer}, {Alam}, {Baeyens}, {Barrado}, {Caceres}, {Carter}, {Casewell}, {Challener}, {Crossfield}, {Decin}, {D{\'e}sert}, {Dobbs-Dixon}, {Dyrek}, {Espinoza}, {Feinstein}, {Gibson}, {Harrington}, {Helling}, {Hu}, {Iro}, {Kempton}, {Kendrew}, {Komacek}, {Krick}, {Lagage}, {Leconte}, {Lendl}, {Lewis}, {Lothringer}, {Malsky}, {Mancini}, {Mansfield}, {Mayne}, {Evans-Soma}, {Molaverdikhani}, {Nikolov}, {Nixon}, {Palle}, {Petit dit de la Roche}, {Piaulet}, {Powell}, {Rackham}, {Schneider}, {Steinrueck}, {Taylor}, {Welbanks}, {Yurchenko}, {Zhang}, \& {Zieba}}]{Bell2024}
{Bell}, T.~J., {Crouzet}, N., {Cubillos}, P.~E., {et~al.} 2024, Nature Astronomy, 8, 879, \dodoi{10.1038/s41550-024-02230-x}

\bibitem[{{Benneke} {et~al.}(2019{\natexlab{a}}){Benneke}, {Wong}, {Piaulet}, {Knutson}, {Lothringer}, {Morley}, {Crossfield}, {Gao}, {Greene}, {Dressing}, {Dragomir}, {Howard}, {McCullough}, {Kempton}, {Fortney}, \& {Fraine}}]{Benneke2019_K218b}
{Benneke}, B., {Wong}, I., {Piaulet}, C., {et~al.} 2019{\natexlab{a}}, \apjl, 887, L14, \dodoi{10.3847/2041-8213/ab59dc}

\bibitem[{{Benneke} {et~al.}(2019{\natexlab{b}}){Benneke}, {Wong}, {Piaulet}, {Knutson}, {Lothringer}, {Morley}, {Crossfield}, {Gao}, {Greene}, {Dressing}, {Dragomir}, {Howard}, {McCullough}, {Kempton}, {Fortney}, \& {Fraine}}]{Benneke2019}
---. 2019{\natexlab{b}}, \apjl, 887, L14, \dodoi{10.3847/2041-8213/ab59dc}

\bibitem[{Benneke {et~al.}(2024)Benneke, Roy, Coulombe, Radica, Piaulet, Ahrer, Pierrehumbert, Krissansen-Totton, Schlichting, Hu, Yang, Christie, Thorngren, Young, Pelletier, Knutson, Miguel, Evans-Soma, Dorn, Gagnebin, Fortney, Komacek, MacDonald, Raul, Cloutier, Acuna, Lafrenière, Cadieux, Doyon, Welbanks, \& Allart}]{benneke_jwst_2024}
Benneke, B., Roy, P.-A., Coulombe, L.-P., {et~al.} 2024, {JWST} {Reveals} {CH}\$\_4\$, {CO}\$\_2\$, and {H}\$\_2\${O} in a {Metal}-rich {Miscible} {Atmosphere} on a {Two}-{Earth}-{Radius} {Exoplanet},  arXiv.
\newblock \url{http://arxiv.org/abs/2403.03325}

\bibitem[{{B{\'e}zard} {et~al.}(2020){B{\'e}zard}, {Charnay}, \& {Blain}}]{Bezard2022}
{B{\'e}zard}, B., {Charnay}, B., \& {Blain}, D. 2020, arXiv e-prints, arXiv:2011.10424, \dodoi{10.48550/arXiv.2011.10424}

\bibitem[{{Bieler} {et~al.}(2015){Bieler}, {Altwegg}, {Balsiger}, {Bar-Nun}, {Berthelier}, {Bochsler}, {Briois}, {Calmonte}, {Combi}, {de Keyser}, {van Dishoeck}, {Fiethe}, {Fuselier}, {Gasc}, {Gombosi}, {Hansen}, {H{\"a}ssig}, {J{\"a}ckel}, {Kopp}, {Korth}, {Le Roy}, {Mall}, {Maggiolo}, {Marty}, {Mousis}, {Owen}, {R{\`e}me}, {Rubin}, {S{\'e}mon}, {Tzou}, {Waite}, {Walsh}, \& {Wurz}}]{Bieler2015}
{Bieler}, A., {Altwegg}, K., {Balsiger}, H., {et~al.} 2015, \nat, 526, 678, \dodoi{10.1038/nature15707}

\bibitem[{{Blain} {et~al.}(2021){Blain}, {Charnay}, \& {B{\'e}zard}}]{Blain2021}
{Blain}, D., {Charnay}, B., \& {B{\'e}zard}, B. 2021, \aap, 646, A15, \dodoi{10.1051/0004-6361/202039072}

\bibitem[{{Borysow} {et~al.}(1988){Borysow}, {Frommhold}, \& {Birnbaum}}]{Borysow1988}
{Borysow}, J., {Frommhold}, L., \& {Birnbaum}, G. 1988, \apj, 326, 509, \dodoi{10.1086/166112}

\bibitem[{Bouanich {et~al.}(1986)Bouanich, Blanquet, Walrand, \& Courtoy}]{ocs_1}
Bouanich, J.-P., Blanquet, G., Walrand, J., \& Courtoy, C.~P. 1986, Journal of Quantitative Spectroscopy and Radiative Transfer, 36, 295, \dodoi{10.1016/0022-4073(86)90053-1}

\bibitem[{Bourque {et~al.}(2021)Bourque, Espinoza, Filippazzo, Fix, King, Martlin, Medina, Batalha, Fox, Fowler, Fraine, Hill, Lewis, Stevenson, Valenti, \& Wakeford}]{bourque2021}
Bourque, M., Espinoza, N., Filippazzo, J., {et~al.} 2021, The Exoplanet Characterization Toolkit (ExoCTK), 1.0.0,  Zenodo, \dodoi{10.5281/zenodo.4556063}

\bibitem[{{Bouwman} {et~al.}(2023){Bouwman}, {Kendrew}, {Greene}, {Bell}, {Lagage}, {Schreiber}, {Dicken}, {Sloan}, {Espinoza}, {Scheithauer}, {Coulais}, {Fox}, {Gastaud}, {Glauser}, {Jones}, {Labiano}, {Lahuis}, {Morrison}, {Murray}, {Mueller}, {Nayak}, {Wright}, {Glasse}, \& {Rieke}}]{Bouwman2023}
{Bouwman}, J., {Kendrew}, S., {Greene}, T.~P., {et~al.} 2023, \pasp, 135, 038002, \dodoi{10.1088/1538-3873/acbc49}

\bibitem[{Bray {et~al.}(2011)Bray, Perrin, Jacquemart, \& Lacome}]{ch3cl_1}
Bray, C., Perrin, A., Jacquemart, D., \& Lacome, N. 2011, Journal of Quantitative Spectroscopy and Radiative Transfer, 112, 2446, \dodoi{10.1016/j.jqsrt.2011.06.018}

\bibitem[{{Buchner}(2021)}]{Buchner2021}
{Buchner}, J. 2021, The Journal of Open Source Software, 6, 3001, \dodoi{10.21105/joss.03001}

\bibitem[{{Buchner} {et~al.}(2014){Buchner}, {Georgakakis}, {Nandra}, {Hsu}, {Rangel}, {Brightman}, {Merloni}, {Salvato}, {Donley}, \& {Kocevski}}]{Buchner2014}
{Buchner}, J., {Georgakakis}, A., {Nandra}, K., {et~al.} 2014, \aap, 564, A125, \dodoi{10.1051/0004-6361/201322971}

\bibitem[{{Bushouse}(2020)}]{Bushouse2020}
{Bushouse}, H. 2020, in Astronomical Society of the Pacific Conference Series, Vol. 527, Astronomical Data Analysis Software and Systems XXIX, ed. R.~{Pizzo}, E.~R. {Deul}, J.~D. {Mol}, J.~{de Plaa}, \& H.~{Verkouter}, 583

\bibitem[{{Castelli} \& {Kurucz}(2003)}]{Castelli2003}
{Castelli}, F., \& {Kurucz}, R.~L. 2003, in IAU Symposium, Vol. 210, Modelling of Stellar Atmospheres, ed. N.~{Piskunov}, W.~W. {Weiss}, \& D.~F. {Gray}, A20.
\newblock \doarXiv{astro-ph/0405087}

\bibitem[{{Catling} {et~al.}(2018){Catling}, {Krissansen-Totton}, {Kiang}, {Crisp}, {Robinson}, {DasSarma}, {Rushby}, {Del Genio}, {Bains}, \& {Domagal-Goldman}}]{Catling2018}
{Catling}, D.~C., {Krissansen-Totton}, J., {Kiang}, N.~Y., {et~al.} 2018, Astrobiology, 18, 709, \dodoi{10.1089/ast.2017.1737}

\bibitem[{{Chubb} {et~al.}(2018){Chubb}, {Naumenko}, {Keely}, {Bartolotto}, {Macdonald}, {Mukhtar}, {Grachov}, {White}, {Coleman}, {Liu}, {Fazliev}, {Polovtseva}, {Horneman}, {Campargue}, {Furtenbacher}, {Cs{\'a}sz{\'a}r}, {Yurchenko}, \& {Tennyson}}]{Chubb2018}
{Chubb}, K.~L., {Naumenko}, O., {Keely}, S., {et~al.} 2018, \jqsrt, 218, 178, \dodoi{10.1016/j.jqsrt.2018.07.012}

\bibitem[{{Cloutier} {et~al.}(2017){Cloutier}, {Astudillo-Defru}, {Doyon}, {Bonfils}, {Almenara}, {Benneke}, {Bouchy}, {Delfosse}, {Ehrenreich}, {Forveille}, {Lovis}, {Mayor}, {Menou}, {Murgas}, {Pepe}, {Rowe}, {Santos}, {Udry}, \& {W{\"u}nsche}}]{cloutier2017}
{Cloutier}, R., {Astudillo-Defru}, N., {Doyon}, R., {et~al.} 2017, \aap, 608, A35, \dodoi{10.1051/0004-6361/201731558}

\bibitem[{{Cloutier} {et~al.}(2019){Cloutier}, {Astudillo-Defru}, {Doyon}, {Bonfils}, {Almenara}, {Bouchy}, {Delfosse}, {Forveille}, {Lovis}, {Mayor}, {Menou}, {Murgas}, {Pepe}, {Santos}, {Udry}, \& {W{\"u}nsche}}]{Cloutier2019_K218b_mass}
---. 2019, \aap, 621, A49, \dodoi{10.1051/0004-6361/201833995}

\bibitem[{{Constantinou} \& {Madhusudhan}(2024)}]{Constantinou2024}
{Constantinou}, S., \& {Madhusudhan}, N. 2024, \mnras, 530, 3252, \dodoi{10.1093/mnras/stae633}

\bibitem[{{Constantinou} {et~al.}(2023){Constantinou}, {Madhusudhan}, \& {Gandhi}}]{Constantinou2023}
{Constantinou}, S., {Madhusudhan}, N., \& {Gandhi}, S. 2023, \apjl, 943, L10, \dodoi{10.3847/2041-8213/acaead}

\bibitem[{{Cooke} \& {Madhusudhan}(2024)}]{Cooke2024}
{Cooke}, G.~J., \& {Madhusudhan}, N. 2024, arXiv e-prints, arXiv:2410.07313, \dodoi{10.48550/arXiv.2410.07313}

\bibitem[{{Court} \& {Sephton}(2012)}]{Court2012}
{Court}, R.~W., \& {Sephton}, M.~A. 2012, \planss, 73, 233, \dodoi{10.1016/j.pss.2012.08.026}

\bibitem[{{Cubillos} {et~al.}(2017){Cubillos}, {Harrington}, {Loredo}, {Lust}, {Blecic}, \& {Stemm}}]{Cubillos2017}
{Cubillos}, P., {Harrington}, J., {Loredo}, T.~J., {et~al.} 2017, \aj, 153, 3, \dodoi{10.3847/1538-3881/153/1/3}

\bibitem[{{Damiano} {et~al.}(2024){Damiano}, {Bello-Arufe}, {Yang}, \& {Hu}}]{damiano_lhs_2024}
{Damiano}, M., {Bello-Arufe}, A., {Yang}, J., \& {Hu}, R. 2024, \apjl, 968, L22, \dodoi{10.3847/2041-8213/ad5204}

\bibitem[{Daumont {et~al.}(2001)Daumont, Auwera, Teffo, Perevalov, \& Tashkun}]{n2o_2}
Daumont, L., Auwera, J., Teffo, J.-L., Perevalov, V., \& Tashkun, S. 2001, Journal of Molecular Spectroscopy, 208, 281, \dodoi{10.1006/jmsp.2001.8400}

\bibitem[{{Domagal-Goldman} {et~al.}(2011{\natexlab{a}}){Domagal-Goldman}, {Meadows}, {Claire}, \& {Kasting}}]{DomagalGoldman2011}
{Domagal-Goldman}, S.~D., {Meadows}, V.~S., {Claire}, M.~W., \& {Kasting}, J.~F. 2011{\natexlab{a}}, Astrobiology, 11, 419, \dodoi{10.1089/ast.2010.0509}

\bibitem[{{Domagal-Goldman} {et~al.}(2011{\natexlab{b}}){Domagal-Goldman}, {Meadows}, {Claire}, \& {Kasting}}]{Domagal-goldman2011}
---. 2011{\natexlab{b}}, Astrobiology, 11, 419, \dodoi{10.1089/ast.2010.0509}

\bibitem[{Dyrek {et~al.}(2024)Dyrek, Min, Decin, Bouwman, Crouzet, Mollière, Lagage, Konings, Tremblin, Güdel, Pye, Waters, Henning, Vandenbussche, Ardevol~Martinez, Argyriou, Ducrot, Heinke, van Looveren, Absil, Barrado, Baudoz, Boccaletti, Cossou, Coulais, Edwards, Gastaud, Glasse, Glauser, Greene, Kendrew, Krause, Lahuis, Mueller, Olofsson, Patapis, Rouan, Royer, Scheithauer, Waldmann, Whiteford, Colina, van Dishoeck, Östlin, Ray, \& Wright}]{dyrek_so2_2024}
Dyrek, A., Min, M., Decin, L., {et~al.} 2024, Nature, 625, 51, \dodoi{10.1038/s41586-023-06849-0}

\bibitem[{{Fayolle} {et~al.}(2017){Fayolle}, {{\"O}berg}, {J{\o}rgensen}, {Altwegg}, {Calcutt}, {M{\"u}ller}, {Rubin}, {van der Wiel}, {Bjerkeli}, {Bourke}, {Coutens}, {van Dishoeck}, {Drozdovskaya}, {Garrod}, {Ligterink}, {Persson}, {Wampfler}, \& {Rosina Team}}]{Fayolle2017}
{Fayolle}, E.~C., {{\"O}berg}, K.~I., {J{\o}rgensen}, J.~K., {et~al.} 2017, Nature Astronomy, 1, 703, \dodoi{10.1038/s41550-017-0237-7}

\bibitem[{{Felton} {et~al.}(2022){Felton}, {Bastelberger}, {Mandt}, {Luspay-Kuti}, {Fauchez}, \& {Domagal-Goldman}}]{Felton2022}
{Felton}, R.~C., {Bastelberger}, S.~T., {Mandt}, K.~E., {et~al.} 2022, Journal of Geophysical Research (Planets), 127, e06853, \dodoi{10.1029/2021JE006853}

\bibitem[{Feroz {et~al.}(2009)Feroz, Hobson, \& Bridges}]{Feroz2009}
Feroz, F., Hobson, M.~P., \& Bridges, M. 2009, Monthly Notices of the Royal Astronomical Society, 398, 1601, \dodoi{10.1111/j.1365-2966.2009.14548.x}

\bibitem[{{Foreman-Mackey} {et~al.}(2017){Foreman-Mackey}, {Agol}, {Ambikasaran}, \& {Angus}}]{Foreman-Mackey2017}
{Foreman-Mackey}, D., {Agol}, E., {Ambikasaran}, S., \& {Angus}, R. 2017, \aj, 154, 220, \dodoi{10.3847/1538-3881/aa9332}

\bibitem[{Foreman-Mackey {et~al.}(2013)Foreman-Mackey, Hogg, Lang, \& Goodman}]{foreman-mackey_emcee_2013}
Foreman-Mackey, D., Hogg, D.~W., Lang, D., \& Goodman, J. 2013, Publications of the Astronomical Society of the Pacific, 125, 306, \dodoi{10.1086/670067}

\bibitem[{{Fukui} {et~al.}(2022)}]{Fukui2022}
{Fukui}, A., {et~al.} 2022, Publ. Astron. Soc. Japan, 74, L1, \dodoi{10.1093/pasj/psab106}

\bibitem[{{Gandhi} \& {Madhusudhan}(2017)}]{Gandhi2017}
{Gandhi}, S., \& {Madhusudhan}, N. 2017, MNRAS, 472, 2334, \dodoi{10.1093/mnras/stx1601}

\bibitem[{{Gandhi} {et~al.}(2020){Gandhi}, {Brogi}, {Yurchenko}, {Tennyson}, {Coles}, {Webb}, {Birkby}, {Guilluy}, {Hawker}, {Madhusudhan}, {Bonomo}, \& {Sozzetti}}]{Gandhi2020}
{Gandhi}, S., {Brogi}, M., {Yurchenko}, S.~N., {et~al.} 2020, \mnras, 495, 224, \dodoi{10.1093/mnras/staa981}

\bibitem[{{Gardner} {et~al.}(2006){Gardner}, {Mather}, {Clampin}, {Doyon}, {Greenhouse}, {Hammel}, {Hutchings}, {Jakobsen}, {Lilly}, {Long}, {Lunine}, {McCaughrean}, {Mountain}, {Nella}, {Rieke}, {Rieke}, {Rix}, {Smith}, {Sonneborn}, {Stiavelli}, {Stockman}, {Windhorst}, \& {Wright}}]{Gardner2006}
{Gardner}, J.~P., {Mather}, J.~C., {Clampin}, M., {et~al.} 2006, \ssr, 123, 485, \dodoi{10.1007/s11214-006-8315-7}

\bibitem[{Glein(2024)}]{glein_geochemical_2024}
Glein, C.~R. 2024, The Astrophysical Journal Letters, 964, L19, \dodoi{10.3847/2041-8213/ad3079}

\bibitem[{{Golebiowski} {et~al.}(2014){Golebiowski}, {de Ghellinck d'Elseghem Vaernewijck}, {Herman}, {Vander Auwera}, \& {Fayt}}]{ocs_2}
{Golebiowski}, D., {de Ghellinck d'Elseghem Vaernewijck}, X., {Herman}, M., {Vander Auwera}, J., \& {Fayt}, A. 2014, \jqsrt, 149, 184, \dodoi{10.1016/j.jqsrt.2014.07.005}

\bibitem[{Gomez {et~al.}(2010)Gomez, Jacquemart, Lacome, \& Mandin}]{c2h2_mir}
Gomez, L., Jacquemart, D., Lacome, N., \& Mandin, J.-Y. 2010, Journal of Quantitative Spectroscopy and Radiative Transfer, 111, 2256, \dodoi{10.1016/j.jqsrt.2010.01.031}

\bibitem[{Gordon {et~al.}(2017)Gordon, Rothman, Hill, Kochanov, Tan, Bernath, Birk, Boudon, Campargue, Chance, Drouin, Flaud, Gamache, Hodges, Jacquemart, Perevalov, Perrin, Shine, Smith, Tennyson, Toon, H.~Tran, Barbe, Csaszar, Devi, Furtenbacher, Harrison, Hartmann, Jolly, Johnson, Karman, Kleiner, Kyuberis, Loos, Lyulin, Massie, Mikhailenko, Moazzen-Ahmadi, Muller, Naumenko, Nikitin, Polyansky, Rey, Rotger, Sharpe, Sung, Starikova, Tashkun, Auwera, Wagner, Wilzewski, Wcislo, Yu, \& Zak}]{HITRAN2016}
Gordon, I., Rothman, L., Hill, C., {et~al.} 2017, Journal of Quantitative Spectroscopy and Radiative Transfer, \dodoi{10.1016/j.jqsrt.2017.06.038}

\bibitem[{{Gordon} {et~al.}(2022){Gordon}, {Rothman}, {Hargreaves}, {Hashemi}, {Karlovets}, {Skinner}, {Conway}, {Hill}, {Kochanov}, {Tan}, {Wcis{\l}o}, {Finenko}, {Nelson}, {Bernath}, {Birk}, {Boudon}, {Campargue}, {Chance}, {Coustenis}, {Drouin}, {Flaud}, {Gamache}, {Hodges}, {Jacquemart}, {Mlawer}, {Nikitin}, {Perevalov}, {Rotger}, {Tennyson}, {Toon}, {Tran}, {Tyuterev}, {Adkins}, {Baker}, {Barbe}, {Can{\`{e}}}, {Cs{'{a}}sz{'{a}}r}, {Dudaryonok}, {Egorov}, {Fleisher}, {Fleurbaey}, {Foltynowicz}, {Furtenbacher}, {Harrison}, {Hartmann}, {Horneman}, {Huang}, {Karman}, {Karns}, {Kassi}, {Kleiner}, {Kofman}, {Kwabia-Tchana}, {Lavrentieva}, {Lee}, {Long}, {Lukashevskaya}, {Lyulin}, {Makhnev}, {Matt}, {Massie}, {Melosso}, {Mikhailenko}, {Mondelain}, {M{"{u}}ller}, {Naumenko}, {Perrin}, {Polyansky}, {Raddaoui}, {Raston}, {Reed}, {Rey}, {Richard}, {T{'{o}}bi{'{a}}s}, {Sadiek}, {Schwenke}, {Starikova}, {Sung}, {Tamassia}, {Tashkun}, {Vander Auwera}, {Vasilenko}, {Vigasin}, {Villanueva}, {Vispoel}, {Wagner},
  {Yachmenev}, \& {Yurchenko}}]{HITRAN2020}
{Gordon}, I.~E., {Rothman}, L.~S., {Hargreaves}, R.~J., {et~al.} 2022, Journal of Quantitative Spectroscopy and Radiative Transfer, 277, 107949, \dodoi{10.1016/j.jqsrt.2021.107949}

\bibitem[{Grant \& Wakeford(2024)}]{Grant2024_exoticLD}
Grant, D., \& Wakeford, H.~R. 2024, Journal of Open Source Software, 9, 6816, \dodoi{10.21105/joss.06816}

\bibitem[{Grant {et~al.}(2023)Grant, Lewis, Wakeford, Batalha, Glidden, Goyal, Mullens, MacDonald, May, Seager, Stevenson, Valenti, Visscher, Alderson, Allen, Cañas, Colón, Clampin, Espinoza, Gressier, Huang, Lin, Long, Louie, Peña-Guerrero, Ranjan, Sotzen, Valentine, Anderson, Balmer, Bellini, Hoch, Kammerer, Libralato, Mountain, Perrin, Pueyo, Rickman, Rebollido, Sohn, Marel, \& Watkins}]{grant_jwst-tst_2023}
Grant, D., Lewis, N.~K., Wakeford, H.~R., {et~al.} 2023, The Astrophysical Journal Letters, 956, L32, \dodoi{10.3847/2041-8213/acfc3b}

\bibitem[{{Greene} {et~al.}(2023){Greene}, {Bell}, {Ducrot}, {Dyrek}, {Lagage}, \& {Fortney}}]{Greene2023}
{Greene}, T.~P., {Bell}, T.~J., {Ducrot}, E., {et~al.} 2023, \nat, 618, 39, \dodoi{10.1038/s41586-023-05951-7}

\bibitem[{{H{\"a}nni} {et~al.}(2024){H{\"a}nni}, {Altwegg}, {Combi}, {Fuselier}, {De Keyser}, {Ligterink}, {Rubin}, \& {Wampfler}}]{Hanni2024}
{H{\"a}nni}, N., {Altwegg}, K., {Combi}, M., {et~al.} 2024, \apj, 976, 74, \dodoi{10.3847/1538-4357/ad8565}

\bibitem[{{Harrison} {et~al.}(2012){Harrison}, {Allen}, \& {Bernath}}]{Harrison2012}
{Harrison}, J.~J., {Allen}, N. D.~C., \& {Bernath}, P.~F. 2012, \jqsrt, 113, 2189, \dodoi{10.1016/j.jqsrt.2012.07.021}

\bibitem[{{He} {et~al.}(2020){He}, {H{\"o}rst}, {Lewis}, {Yu}, {Moses}, {McGuiggan}, {Marley}, {Kempton}, {Moran}, {Morley}, \& {Vuitton}}]{He2020}
{He}, C., {H{\"o}rst}, S.~M., {Lewis}, N.~K., {et~al.} 2020, Nature Astronomy, 4, 986, \dodoi{10.1038/s41550-020-1072-9}

\bibitem[{{Holmberg} \& {Madhusudhan}(2023)}]{holmberg2023}
{Holmberg}, M., \& {Madhusudhan}, N. 2023, \mnras, 524, 377, \dodoi{10.1093/mnras/stad1580}

\bibitem[{Holmberg \& Madhusudhan(2024)}]{holmberg_possible_2024}
Holmberg, M., \& Madhusudhan, N. 2024, Astronomy and Astrophysics, 683, L2, \dodoi{10.1051/0004-6361/202348238}

\bibitem[{{Holmberg} \& {Madhusudhan}(2024)}]{Holmberg2024}
{Holmberg}, M., \& {Madhusudhan}, N. 2024, \aap, 683, L2, \dodoi{10.1051/0004-6361/202348238}

\bibitem[{Horne(1986)}]{horne_optimal_1986}
Horne, K. 1986, Publications of the Astronomical Society of the Pacific, 98, 609, \dodoi{10.1086/131801}

\bibitem[{{Hu}(2021)}]{Hu_photo21}
{Hu}, R. 2021, \apj, 921, 27, \dodoi{10.3847/1538-4357/ac1789}

\bibitem[{{Innes} {et~al.}(2023){Innes}, {Tsai}, \& {Pierrehumbert}}]{innes_runaway_2023}
{Innes}, H., {Tsai}, S.-M., \& {Pierrehumbert}, R.~T. 2023, \apj, 953, 168, \dodoi{10.3847/1538-4357/ace346}

\bibitem[{{Kawauchi} {et~al.}(2022)}]{Kawauchi2022}
{Kawauchi}, K., {et~al.} 2022, Astron. Astrophys., 666, A4, \dodoi{10.1051/0004-6361/202243381}

\bibitem[{{Kempton} {et~al.}(2023){Kempton}, {Zhang}, {Bean}, {Steinrueck}, {Piette}, {Parmentier}, {Malsky}, {Roman}, {Rauscher}, {Gao}, {Bell}, {Xue}, {Taylor}, {Savel}, {Arnold}, {Nixon}, {Stevenson}, {Mansfield}, {Kendrew}, {Zieba}, {Ducrot}, {Dyrek}, {Lagage}, {Stassun}, {Henry}, {Barman}, {Lupu}, {Malik}, {Kataria}, {Ih}, {Fu}, {Welbanks}, \& {McGill}}]{Kempton2023}
{Kempton}, E. M.~R., {Zhang}, M., {Bean}, J.~L., {et~al.} 2023, \nat, 620, 67, \dodoi{10.1038/s41586-023-06159-5}

\bibitem[{{Kendrew} {et~al.}(2015){Kendrew}, {Scheithauer}, {Bouchet}, {Amiaux}, {Azzollini}, {Bouwman}, {Chen}, {Dubreuil}, {Fischer}, {Glasse}, {Greene}, {Lagage}, {Lahuis}, {Ronayette}, {Wright}, \& {Wright}}]{Kendrew2015}
{Kendrew}, S., {Scheithauer}, S., {Bouchet}, P., {et~al.} 2015, \pasp, 127, 623, \dodoi{10.1086/682255}

\bibitem[{{Khare} {et~al.}(1978){Khare}, {Sagan}, {Bandurski}, \& {Nagy}}]{Khare1978}
{Khare}, B.~N., {Sagan}, C., {Bandurski}, E.~L., \& {Nagy}, B. 1978, Science, 199, 1199, \dodoi{10.1126/science.199.4334.1199}

\bibitem[{{Kipping}(2013)}]{Kipping2013}
{Kipping}, D.~M. 2013, \mnras, 435, 2152, \dodoi{10.1093/mnras/stt1435}

\bibitem[{Kochanov {et~al.}(2019)Kochanov, Gordon, Rothman, Shine, Sharpe, Johnson, Wallington, Harrison, Bernath, Birk, Wagner, Bris, Bravo, \& Hill}]{dms_cs2_1}
Kochanov, R., Gordon, I., Rothman, L., {et~al.} 2019, Journal of Quantitative Spectroscopy and Radiative Transfer, \dodoi{10.1016/j.jqsrt.2019.04.001}

\bibitem[{Kreidberg(2015)}]{kreidberg_batman_2015}
Kreidberg, L. 2015, Publications of the Astronomical Society of the Pacific, 127, 1161, \dodoi{10.1086/683602}

\bibitem[{{Krissansen-Totton} {et~al.}(2018){Krissansen-Totton}, {Olson}, \& {Catling}}]{Krissansen-Totton2018}
{Krissansen-Totton}, J., {Olson}, S., \& {Catling}, D.~C. 2018, Science Advances, 4, eaao5747, \dodoi{10.1126/sciadv.aao5747}

\bibitem[{Leconte {et~al.}(2024)Leconte, Spiga, Clément, Guerlet, Selsis, Milcareck, Cavalié, Moreno, Lellouch, Carrión-González, Charnay, \& Lefèvre}]{leconte_3d_2024}
Leconte, J., Spiga, A., Clément, N., {et~al.} 2024, Astronomy \& Astrophysics, 686, A131, \dodoi{10.1051/0004-6361/202348928}

\bibitem[{{Leung} {et~al.}(2022){Leung}, {Schwieterman}, {Parenteau}, \& {Fauchez}}]{Leung2022}
{Leung}, M., {Schwieterman}, E.~W., {Parenteau}, M.~N., \& {Fauchez}, T.~J. 2022, \apj, 938, 6, \dodoi{10.3847/1538-4357/ac8799}

\bibitem[{{Li} {et~al.}(2015){Li}, {Gordon}, {Rothman}, {Tan}, {Hu}, {Kassi}, {Campargue}, \& {Medvedev}}]{Li2015}
{Li}, G., {Gordon}, I.~E., {Rothman}, L.~S., {et~al.} 2015, \apjs, 216, 15, \dodoi{10.1088/0067-0049/216/1/15}

\bibitem[{{Lim} {et~al.}(2023){Lim}, {Benneke}, {Doyon}, {MacDonald}, {Piaulet}, {Artigau}, {Coulombe}, {Radica}, {L'Heureux}, {Albert}, {Rackham}, {de Wit}, {Salhi}, {Roy}, {Flagg}, {Fournier-Tondreau}, {Taylor}, {Cook}, {Lafreni{\`e}re}, {Cowan}, {Kaltenegger}, {Rowe}, {Espinoza}, {Dang}, \& {Darveau-Bernier}}]{Lim2023}
{Lim}, O., {Benneke}, B., {Doyon}, R., {et~al.} 2023, \apjl, 955, L22, \dodoi{10.3847/2041-8213/acf7c4}

\bibitem[{{Luu} {et~al.}(2024){Luu}, {Yu}, {Glein}, {Innes}, {Aguichine}, {Krissansen-Totton}, {Moses}, {Tsai}, {Zhang}, {Truong}, \& {Fortney}}]{Luu2024}
{Luu}, C.~N., {Yu}, X., {Glein}, C.~R., {et~al.} 2024, \apjl, 977, L51, \dodoi{10.3847/2041-8213/ad9eb1}

\bibitem[{{Madhusudhan} {et~al.}(2023{\natexlab{a}}){Madhusudhan}, {Moses}, {Rigby}, \& {Barrier}}]{Madhusudhan_chem_2023}
{Madhusudhan}, N., {Moses}, J.~I., {Rigby}, F., \& {Barrier}, E. 2023{\natexlab{a}}, Faraday Discussions, 245, 80, \dodoi{10.1039/D3FD00075C}

\bibitem[{Madhusudhan {et~al.}(2020)Madhusudhan, Nixon, Welbanks, Piette, \& Booth}]{madhu2020}
Madhusudhan, N., Nixon, M.~C., Welbanks, L., Piette, A. A.~A., \& Booth, R.~A. 2020, The Astrophysical Journal, 891, L7, \dodoi{10.3847/2041-8213/ab7229}

\bibitem[{{Madhusudhan} {et~al.}(2021){Madhusudhan}, {Piette}, \& {Constantinou}}]{Madhusudhan2021}
{Madhusudhan}, N., {Piette}, A. A.~A., \& {Constantinou}, S. 2021, \apj, 918, 1, \dodoi{10.3847/1538-4357/abfd9c10.48550/arXiv.2108.10888}

\bibitem[{{Madhusudhan} {et~al.}(2023{\natexlab{b}}){Madhusudhan}, {Sarkar}, {Constantinou}, {Holmberg}, {Piette}, \& {Moses}}]{Madhusudhan_2023_K218b}
{Madhusudhan}, N., {Sarkar}, S., {Constantinou}, S., {et~al.} 2023{\natexlab{b}}, \apjl, 956, L13, \dodoi{10.3847/2041-8213/acf577}

\bibitem[{Madhusudhan {et~al.}(2023)Madhusudhan, Sarkar, Constantinou, Holmberg, Piette, \& Moses}]{madhusudhan_carbon-bearing_2023}
Madhusudhan, N., Sarkar, S., Constantinou, S., {et~al.} 2023, The Astrophysical Journal Letters, 956, L13, \dodoi{10.3847/2041-8213/acf577}

\bibitem[{Madhusudhan \& Seager(2009)}]{Madhusudhan2009}
Madhusudhan, N., \& Seager, S. 2009, \apj, 707, 24, \dodoi{10.1088/0004-637x/707/1/24}

\bibitem[{{May} {et~al.}(2023){May}, {MacDonald}, {Bennett}, {Moran}, {Wakeford}, {Peacock}, {Lustig-Yaeger}, {Highland}, {Stevenson}, {Sing}, {Mayorga}, {Batalha}, {Kirk}, {L{\'o}pez-Morales}, {Valenti}, {Alam}, {Alderson}, {Fu}, {Gonzalez-Quiles}, {Lothringer}, {Rustamkulov}, \& {Sotzen}}]{May2023}
{May}, E.~M., {MacDonald}, R.~J., {Bennett}, K.~A., {et~al.} 2023, \apjl, 959, L9, \dodoi{10.3847/2041-8213/ad054f}

\bibitem[{{Meadows} {et~al.}(2022){Meadows}, {Graham}, {Abrahamsson}, {Adam}, {Amador-French}, {Arney}, {Barge}, {Barlow}, {Berea}, {Bose}, {Bower}, {Chan}, {Cleaves}, {Corpolongo}, {Currie}, {Domagal-Goldman}, {Dong}, {Eigenbrode}, {Enright}, {Fauchez}, {Fisk}, {Fricke}, {Fujii}, {Gangidine}, {Gezer}, {Glavin}, {Grenfell}, {Harman}, {Hatzenpichler}, {Hausrath}, {Henderson}, {Johnson}, {Jones}, {Hamilton}, {Hickman-Lewis}, {Jahnke}, {Kacar}, {Kopparapu}, {Kempes}, {Kish}, {Krissansen-Totton}, {Leavitt}, {Komatsu}, {Lichtenberg}, {Lindsay}, {Maggiori}, {Des Marais}, {Mathis}, {Morono}, {Neveu}, {Ni}, {Nixon}, {Olson}, {Parenteau}, {Perl}, {Quinn}, {Raj}, {Rodriguez}, {Rutter}, {Sandora}, {Schmidt}, {Schwieterman}, {Segura}, {Sekerci}, {Seyler}, {Smith}, {Soares}, {Som}, {Suzuki}, {Teece}, {Weber}, {Wolfe-Simon}, {Wong}, {Yano}, \& {Young}}]{Meadows2022}
{Meadows}, V., {Graham}, H., {Abrahamsson}, V., {et~al.} 2022, arXiv e-prints, arXiv:2210.14293, \dodoi{10.48550/arXiv.2210.14293}

\bibitem[{{Mikal-Evans} {et~al.}(2023)}]{Evans2023}
{Mikal-Evans}, T., {et~al.} 2023, Astron. J., 165, 84, \dodoi{10.3847/1538-3881/aca90b}

\bibitem[{{Montet} {et~al.}(2015){Montet}, {Morton}, {Foreman-Mackey}, {Johnson}, {Hogg}, {Bowler}, {Latham}, {Bieryla}, \& {Mann}}]{Montet2015}
{Montet}, B.~T., {Morton}, T.~D., {Foreman-Mackey}, D., {et~al.} 2015, \apj, 809, 25, \dodoi{10.1088/0004-637X/809/1/25}

\bibitem[{{Moran} {et~al.}(2023){Moran}, {Stevenson}, {Sing}, {MacDonald}, {Kirk}, {Lustig-Yaeger}, {Peacock}, {Mayorga}, {Bennett}, {L{\'o}pez-Morales}, {May}, {Rustamkulov}, {Valenti}, {Adams Redai}, {Alam}, {Batalha}, {Fu}, {Gonzalez-Quiles}, {Highland}, {Kruse}, {Lothringer}, {Ortiz Ceballos}, {Sotzen}, \& {Wakeford}}]{moran_high_2023}
{Moran}, S.~E., {Stevenson}, K.~B., {Sing}, D.~K., {et~al.} 2023, \apjl, 948, L11, \dodoi{10.3847/2041-8213/accb9c}

\bibitem[{{Morrison} {et~al.}(2023){Morrison}, {Dicken}, {Argyriou}, {Ressler}, {Gordon}, {Regan}, {Cracraft}, {Rieke}, {Engesser}, {Alberts}, {Alvarez-Marquez}, {Colbert}, {Fox}, {Gasman}, {Law}, {Garcia Marin}, {G{\'a}sp{\'a}r}, {Guillard}, {Kendrew}, {Labiano}, {Laine}, {Noriega-Crespo}, {Shivaei}, \& {Sloan}}]{Morrison2023}
{Morrison}, J.~E., {Dicken}, D., {Argyriou}, I., {et~al.} 2023, \pasp, 135, 075004, \dodoi{10.1088/1538-3873/acdea6}

\bibitem[{M\"{u}ller {et~al.}(2005)M\"{u}ller, Schl\"{o}der, Stutzki, \& Winnewisser}]{ocs_3}
M\"{u}ller, H., Schl\"{o}der, F., Stutzki, J., \& Winnewisser, G. 2005, Journal of Molecular Structure, 742, 215, \dodoi{10.1016/j.molstruc.2005.01.027}

\bibitem[{Nikitin {et~al.}(2016)Nikitin, Dmitrieva, \& Gordon}]{ch3cl_2}
Nikitin, A., Dmitrieva, T., \& Gordon, I. 2016, Journal of Quantitative Spectroscopy and Radiative Transfer, 177, 49, \dodoi{10.1016/j.jqsrt.2016.03.007}

\bibitem[{{Orton} {et~al.}(2007){Orton}, {Gustafsson}, {Burgdorf}, \& {Meadows}}]{orton2007}
{Orton}, G.~S., {Gustafsson}, M., {Burgdorf}, M., \& {Meadows}, V. 2007, \icarus, 189, 544, \dodoi{10.1016/j.icarus.2007.02.003}

\bibitem[{{Piaulet} {et~al.}(2023)}]{Piaulet2023}
{Piaulet}, C., {et~al.} 2023, Nat. Astron., 7, 206, \dodoi{10.1038/s41550-022-01835-4}

\bibitem[{{Pierrehumbert}(2023)}]{Pierrehumbert2023}
{Pierrehumbert}, R.~T. 2023, \apj, 944, 20, \dodoi{10.3847/1538-4357/acafdf}

\bibitem[{Piette \& Madhusudhan(2020)}]{piette_temperature_2020}
Piette, A. A.~A., \& Madhusudhan, N. 2020, The Astrophysical Journal, 904, 154, \dodoi{10.3847/1538-4357/abbfb1}

\bibitem[{{Pinhas} {et~al.}(2019){Pinhas}, {Madhusudhan}, {Gandhi}, \& {MacDonald}}]{pinhas2019}
{Pinhas}, A., {Madhusudhan}, N., {Gandhi}, S., \& {MacDonald}, R. 2019, \mnras, 482, 1485, \dodoi{10.1093/mnras/sty2544}

\bibitem[{{Powell} {et~al.}(2024){Powell}, {Feinstein}, {Lee}, {Zhang}, {Tsai}, {Taylor}, {Kirk}, {Bell}, {Barstow}, {Gao}, {Bean}, {Blecic}, {Chubb}, {Crossfield}, {Jordan}, {Kitzmann}, {Moran}, {Morello}, {Moses}, {Welbanks}, {Yang}, {Zhang}, {Ahrer}, {Bello-Arufe}, {Brande}, {Casewell}, {Crouzet}, {Cubillos}, {Demory}, {Dyrek}, {Flagg}, {Hu}, {Inglis}, {Jones}, {Kreidberg}, {L{\'o}pez-Morales}, {Lagage}, {Meier Vald{\'e}s}, {Miguel}, {Parmentier}, {Piette}, {Rackham}, {Radica}, {Redfield}, {Stevenson}, {Wakeford}, {Aggarwal}, {Alam}, {Batalha}, {Batalha}, {Benneke}, {Berta-Thompson}, {Brady}, {Caceres}, {Carter}, {D{\'e}sert}, {Harrington}, {Iro}, {Line}, {Lothringer}, {MacDonald}, {Mancini}, {Molaverdikhani}, {Mukherjee}, {Nixon}, {Oza}, {Palle}, {Rustamkulov}, {Sing}, {Steinrueck}, {Venot}, {Wheatley}, \& {Yurchenko}}]{Powell2024}
{Powell}, D., {Feinstein}, A.~D., {Lee}, E. K.~H., {et~al.} 2024, \nat, 626, 979, \dodoi{10.1038/s41586-024-07040-9}

\bibitem[{{Raulin} \& {Toupance}(1975)}]{Raulin1975}
{Raulin}, F., \& {Toupance}, G. 1975, Origins of Life, 6, 91, \dodoi{10.1007/BF01372393}

\bibitem[{{Reed} {et~al.}(2024){Reed}, {Shearer}, {McGlynn}, {Wing}, {Tolbert}, \& {Browne}}]{Reed2024}
{Reed}, N.~W., {Shearer}, R.~L., {McGlynn}, S.~E., {et~al.} 2024, \apjl, 973, L38, \dodoi{10.3847/2041-8213/ad74da}

\bibitem[{{Reed} \& {Hodges}(2015)}]{Reed2015}
{Reed}, Z.~D., \& {Hodges}, J.~T. 2015, \jqsrt, 159, 87, \dodoi{10.1016/j.jqsrt.2015.03.010}

\bibitem[{R\'{e}galia-Jarlot {et~al.}(2002)R\'{e}galia-Jarlot, Hamdouni, Thomas, der Heyden, \& Barbe}]{ocs_7}
R\'{e}galia-Jarlot, L., Hamdouni, A., Thomas, X., der Heyden, P.~V., \& Barbe, A. 2002, Journal of Quantitative Spectroscopy and Radiative Transfer, 74, 455, \dodoi{10.1016/S0022-4073(01)00267-9}

\bibitem[{{Richard} {et~al.}(2012){Richard}, {Gordon}, {Rothman}, {Abel}, {Frommhold}, {Gustafsson}, {Hartmann}, {Hermans}, {Lafferty}, {Orton}, {Smith}, \& {Tran}}]{richard2012}
{Richard}, C., {Gordon}, I.~E., {Rothman}, L.~S., {et~al.} 2012, \jqsrt, 113, 1276, \dodoi{10.1016/j.jqsrt.2011.11.004}

\bibitem[{{Rigby} {et~al.}(2024){Rigby}, {Pica-Ciamarra}, {Holmberg}, {Madhusudhan}, {Constantinou}, {Schaefer}, {Deng}, {Lee}, \& {Moses}}]{Rigby_towards}
{Rigby}, F.~E., {Pica-Ciamarra}, L., {Holmberg}, M., {et~al.} 2024, \apj, 975, 101, \dodoi{10.3847/1538-4357/ad6c38}

\bibitem[{{Rothman} {et~al.}(2010){Rothman}, {Gordon}, {Barber}, {Dothe}, {Gamache}, {Goldman}, {Perevalov}, {Tashkun}, \& {Tennyson}}]{Rothman2010}
{Rothman}, L.~S., {Gordon}, I.~E., {Barber}, R.~J., {et~al.} 2010, \jqsrt, 111, 2139, \dodoi{10.1016/j.jqsrt.2010.05.001}

\bibitem[{{Rubin} {et~al.}(2019){Rubin}, {Altwegg}, {Balsiger}, {Berthelier}, {Combi}, {De Keyser}, {Drozdovskaya}, {Fiethe}, {Fuselier}, {Gasc}, {Gombosi}, {H{\"a}nni}, {Hansen}, {Mall}, {R{\`e}me}, {Schroeder}, {Schuhmann}, {S{\'e}mon}, {Waite}, {Wampfler}, \& {Wurz}}]{Rubin2019}
{Rubin}, M., {Altwegg}, K., {Balsiger}, H., {et~al.} 2019, \mnras, 489, 594, \dodoi{10.1093/mnras/stz2086}

\bibitem[{{Sagan} \& {Khare}(1971)}]{Sagan1971}
{Sagan}, C., \& {Khare}, B.~N. 1971, \apj, 168, 563, \dodoi{10.1086/151109}

\bibitem[{{Sarkar} {et~al.}(2024){Sarkar}, {Madhusudhan}, {Constantinou}, \& {Holmberg}}]{Sarkar2024}
{Sarkar}, S., {Madhusudhan}, N., {Constantinou}, S., \& {Holmberg}, M. 2024, \mnras, 531, 2731, \dodoi{10.1093/mnras/stae1230}

\bibitem[{{Sarkis} {et~al.}(2018){Sarkis}, {Henning}, {K{\"u}rster}, {Trifonov}, {Zechmeister}, {Tal-Or}, {Anglada-Escud{\'e}}, {Hatzes}, {Lafarga}, {Dreizler}, {Ribas}, {Caballero}, {Reiners}, {Mallonn}, {Morales}, {Kaminski}, {Aceituno}, {Amado}, {B{\'e}jar}, {Hagen}, {Jeffers}, {Quirrenbach}, {Launhardt}, {Marvin}, \& {Montes}}]{sarkis2018}
{Sarkis}, P., {Henning}, T., {K{\"u}rster}, M., {et~al.} 2018, \aj, 155, 257, \dodoi{10.3847/1538-3881/aac108}

\bibitem[{{Scarsdale} {et~al.}(2024){Scarsdale}, {Wogan}, {Wakeford}, {Wallack}, {Batalha}, {Alderson}, {Aguichine}, {Wolfgang}, {Teske}, {Moran}, {L{\'o}pez-Morales}, {Kirk}, {Gordon}, {Gao}, {Batalha}, {Alam}, \& {Adams Redai}}]{Scarsdale2024}
{Scarsdale}, N., {Wogan}, N., {Wakeford}, H.~R., {et~al.} 2024, \aj, 168, 276, \dodoi{10.3847/1538-3881/ad73cf}

\bibitem[{{Scheucher} {et~al.}(2020){Scheucher}, {Wunderlich}, {Grenfell}, {Godolt}, {Schreier}, {Kappel}, {Haus}, {Herbst}, \& {Rauer}}]{Scheucher2020}
{Scheucher}, M., {Wunderlich}, F., {Grenfell}, J.~L., {et~al.} 2020, ApJ, 898, 44

\bibitem[{{Schlawin} {et~al.}(2024){Schlawin}, {Mukherjee}, {Ohno}, {Bell}, {Beatty}, {Greene}, {Line}, {Challener}, {Parmentier}, {Fortney}, {Rauscher}, {Wiser}, {Welbanks}, {Murphy}, {Edelman}, {Batalha}, {Moran}, {Mehta}, \& {Rieke}}]{Schlawin2024}
{Schlawin}, E., {Mukherjee}, S., {Ohno}, K., {et~al.} 2024, \aj, 168, 104, \dodoi{10.3847/1538-3881/ad58e0}

\bibitem[{{Schwieterman} \& {Leung}(2024)}]{Schwieterman2024}
{Schwieterman}, E.~W., \& {Leung}, M. 2024, Reviews in Mineralogy and Geochemistry, 90, 465, \dodoi{10.2138/rmg.2024.90.13}

\bibitem[{{Schwieterman} {et~al.}(2018){Schwieterman}, {Kiang}, {Parenteau}, {Harman}, {DasSarma}, {Fisher}, {Arney}, {Hartnett}, {Reinhard}, {Olson}, {Meadows}, {Cockell}, {Walker}, {Grenfell}, {Hegde}, {Rugheimer}, {Hu}, \& {Lyons}}]{Schwieterman2018}
{Schwieterman}, E.~W., {Kiang}, N.~Y., {Parenteau}, M.~N., {et~al.} 2018, Astrobiology, 18, 663, \dodoi{10.1089/ast.2017.1729}

\bibitem[{{Seager} {et~al.}(2013{\natexlab{a}}){Seager}, {Bains}, \& {Hu}}]{Seager2013a}
{Seager}, S., {Bains}, W., \& {Hu}, R. 2013{\natexlab{a}}, \apj, 775, 104, \dodoi{10.1088/0004-637X/775/2/104}

\bibitem[{{Seager} {et~al.}(2013{\natexlab{b}}){Seager}, {Bains}, \& {Hu}}]{Seager2013b}
---. 2013{\natexlab{b}}, \apj, 777, 95, \dodoi{10.1088/0004-637X/777/2/95}

\bibitem[{{Segura} {et~al.}(2005){Segura}, {Kasting}, {Meadows}, {Cohen}, {Scalo}, {Crisp}, {Butler}, \& {Tinetti}}]{segura2005}
{Segura}, A., {Kasting}, J.~F., {Meadows}, V., {et~al.} 2005, Astrobiology, 5, 706, \dodoi{10.1089/ast.2005.5.706}

\bibitem[{Sharpe {et~al.}(2004)Sharpe, Johnson, Sams, Chu, Rhoderick, \& Johnson}]{dms_cs2_2}
Sharpe, S.~W., Johnson, T.~J., Sams, R.~L., {et~al.} 2004, Applied Spectroscopy, 58, 1452, \dodoi{10.1366/0003702042641281}

\bibitem[{Shorttle {et~al.}(2024)Shorttle, Jordan, Nicholls, Lichtenberg, \& Bower}]{shorttle_distinguishing_2024}
Shorttle, O., Jordan, S., Nicholls, H., Lichtenberg, T., \& Bower, D.~J. 2024, The Astrophysical Journal Letters, 962, L8, \dodoi{10.3847/2041-8213/ad206e}

\bibitem[{{Skilling}(2004)}]{Skilling2004}
{Skilling}, J. 2004, in American Institute of Physics Conference Series, Vol. 735, Bayesian Inference and Maximum Entropy Methods in Science and Engineering: 24th International Workshop on Bayesian Inference and Maximum Entropy Methods in Science and Engineering, ed. R.~{Fischer}, R.~{Preuss}, \& U.~V. {Toussaint}, 395--405, \dodoi{10.1063/1.1835238}

\bibitem[{{Sousa-Silva} {et~al.}(2020){Sousa-Silva}, {Seager}, {Ranjan}, {Petkowski}, {Zhan}, {Hu}, \& {Bains}}]{sousa-silva2020}
{Sousa-Silva}, C., {Seager}, S., {Ranjan}, S., {et~al.} 2020, Astrobiology, 20, 235, \dodoi{10.1089/ast.2018.1954}

\bibitem[{Sung {et~al.}(2009)Sung, Toth, Brown, \& Crawford}]{ocs_5}
Sung, K., Toth, R., Brown, L., \& Crawford, T. 2009, Journal of Quantitative Spectroscopy and Radiative Transfer, 110, 2082, \dodoi{10.1016/j.jqsrt.2009.05.013}

\bibitem[{{Tashkun} {et~al.}(2015){Tashkun}, {Perevalov}, {Gamache}, \& {Lamouroux}}]{Tashkun2015}
{Tashkun}, S.~A., {Perevalov}, V.~I., {Gamache}, R.~R., \& {Lamouroux}, J. 2015, \jqsrt, 152, 45, \dodoi{10.1016/j.jqsrt.2014.10.017}

\bibitem[{Toth {et~al.}(2010)Toth, Sung, Brown, \& Crawford}]{ocs_6}
Toth, R., Sung, K., Brown, L., \& Crawford, T. 2010, Journal of Quantitative Spectroscopy and Radiative Transfer, 111, 1193, \dodoi{10.1016/j.jqsrt.2009.10.014}

\bibitem[{{Tsai} {et~al.}(2021){Tsai}, {Innes}, {Lichtenberg}, {Taylor}, {Malik}, {Chubb}, \& {Pierrehumbert}}]{Tsai2021}
{Tsai}, S.-M., {Innes}, H., {Lichtenberg}, T., {et~al.} 2021, \apjl, 922, L27, \dodoi{10.3847/2041-8213/ac399a}

\bibitem[{{Tsai} {et~al.}(2024){Tsai}, {Innes}, {Wogan}, \& {Schwieterman}}]{Tsai24_Bio}
{Tsai}, S.-M., {Innes}, H., {Wogan}, N.~F., \& {Schwieterman}, E.~W. 2024, \apjl, 966, L24, \dodoi{10.3847/2041-8213/ad3801}

\bibitem[{{Tsiaras} {et~al.}(2016){Tsiaras}, {Waldmann}, {Rocchetto}, {Varley}, {Morello}, {Damiano}, \& {Tinetti}}]{tsiaras_2016_pylightcurve}
{Tsiaras}, A., {Waldmann}, I.~P., {Rocchetto}, M., {et~al.} 2016, {pylightcurve: Exoplanet lightcurve model}, Astrophysics Source Code Library, record ascl:1612.018

\bibitem[{{Tsiaras} {et~al.}(2019){Tsiaras}, {Waldmann}, {Tinetti}, {Tennyson}, \& {Yurchenko}}]{Tsiaras2019}
{Tsiaras}, A., {Waldmann}, I.~P., {Tinetti}, G., {Tennyson}, J., \& {Yurchenko}, S.~N. 2019, Nature Astronomy, 451

\bibitem[{{Underwood} {et~al.}(2016){Underwood}, {Tennyson}, {Yurchenko}, {Huang}, {Schwenke}, {Lee}, {Clausen}, \& {Fateev}}]{Underwood2016}
{Underwood}, D.~S., {Tennyson}, J., {Yurchenko}, S.~N., {et~al.} 2016, \mnras, 459, 3890, \dodoi{10.1093/mnras/stw849}

\bibitem[{Vehtari {et~al.}(2017)Vehtari, Gelman, \& Gabry}]{Vehtari2017}
Vehtari, A., Gelman, A., \& Gabry, J. 2017, Statistics and computing, 27, 1413

\bibitem[{{VonNiederhausern} {et~al.}(2006){VonNiederhausern}, {Wilson}, \& {Giles}}]{VonNiederhausern2006}
{VonNiederhausern}, D.~M., {Wilson}, G.~M., \& {Giles}, N.~F. 2006, J. Chem. Eng. Data, 51, 1990

\bibitem[{{Vuitton} {et~al.}(2021){Vuitton}, {Moran}, {He}, {Wolters}, {Flandinet}, {Orthous-Daunay}, {Moses}, {Valenti}, {Lewis}, \& {H{\"o}rst}}]{Vuitton2021}
{Vuitton}, V., {Moran}, S.~E., {He}, C., {et~al.} 2021, \psj, 2, 2, \dodoi{10.3847/PSJ/abc558}

\bibitem[{{Wallack} {et~al.}(2024){Wallack}, {Batalha}, {Alderson}, {Scarsdale}, {Adams Redai}, {Aguichine}, {Alam}, {Gao}, {Wolfgang}, {Batalha}, {Kirk}, {L{\'o}pez-Morales}, {Moran}, {Teske}, {Wakeford}, \& {Wogan}}]{Wallack2024}
{Wallack}, N.~L., {Batalha}, N.~E., {Alderson}, L., {et~al.} 2024, \aj, 168, 77, \dodoi{10.3847/1538-3881/ad3917}

\bibitem[{{Welbanks} {et~al.}(2023){Welbanks}, {McGill}, {Line}, \& {Madhusudhan}}]{Welbanks2023}
{Welbanks}, L., {McGill}, P., {Line}, M., \& {Madhusudhan}, N. 2023, \aj, 165, 112, \dodoi{10.3847/1538-3881/acab67}

\bibitem[{{Welbanks} {et~al.}(2024){Welbanks}, {Bell}, {Beatty}, {Line}, {Ohno}, {Fortney}, {Schlawin}, {Greene}, {Rauscher}, {McGill}, {Murphy}, {Parmentier}, {Tang}, {Edelman}, {Mukherjee}, {Wiser}, {Lagage}, {Dyrek}, \& {Arnold}}]{Welbanks2024}
{Welbanks}, L., {Bell}, T.~J., {Beatty}, T.~G., {et~al.} 2024, \nat, 630, 836, \dodoi{10.1038/s41586-024-07514-w}

\bibitem[{{Winn} {et~al.}(2007){Winn}, {Holman}, {Bakos}, {P{\'a}l}, {Johnson}, {Williams}, {Shporer}, {Mazeh}, {Fernandez}, {Latham}, \& {Gillon}}]{Winn2007}
{Winn}, J.~N., {Holman}, M.~J., {Bakos}, G.~{\'A}., {et~al.} 2007, \aj, 134, 1707, \dodoi{10.1086/521599}

\bibitem[{{Wogan} {et~al.}(2024){Wogan}, {Batalha}, {Zahnle}, {Krissansen-Totton}, {Tsai}, \& {Hu}}]{wogan_jwst_2024}
{Wogan}, N.~F., {Batalha}, N.~E., {Zahnle}, K.~J., {et~al.} 2024, {JWST Observations of K2-18b Can Be Explained by a Gas-rich Mini-Neptune with No Habitable Surface}, \dodoi{10.3847/2041-8213/ad2616}

\bibitem[{{Yang} \& {Hu}(2024)}]{Yang_Hu_24}
{Yang}, J., \& {Hu}, R. 2024, \apjl, 971, L48, \dodoi{10.3847/2041-8213/ad6b25}

\bibitem[{{Yu} {et~al.}(2021){Yu}, {Moses}, {Fortney}, \& {Zhang}}]{Yu2021}
{Yu}, X., {Moses}, J.~I., {Fortney}, J.~J., \& {Zhang}, X. 2021, \apj, 914, 38, \dodoi{10.3847/1538-4357/abfdc7}

\bibitem[{{Yurchenko} {et~al.}(2011){Yurchenko}, {Barber}, \& {Tennyson}}]{Yurchenko2011}
{Yurchenko}, S.~N., {Barber}, R.~J., \& {Tennyson}, J. 2011, \mnras, 413, 1828, \dodoi{10.1111/j.1365-2966.2011.18261.x}

\bibitem[{{Yurchenko} \& {Tennyson}(2014)}]{Yurchenko2014}
{Yurchenko}, S.~N., \& {Tennyson}, J. 2014, \mnras, 440, 1649, \dodoi{10.1093/mnras/stu326}

\bibitem[{{Zahnle}(1996)}]{zahnle1996}
{Zahnle}, K. 1996, in IAU Colloq. 156: The Collision of Comet Shoemaker-Levy 9 and Jupiter, ed. K.~S. {Noll}, H.~A. {Weaver}, \& P.~D. {Feldman}, 183--212

\end{thebibliography}

\bibliographystyle{aasjournal}

\end{document}